\documentclass{vldb}
\usepackage{xcolor}
\usepackage{graphicx}
\usepackage{float}
\usepackage{xspace}
\usepackage{textcomp}
\usepackage{times}
\usepackage{mathptmx}
\usepackage[font={it}]{caption}
\usepackage[T1]{fontenc}
\usepackage{listings}
\usepackage{subcaption}
\usepackage{wrapfig}
\usepackage{balance}
\usepackage{paralist}

\usepackage{soul}
\usepackage{relsize}
\usepackage{makecell}
\usepackage{framed}
\usepackage{adjustbox}
\usepackage{pgfplots}
\usepackage{etoolbox}
\usepackage{xifthen}
\usepackage{quoting}
\usepackage{newfloat}
\usepackage{array}
\usepackage[normalem]{ulem}
\usepackage{tikz,array,calc}
\usepackage{cite}

\usetikzlibrary{decorations.pathreplacing}

\quotingsetup{vskip=.4em}
\newcolumntype{L}[1]{>{\raggedright\let\newline\\\arraybackslash\hspace{0pt}}m{#1}}
\newcolumntype{C}[1]{>{\centering\let\newline\\\arraybackslash\hspace{0pt}}m{#1}}
\newcolumntype{R}[1]{>{\raggedleft\let\newline\\\arraybackslash\hspace{0pt}}m{#1}}

\newcommand{\mc}[1]{(Metareview)}

\newcommand\cout{\bgroup\markoverwith{\textcolor{blue}{\rule[0.5ex]{2pt}{1pt}}}\ULon}

\newcommand{\agp}[1]{\noindent{\textcolor{blue}{Aditya: #1}}}
\newcommand{\tar}[1]{}

\newcommand{\alkim}[1]{}

\newcommand{\later}[1]{}

\newcommand{\papertext}[1]{{}}
\newcommand{\techreport}[1]{{#1}}
\newif\iftechreportcode

\newtoggle{paper}
\toggletrue{paper}

\renewcommand{\paragraph}[1]{\vspace{2pt} \noindent {\bf #1}}
\newcommand{\paraem}[1]{\vspace{0.5pt} \noindent {\bf {\em #1}}}

\newcommand{\anon}[1]{X}

\newcommand{\IN}{$<$--\xspace}
\newcommand{\TO}{--$>$\xspace}

\newcommand{\noopt}{{\sc No-Opt}\xspace}
\newcommand{\para}{{\sc Parallel}\xspace}
\newcommand{\spec}{{\sc Speculate}\xspace}
\newcommand{\fuse}{{\sc SmartFuse}\xspace}

\newcommand{\zv}{{\sf zenvisage}\xspace}
\newcommand{\zq}{{\sf ZQL}\xspace}
\newcommand{\sql}{{\sf SQL}\xspace}

\newcommand{\ggplot}{{\sf ggplot}\xspace}

\newcommand{\vzexp}{visual exploration\xspace}

\newcommand{\VzExp}{Visual Exploration\xspace}
\newcommand{\vzcmp}{\vzexp complete\xspace}
\newcommand{\VzCmpness}{\VzExp Completeness\xspace}
\newcommand{\vzalg}{\vzexp algebra\xspace}

\newcommand{\VzAlg}{\VzExp Algebra\xspace}
\newcommand{\vzlan}{\vzexp language\xspace}
\newcommand{\vzlans}{\vzexp languages\xspace}
\newcommand{\vzuni}{visual universe\xspace}
\newcommand{\vzgrp}{visual group\xspace}
\newcommand{\vzgrps}{visual groups\xspace}
\newcommand{\vzchn}{visual source\xspace}
\newcommand{\vzchns}{visual sources\xspace}

\newcommand{\wild}{$*$\xspace}

\newcommand{\vzsel}{\sigma^v}
\newcommand{\vzsort}{\tau^v}
\newcommand{\vzlimit}{\mu^v}

\newcommand{\vzdedup}{\delta^v}

\newcommand{\vzunion}{\cup^v}
\newcommand{\vzint}{\cap^v}
\newcommand{\vzdiff}{\setminus^v}
\newcommand{\vzswap}{\beta^v}
\newcommand{\vzdist}{\phi^v}
\newcommand{\vzfind}{\eta^v}

\newcommand{\zqname}{Name\xspace}
\newcommand{\zqproc}{Process\xspace}

\newcommand{\calR}{\mathcal{R}}

\newcommand{\calV}{\mathcal{V}}
\newcommand{\calX}{\mathcal{X}}
\newcommand{\calY}{\mathcal{Y}}

\newcommand{\cnode}[1][]{%
  \begingroup%
  \def\cnode{$c$-node}%
  \ifthenelse{\isin{s}{#1}}{%
    {\cnode}s%
    }{%
    \cnode%
  }%
  \endgroup%
}
\newcommand{\pnode}[1][]{%
  \begingroup%
  \def\pnode{$p$-node}%
  \ifthenelse{\isin{s}{#1}}{%
    {\pnode}s%
    }{%
    \pnode%
  }%
  \endgroup%
}

\newcommand{\axvar}[1][]{%
  \begingroup%
  \ifthenelse{\isin{c}{#1}}{%
    \def\axvar{Axis variable}%
    }{%
    \ifthenelse{\isin{C}{#1}}{%
      \def\axvar{Axis Variable}%
      }{%
      \ifthenelse{\isin{a}{#1}}{%
        \def\axvar{axvar}%
        }{
        \ifthenelse{\isin{A}{#1}}{%
          \def\axvar{av}%
          }{%
          \def\axvar{axis variable}%
        }%
      }%
    }%
  }%
  \ifthenelse{\isin{s}{#1}}{%
    {\axvar}s%
    }{%
    {\axvar}%
  }%
  \endgroup%
}

\newcommand{\nmvar}[1][]{%
  \begingroup%
  \ifthenelse{\isin{c}{#1}}{%
    \def\nmvar{Name variable}%
    }{%
    \ifthenelse{\isin{C}{#1}}{%
      \def\nmvar{Name Variable}%
      }{%
      \ifthenelse{\isin{a}{#1}}{%
        \def\nmvar{nmvar}%
        }{%
        \def\nmvar{name variable}%
      }%
    }%
  }%
  \ifthenelse{\isin{s}{#1}}{%
    {\nmvar}s%
    }{%
    \nmvar%
  }%
  \endgroup%
}

\newcommand{\aggr}[1][]{%
  \begingroup%
  \ifthenelse{\isin{c}{#1}}{%
    \def\nmvar{Aggregator}%
    }{%
    \ifthenelse{\isin{C}{#1}}{%
      \def\nmvar{Aggregator}%
      }{%
      \def\nmvar{aggregator}%
    }%
  }%
  \ifthenelse{\isin{s}{#1}}{%
    {\nmvar}s%
    }{%
    \nmvar%
  }%
  \endgroup%
}

\DeclareMathAlphabet{\mathcal}{OMS}{cmsy}{m}{n}

\newenvironment{denselist}{
    \begin{list}{\small{$\bullet$}}%
    {\setlength{\itemsep}{0ex} \setlength{\topsep}{0ex}
    \setlength{\parsep}{0pt} \setlength{\itemindent}{0pt}
    \setlength{\leftmargin}{1.5em}
    \setlength{\partopsep}{0pt}}}%
{\end{list}}

\begin{document}

\newtheorem{definition}{Definition}[section]
\newtheorem{proposition}[definition]{Proposition}
\newtheorem{lemma}[definition]{Lemma}
\newtheorem{remark}[definition]{Remark}
\newtheorem{corollary}[definition]{Corollary}
\newtheorem{claim}[definition]{Claim}
\newtheorem{theorem}[definition]{Theorem}
\newtheorem{heuristic}[definition]{Heuristic}
\newtheorem{example}[definition]{Example}
\newtheorem{dimension}{Dimension}
\newcounter{prob}
\newtheorem{problem}[prob]{Problem}
\newtheorem{simplealg}{Algorithm}
\numberofauthors{1}

\renewcommand{\baselinestretch}{0.97}

\author{
} 

\title{Effortless Data Exploration with {\Huge  \textbf{\sf zenvisage}}: \\ An Expressive and Interactive Visual Analytics System}

\author{
   \begin{tabular*}{\textwidth}{ccccc}
     \hspace{0pt}Tarique Siddiqui$^1$\thanks{\scriptsize Both authors contributed equally
     to this work.} &
     \hspace{0pt}Albert Kim$^2\footnotemark[1]$ &
     \hspace{0pt}John Lee$^1$ &
     \hspace{0pt}Karrie Karahalios$^{1,3}$ &
     \hspace{0pt}Aditya Parameswaran$^1$ 
   \end{tabular*}\\[2pt]
   \begin{tabular}{cc}
     $^1$\affaddr{University of Illinois, Urbana-Champaign (UIUC)} \ \ \  $^3$\affaddr{Adobe Inc.}&
     $^2$\affaddr{MIT} \\
    \affaddr{\{tsiddiq2,lee98,kkarahal,adityagp\}@illinois.edu} &
     \affaddr{alkim@csail.mit.edu} \\
 \end{tabular}
}

\maketitle
\setcounter{page}{1}


\begin{abstract}
Data visualization is by far the most commonly used mechanism to explore
and extract insights from datasets, especially by novice data scientists.
And yet, current visual analytics tools are rather limited in their ability
to operate on collections of 
visualizations---by composing, filtering, comparing, and sorting them---to
find those that depict desired trends or patterns. 
The process of visual data exploration remains a 
tedious process of {\em trial-and-error}.
We propose \zv, a visual analytics platform for effortlessly finding desired visual patterns 
from large datasets. We introduce \zv's general purpose visual exploration language, \zq 
("zee-quel") for specifying the desired visual patterns, drawing from use-cases in a variety of domains, including biology, mechanical engineering, climate science,
and commerce. 
We formalize the expressiveness of
\zq via a \vzalg---an algebra on collections of visualizations---and 
demonstrate that \zq is as expressive as that algebra. 
\zv exposes an interactive front-end 
that supports the issuing 
of \zq queries, and also supports 
interactions that are ``short-cuts'' to certain
commonly used \zq queries.
To execute these queries, 
\zv uses a novel \zq graph-based query optimizer that leverages a suite of 
optimizations tailored to the goal of processing collections of visualizations
in certain pre-defined ways. 
Lastly, a user survey and study demonstrates that
data scientists are able to effectively use \zv to 
eliminate error-prone and tedious exploration
and directly identify desired visualizations.
\end{abstract}


\section{Introduction}
\label{sec:intro}
Interactive visualization tools, such as Tableau~\cite{tableau}
and Spotfire~\cite{spotfire}, have paved the way for the 
democratization of data exploration and data science. 
These tools have witnessed an ever-expanding user base---as a concrete example, 
Tableau's revenues last year were in the hundreds of millions of US Dollars and is 
expected to reach tens of billions soon~\cite{tableau-increases}.
Using such tools, or even tools like Microsoft Excel, the standard 
data analysis recipe is as follows:
the data scientists load a dataset into the tool,  
select visualizations to examine, 
study the results, and then repeat the process
until they find ones that match
their desired pattern or need. 
Thus, using this repeated process of manual examination, or {\em trial-and-error},
data scientists are able to formulate and test hypothesis, 
and derive insights. 
The key premise of this work is that 
 to find
desired patterns in datasets, 
{\bf \em manual examination of each visualization in a collection  is 
simply unsustainable}, especially
on large, complex datasets.
Even on moderately sized datasets, a data scientist may need to examine as many as 
tens of thousands of visualizations, all to test a single hypothesis, 
a severe impediment to data exploration.

To illustrate, we describe the challenges of 
several collaborator groups who have been hobbled by the ineffectiveness of current 
data exploration tools:

\paraem{Case Study 1: Engineering Data Analysis.}
Battery scientists at Carnegie Mellon University
perform visual exploration of datasets of solvent
properties to design better batteries.
A specific task may involve finding solvents
with desired behavior: e.g.,
those whose 
solvation energy
of Li$^+$ vs.~the boiling point is a 
{\em roughly increasing trend}. 
To do this using current tools, 
these scientists manually examine
the plot of Li$^+$ solvation energy vs.~boiling point for each of the thousands of solvents, to find those that match 
the desired pattern of a roughly increasing trend.

\paraem{Case Study 2: Advertising Data Analysis.}
Advertisers at ad analytics
firm Turn, Inc., 
often examine their portfolio of advertisements to see
if their campaigns are performing as expected.
For instance, an advertiser may be interested in seeing if there are any keywords
that are {\em behaving unusually} with respect to other keywords in Asia---for example,
maybe most keywords have a specific trend for 
click-through rates (CTR) over time, 
while a small number of them
have a different trend.
To do this using the current tools available at Turn, 
the advertiser needs to manually examine the plots of 
CTR over time for each 
keyword (thousands of such plots),
and remember what are the typical trends. 

\paraem{Case Study 3: Genomic Data Analysis.}
Clinical researchers at the NIH-funded genomics center
at UIUC and Mayo Clinic 
are interested in studying 
data from clinical trials.
One such task involves finding pairs of genes
that {\em visually explain the differences} 
in clinical trial outcomes (positive vs.~negative)---visualized 
via a scatterplot with the x- and y- axes each referring
to a gene, and each outcome depicted as a point 
in the scatterplot---with the positive outcomes
depicted in one color, and the negative ones as another.
Current tools require the researchers to
generate and manually evaluate tens of thousands of scatter plots of pairs of
genes to determine whether
the outcomes can be clearly
distinguished in the scatter plot.

\techreport{
\paraem{Case Study 4: Environmental Data Analysis.}
Climate scientists at the National Center for
Supercomputing Applications at Illinois are interested
in studying the nutrient and water property readings on sensors within buoys
at various locations in the Great Lakes. 
Often, they find that a sensor is displaying unusual behavior
for a specific property, and want
to figure out {\em what is different} about this sensor relative to others,
and if other properties for this sensor are {\em showing similar behavior}.
In either case, the scientists would need to separately examine each
property for each sensor (in total 100s of thousands of visualizations)
to identify explanations or similarities.

\paraem{Case Study 5: Server Monitoring Analysis.}
 The server monitoring team at Facebook has noticed a spike in the
per-query response time for Image Search in Russia on August 15, after which the response time flattened out.
The team would like to identify if there are {\em other attributes 
that have a similar behavior} with
per-query response time, which may indicate the reason for the spike and subsequent flattening.
To do this, the server monitoring team generates visualizations for
different metrics as a function of the date, and assess if any of them has similar behavior
to the response time for Image Search. Given that the number of metrics is likely 
in the thousands, this takes a very long time.

\paraem{Case Study 6: Mobile App Analysis.} 
The complaints section of the Google mobile platform team 
have noticed that a certain mobile app has 
received many complaints.
They would like to figure out {\em what is different} about this app relative to others.
To do this, they need to plot various metrics for this app to figure out 
why it is behaving anomalously. 
For instance, they may look at network traffic generated by this app over time,
or at the distribution of energy consumption across different users.
In all of these cases, the team would need to generate several visualizations manually 
and browse through all of them in the hope of finding what could be the issues with the app.
}

\vspace{2pt}
\noindent 
Thus, in these examples, the recurring theme is 
the manual examination of a large number of generated visualizations for a specific visual pattern.
Indeed, we have found that in these scenarios\papertext{, as well as others that we have encountered 
via other collaborators---in {\em climate science, server monitoring, and mobile app analysis}}---data exploration 
can be a tedious and time-consuming process with current
visualization tools.

\paragraph{Key Insight.} The goal of this paper is to develop \zv, a visual analytics
system that can {\bf \em automate the search for desired
visual patterns}.
Our key insight in developing \zv is that the data exploration 
needs in all of these scenarios 
can be captured within a common set of operations
on collections of visualizations.
These operations include:
{\em composing}  
collections of visualizations,
{\em filtering} visualizations
based on some conditions,
{\em comparing} visualizations,
and {\em sorting} them based on some condition.
The conditions include 
similarity or dissimilarity to a specific pattern, 
``typical'' or anomalous behavior,
or the ability to provide 
explanatory or discriminatory power.
These operations and conditions form the kernel of a 
new data exploration language, \zq ("zee-quel"), that forms
the foundation upon which  \zv is built.

\paragraph{Key Challenges.} 
We encountered many challenges in building the \zv visual analytics platform, 
a substantial advancement over manually-intensive visualization
tools like Tableau and Spotfire; these tools
enable the examination of {\em one visualization at a time},
without the ability to automatically identify
relevant visualizations from a collection of visualizations.

First, there were many challenges in developing \zq,
the underlying query language for \zv.
Unlike relational query languages that operate
directly on data, \zq operates on collections of visualizations,
which are themselves aggregate queries on data. 
\techreport{Thus, in a sense \zq is a query language that operates
on other queries as a first class citizen.}
This leads to a number of challenges that are not addressed
in a relational query language context. For example,
we had to develop a natural way to users to specify
a collection of visualizations to operate on, without
having to explicitly list them;
even though the criteria on which the visualizations
were compared varied widely, we had to develop a 
small number of general mechanisms that capture all of these
criteria.
Often, the visualizations that we operated on had
to be modified in various ways---e.g., we
might be interested in visualizing the sales of
a product whose profits have been dropping---composing
these visualizations from existing ones is not straightforward.
Lastly, drilling down into specific visualizations
from a collection also required special care.
Our \zq language is a synthesis of desiderata
after discussions with data scientists
from a variety of domains, and has been under
development for the past two years. 
To further show that \zq is {\em complete} under
a new \vzalg that we develop involved additional challenges.

Second, in terms of front-end development, 
\zv, as an interactive analytics tool,
needs to support the ability for users to 
interactively specify \zq queries---specifically, 
interactive short-cuts for commonly used \zq queries, 
as well as the ability to pose extended \zq queries
for more complex needs.
Identifying common interaction ``idioms'' for these
needs took many months. 

Third,  
an important challenge in building \zv is
the back-end that supports the execution of \zq.
A single \zq query can lead to the generation of 10,000s of 
visualizations---executing each one 
independently as an aggregate query,
would take several hours, rendering the 
tool somewhat useless.
\techreport{(As it turns out, this time would be what an analyst aiming
to discover the same pattern would have to spend
with present visualization tools,
so the naive automation may still help reducing the amount of
manual effort.) }%
\zv's query optimizer 
operates as a wrapper
over any traditional relational database system. 
This query optimizer compiles \zq queries
down to a directed acyclic graph of 
operations on collections of visualizations,
followed with the optimizer
using 
a combination of intelligent speculation and combination, 
to issue queries to the underlying database. 
We also demonstrate that the underlying problem
is {\sc NP-Hard}.
Our query optimizer leads
to substantial improvements over the naive
schemes adopted within relational 
database systems for multi-query optimization.

\techreport{
\paragraph{Related Work.}
There are a number of
tools one could use for interactive analysis;
here, we briefly describe why those tools are inadequate 
for the important need of 
automating the search
for desired visual insights.
We describe related work in detail in Section~\ref{sec:related}.
To start, visualization tools like
Tableau and Spotfire only generate and provide one visualization
at a time, 
while \zv analyzes collections of visualizations at a time,
and identifies relevant ones from that collection---making it
substantially more powerful.
While we do use relational database systems
as a computation layer,
it is 
cumbersome to near-impossible to express
these user needs in \sql. 
As an example, finding visualizations of 
solvents for whom a given property
follows a roughly increasing trend is impossible to write
within native SQL, and would require custom UDFs---these UDFs 
would need to be hand-written for every \zq query.
\techreport{Similarly, finding visualizations of 
keywords where CTR over time in Asia is behaving
unusually with respect to other keywords 
is challenging to write within SQL. }%
For the small space of queries 
where it is possible to write
the queries within \sql 
these queries require non-standard constructs,
and are both complex and cumbersome to write,
even for expert \sql users, 
and are optimized very poorly
(see Section~\ref{sec:related}).
\techreport{It is also much more natural 
for end-users to operate directly on visualizations
than on data. 
Indeed, users who have never programmed or written
SQL before find it easy to understand and write 
a subset of \zq queries, as we
will show subsequently.} 
Statistical, 
data mining, and machine learning
 certainly provide functionality beyond \zv
in supporting prediction
and statistics; 
these functionalities are exposed as ``one-click'' algorithms
that can be applied on data.
However, no functionality is provided for searching for desired
patterns; no querying functionality beyond the one-click algorithms,
and no optimization. To use such tools for \zq, many
lines of code and hand-optimization is needed.
\techreport{As such, these tools are beyond the reach of
novice data scientists who simply want to explore 
and visualize their datasets.}
}

\paragraph{Outline.}
We first describe our
query language for \zv, \zq (Section~\ref{sec:querylanguage})%
\techreport{, and formalize the notion of a {\em \vzalg}, an analog of relational
algebra, describing a core set of capabilities for any 
language that supports visual data exploration and
demonstrate that \zq is {\em complete} in that
it subsumes these capabilities (Section~\ref{sec:formal-expr})}.
We then describe the graph-based query translator and optimizer for \zq
(Section~\ref{subsec:query_execution}).
Next, our initial prototype of \zv is presented (Section~\ref{sec:architecture}).
We also describe our performance experiments
(Section~\ref{sec:experiments}), and
present a user survey and study focused on evaluating the 
effectiveness and usability of \zv (Section~\ref{sec:userstudy}).
\techreport{In the appendix, we present additional details of 
our query language,
along with complete examples, 
and additional details on user study.}%
\papertext{%
Lastly, we describe how \zv differs from related work (Section~\ref{sec:related}).

In our extended technical report~\cite{techreport}, we provide
additional details that we weren't able to fit into the paper.
In particular, we formalize
the notion of a {\em \vzalg}, an analog of relational
algebra, describing a core set of capabilities for any language that supports visual
 data exploration, and demonstrate that \zq is {\em complete} in that
it subsumes these capabilities.
 We also provide additional details of our query language.}

\iftoggle{paper}{
\vspace{-6pt}
\section{Query Language}
\label{sec:querylanguage}
\vspace{-2pt}
\zv's query language, \zq, provides users with a powerful mechanism to operate 
on collections of visualizations. 
In fact, \zq treats visualizations as a {\em first-class citizen}, 
enabling users to operate at a high level on collections of visualizations
much like one would operate on relational data with \sql.
For example, a user may want to filter out all visualizations 
where the visualization shows a roughly decreasing trend from a collection,
or 
a user may want to create a collection of visualizations which are most
similar to a visualization of interest. 
Regardless of the query, \zq
provides an intuitive, yet flexible specification mechanism for
users to express the desired patterns of interest (in other words, their {\em exploration needs})
using a small number
of \zq lines. 
Overall, \zq provides users the ability to compose collections of visualizations,
filter them, and sort and compare them in various ways. 

\zq draws heavy inspiration from the
Query by Example (QBE) language~\cite{qbe} and uses a similar table-based
specification interface. 
Although \zq components are not fundamentally
tied to the tabular interface, we found that our end-users felt
more at home with it; many of them are
non-programmers who are used to spreadsheet tools like Microsoft
Excel.
Users may either directly write \zq, or they may use the \zv front-end, which
supports interactions that are transformed internally into \zq.

We now provide a formal introduction to \zq in the rest of this section. We introduce
many sample queries to make it easy to follow along, and we use a relatable
fictitious product sales-based dataset throughout this paper in our
query examples---we will reveal attributes of
this dataset as we go along.

\vspace{-4pt}
\subsection{Formalization}
\label{subsec:formalization}
\vspace{-3pt}
For describing \zq, we assume that we are operating on a single relation
or a star schema where the attributes are unique
(barring key-foreign key joins), allowing \zq to seamlessly support
  natural joins.
In general, \zq could be applied to arbitrary collections
of relations by letting the user precede an attribute $A$
with the relation name $R$, e.g., $R.A$.
For ease of exposition, we focus on the single relation case.

\vspace{-2pt}
\subsubsection{Overview}
\vspace{-3pt}
\begin{figure}
\vspace{-10pt}
  \centering
  \begin{tikzpicture}[scale=0.6]
    \begin{axis}[
        ylabel=Sales (million \$),
        ylabel near ticks,
        yscale=0.3,
        ybar,
        /pgf/number format/1000 sep={},
        xtick pos =left,
        ytick pos=left
      ]
      \addplot
        coordinates {
          (2012, 50)
          (2013, 33)
          (2014, 40)
          (2015, 50)
          (2016, 70)
        };
    \end{axis}
  \end{tikzpicture}
  \vspace{-12pt}
  \caption{Sales over year visualization for the product chair.}
  \vspace{-18pt}
  \label{fig:sample-viz}
\end{figure}
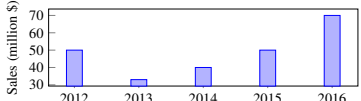

\paragraph{The concept of visualizations.} We start by defining the notion of a visualization.
We use a sample visualization in Figure~\ref{fig:sample-viz} to guide our discussion.
Of course, different visual
analysis tasks may require different types of visualizations (instead of bar charts, 
we may want scatter plots or trend lines), but across all types,
a visualization is defined by the following five main components:
\begin{inparaenum}[(i)]
\item the x-axis attribute,
\item the y-axis attribute,
\item the subset of data used,
\item the type of visualization (e.g., bar chart, scatter plot), and
\item the binning and aggregation functions for the x- and y- axes.
\end{inparaenum}

\begin{table}[!]
\vspace{-6pt}
  \centering\scriptsize
  {\tt
    \begin{tabular}{|r|l|l|l|l|}
      \hline
      \zqname & X & Y & Z & Viz \\ \hline
      *f1 & `year' & `sales' & `product'.`chair' & bar.(y=agg(`sum')) \\ \hline
    \end{tabular}
  }
  \vspace{-7pt}
  \caption{Query for the bar chart of sales over year for the
    product chair.}
  \label{tab:overview-ex}
  \vspace{-5pt}
\end{table}

\begin{table}[!]
\vspace{-6pt}
  \centering\scriptsize
  {\tt
    \begin{tabular}{|r|l|l|l|l|}
      \hline
      \zqname & X & Y & Z & Viz \\ \hline
      *f1 & `year' & `sales' & `product'.* & bar.(y=agg(`sum')) \\ \hline
    \end{tabular}
  }
  \vspace{-6pt}
  \caption{Query for the bar chart of sales over year for
  each product.}
  \label{tab:overview-ex2}
  \vspace{-18pt}
\end{table}

\paragraph{Visualization collections in \zq:} \zq has four columns to support the specification of
visualizations that the five aforementioned components map into:
\begin{inparaenum}[(i)]
\item {\em \textbf{X}},
\item {\em \textbf{Y}},
\item {\em \textbf{Z}}, and
\item {\em \textbf{Viz}}.
\end{inparaenum}

Table~\ref{tab:overview-ex} gives an example of a valid \zq query that uses
these columns to specify a bar chart visualization of overall sales
over the years for the product chair 
(i.e., the visualization in
Figure~\ref{fig:sample-viz})---ignore the Name column for now. 
The details for each of these columns 
are presented subsequently.
In short, the x-axis (\texttt{X}) is the attribute \texttt{year},
the y-axis (\texttt{Y}) is the attribute \texttt{sales},
and the subset of data (\texttt{Z}) is the product chair,
while the type of visualization is a bar chart (\texttt{bar}),
and the binning and aggregation functions indicate that the y axis is an aggregate (\texttt{agg}) --- the sum of sales.

In addition to specifying a single visualization, users may often want to
retrieve multiple visualizations. \zq supports this in two ways. Users
may use multiple rows, and specify one visualization per row. The user may also
specify a \emph{collection} of visualizations in a single row by iterating over
a collection of values for one of the X, Y, Z, and Viz columns.
Table~\ref{tab:overview-ex2} gives an example of how one may iterate over all products
(using the notation * to indicate that the attribute product can take
on all possible values),
returning a separate sales bar chart for each product.

\paragraph{High-level structure of \zq.} Starting from these two examples, we can now move onto
the general structure of \zq queries.
Overall, each \zq query consists of multiple rows, where each row operates on collections of visualizations.
Each row contains three sets of columns, as depicted in Table~\ref{tab:zq-breakdown}:
(i) the first column corresponds to an identifier for the visualization collection,
(ii) the second set of columns defines the visualization collection,
while (iii) the last column corresponds to some operation on 
the visualization collection.
All columns can be left empty if needed
(in such cases, to save space, for convenience, we do not display these columns in our paper). 
For example, the last column may be empty if no operation is to be 
performed, like it was in Table~\ref{tab:overview-ex} and~\ref{tab:overview-ex2}.
We have already discussed (ii); now we will briefly discuss (i) and (iii), corresponding
to {\em \textbf{\zqname}} and {\em \textbf{\zqproc}} respectively.

\paragraph{Identifiers and operations in \zq.} 
The {\em \textbf{\zqproc}} column allows the user to operate on the defined collections of
visualizations, applying high-level filtering, sorting, and comparison. 
The {\em \textbf{\zqname}} column provides a way to label and combine
specified
collections of
visualizations, so users may refer to them in the
\zqproc column. 
Thus, by repeatedly using the X, Y, Z,
and Viz columns to compose visualizations and the \zqproc column to process
those visualizations, the user is able derive the exact set of visualizations
she is looking for.
Note that the result of a \zq query is the \emph{data} used to
generate visualizations. The \zv front-end then uses this data to render the
visualizations for the user to peruse.
\vspace{-2pt}
\begin{table}
\vspace{-15pt}
  \centering\scriptsize
  {\tt
    \begin{tabular}{|r|l|l|l|l|r|}
      \hline
      \zqname & X & Y & Z & Viz & \zqproc \\ \hline
      \vspace{10pt}
      $\overbrace{\textrm{Identifier}}$ & \multicolumn{4}{l|}{$\overbrace{\textrm{Visualization Collection}}$} & $\overbrace{\textrm{Operation}}$ \\
    \end{tabular}
  }
  \vspace{-15pt}
  \caption{\zq query structure.}
  \label{tab:zq-breakdown}
  \vspace{-15pt}
\end{table}

\vspace{-5pt}
\subsubsection{X, Y, and Z}
The X and Y columns specify the attributes used for the x- and y- axes. 
For example, Table~\ref{tab:overview-ex} dictates that the returned
visualization should
have `year' for its x-axis and `sales' for its y-axis.
As mentioned, the user
may also specify a collection of values for the X and Y columns if they wish to
refer to a collection of visualizations in one \zq row. Table~\ref{tab:xy-coll}
refers the collection of both sales-over-years and profit-over-years bar
charts for the chair---the missing values in this query (``...'') are the same as 
Table~\ref{tab:overview-ex}.
As we can see, a collection is constructed using \{\}. If the user wishes to denote
all possible values, the shorthand * symbol may be used, as is shown by
Table~\ref{tab:overview-ex2}. In the case that multiple columns contain
collections, a Cartesian product is performed, and visualizations for every
combination of values is returned. For example, Table~\ref{tab:xy-coll2} would
return the collection of visualizations with specifications:
{\tt \{\mbox{(X: `year', Y: `sales')}, (X: `year', Y: `profit'), (X: `month', Y:
`sales'), (X: `month', Y: `profit')\}}. \techreport{Additionally, \zq allows composing 
multiple attributes in the X and Y columns by supporting Polaris table 
algebra [4] over the operators: +, x, / (Appendix
\ref{querylang-appendix}).}

\begin{table}[!t]
  \vspace{-8pt}
  \centering\scriptsize
  \begin{adjustbox}{center}
    {\tt
      \begin{tabular}{|r|l|l|l|l|}
        \hline
        \zqname & X & Y & Z & Viz \\ \hline
        ... & ... & \{`sales', `profit'\} & ... & ... \\ \hline
      \end{tabular}
    }
  \end{adjustbox}
  \vspace{-8pt}
  \caption{Query for the sales and profit bar charts
  for the product chair (missing values are the same as that in Table~\ref{tab:overview-ex})}
  \label{tab:xy-coll}
\end{table}

\begin{table}[!t]
\vspace{-6pt}
  \centering\scriptsize
  \begin{adjustbox}{center}
    {\tt
      \begin{tabular}{|r|l|l|l|l|}
        \hline
        \zqname & X & Y & Z & Viz \\ \hline
        ... & \{`year', `month'\} & \{`sales', `profit'\} & ... & ... \\ \hline
      \end{tabular}
    }
  \end{adjustbox}
  \vspace{-8pt}
  \caption{Query for the sales and profit bar charts
    over years and months for chairs (missing values are the same as in Table~\ref{tab:overview-ex}).}
  \label{tab:xy-coll2}
  \vspace{-7pt}
\end{table}

\begin{table}[!t]
  \centering\scriptsize
  \begin{adjustbox}{center}
    {\tt
      \begin{tabular}{|r|l|l|l|l|l|}
        \hline
        \zqname & X & Y & Z & Z2 & Viz \\ \hline
        ... & ... & ... & ... & `location'.`US' & ... \\ \hline
      \end{tabular}
    }
  \end{adjustbox}
  \vspace{-8pt}
  \caption{Query which returns the overall sales bar chart
  for the chairs in US (all missing values are the same as that in Table~\ref{tab:overview-ex}).}
  \label{tab:z-mul-col}
  \vspace{-14pt}
  \end{table}

With the Z column, the user can select which subset of the data they wish to
construct their visualizations from.
\zq uses the
\mbox{\emph{$\langle$attribute$\rangle$.$\langle$attribute-value$\rangle$}}
notation
to denote
the selection of data.
Consequently,
the query in Table~\ref{tab:overview-ex}
declares that
the user wishes to retrieve the sales bar chart only for the
chair product.
\techreport{Note that unlike the X and Y columns, both the attribute and the
attribute value must be specified for the Z column; otherwise, a proper subset
of the data would not be identified.}%
Collections are allowed for both the attribute and the attribute value in the Z
column. Table~\ref{tab:overview-ex2} shows an example of using the * shorthand
to specify a collection of bar charts, one for each product.
A Z column which has a collection over attributes might look like:
\mbox{\tt \{`location', `product'\}.*}
(i.e., a visualization for every product and a visualization for every
location).
In addition, the Z column allows users to specify predicate constraints
using syntax like \mbox{\tt `weight'.[? < 10]}; this
specifies all items whose weight is less than 10 lbs. To evaluate, the ? is
replaced with the attribute and the resulting expression is passed to \sql's
{\tt WHERE} clause.
\techreport{The predicate constraint syntax has an analogous predicate collection syntax,
which creates a collection of the values which satisfy the condition.
\mbox{\tt `weight'.[? < 10]} specifies that the resulting visualizations must
only contains items with less than 10 lbs,  \mbox{\tt `weight'.\{? < 10\}}
creates a collection of values, one for each item which is less than 10 lbs. }%

\zq supports multiple constraints on different attributes through the use of multiple Z
columns. In addition to the basic Z column, the user may choose to add Z2,
Z3, ... columns depending on how many constraints she requires.
Table~\ref{tab:z-mul-col} gives an example of a query which looks at sales
plots for chairs only in the US. Note that Z columns are combined using
conjunctive semantics.
\vspace{-4pt}
\subsubsection{Viz}
  \vspace{-2pt}
The Viz column decides the visualization type, binning, and aggregation
functions for the row.
Elements in this column have the format:
\mbox{\emph{$\langle$type$\rangle$.$\langle$bin+aggr$\rangle$}}. 
All examples so far have been bar charts with no binning and {\tt SUM}
aggregation for the y-axis, but other variants
are supported.
The visualization types are derived from the Grammar of
Graphics~\cite{grammar-of-graphics}
specification language, so all plots from the geometric transformation layer of
\ggplot~\cite{ggplot} (the tool that implements Grammar of Graphics) are supported.
For instance, scatter plots are
requested with {\tt point} and heat maps with {\tt bin2d}.
As for binning, binning based on bin width ({\tt bin}) and number of bins
({\tt nbin}) are supported for numerical attributes---we may want to use 
binning, for example, when we are plotting the total number of products whose
prices lie within 0-10, 10-20, and so on. 

Finally, \zq supports all
the basic \sql aggregation functions such as {\tt AVG}, {\tt COUNT}, and {\tt
MAX}. Table~\ref{tab:viz} is an example of a query which uses a different
visualization type, heat map, and creates 20 bins for both x- and y- axes.

\begin{table}[!t]
  \centering\scriptsize
  \begin{adjustbox}{center}
    {\tt
      \begin{tabular}{|r|l|l|l|}
        \hline
        \zqname & X & Y & Viz \\ \hline
        *f1 & `weight' & `sales' & bin2d.(x=nbin(20), y=nbin(20)) \\ \hline
      \end{tabular}
    }
  \end{adjustbox}
  \vspace{-8pt}
  \caption{Query which returns the heat map of sales vs.~weights across all transactions.}
  \label{tab:viz}
    \vspace{-10pt}
\end{table}

\begin{table}[!t]
  \centering\scriptsize
  \begin{adjustbox}{center}
    {\tt
      \begin{tabular}{|r|l|l|l|}
        \hline
        \zqname & X & Y & Z \\ \hline
        f1 & `year' & `sales' & `product'.`chair' \\ \hline
        f2 & `year' & `profit' & `location'.`US' \\ \hline
        *f3 \IN f1 + f2 & & & `weight'.[? < 10] \\ \hline
      \end{tabular}
    }
  \end{adjustbox}
  \vspace{-8pt}
  \caption{Query which returns the sales for chairs or profits for US
    visualizations for all items less than 10 lbs.}
  \label{tab:name-plus}
      \vspace{-15pt}
\end{table}

\techreport{Like the earlier columns, the Viz column also allows collections of
values. Similar to the Z column, collections may be specified
for both the visualization type or the binning and aggregation. If the user wants to
view the same data binned at different granularities, she might specify a bar
chart with several different bin widths: \mbox{\tt bar.(x=\{bin(1), bin(5),
bin(10)\}, y=agg(`sum'))}. On the other hand, if the user wishes to view the
same data in different visualizations, she might write:
\mbox{\tt \{bar.(y=agg(`sum')), point.()\}}.}

The Viz column allows users powerful control over the structure of
the rendered visualization. However, there has been work from the visualization community
which automatically tries to determine the most appropriate visualization type,
binning, and aggregation for a
dataset based on the x- and y- axis
attributes~\cite{wongvoyager2015,Mackinlay:1986:ADG:22949.22950}. 
Thus, we can frequently leave the Viz
column blank and \zv will use these rules of thumb to automatically decide the appropriate setting for us. With this in
mind, we omit the Viz column from the remaining examples with the assumption
that \zv will determine the ``{best}'' visualization structure for us.

\vspace{-5pt}
\subsubsection{\zqname}
\vspace{-2pt}
Together, the values in the X, Y, Z, and Viz columns of each row specify a
collection of visualizations.  The \zqname column allows us to label these
collections so that they can be referred to be in the \zqproc column. For
example, {\tt f1} is the label or identifier given to the collection of sales bar charts in
Table~\ref{tab:overview-ex2}. The * in front of {\tt f1} signifies that 
the collection is an output collection; that is, \zq should return this
collection of visualizations to the user.

However, not all rows need to have a * associated with their \zqname identifier.
A user may define intermediate
collections of visualizations if she wishes to further process them in the \zqproc
column before returning the final results. 
In the case of Table~\ref{tab:name-plus},
{\tt f1} and {\tt f2} are examples of intermediate collections. 

Also in Table~\ref{tab:name-plus}, we have an example of how
the \zqname column allows us to perform high-level set-like operations
to combine visualization collections directly.
For example, {\tt f3 \IN f1 + f2} assigns {\tt f3} to the
collection which includes all visualizations in {\tt f1} and {\tt f2} (similar
to set union). 
This can be useful if the user wishes to combine
variations of values without considering the full Cartesian product.
In our example in Table~\ref{tab:name-plus}, the user is able to combine the sales for chairs plots
with the profits for the US plots without also having to consider the sales for the
US plots or the profits for chairs plots; she would have to do so if she had used
the specification: {\tt \mbox{(Y: \{`sales', `profit'\}}, \mbox{Z: \{`product'.`chair',
`location'.`US'\}})}. 

An interesting aspect of Table~\ref{tab:name-plus}
is that the X and Y columns of the third row are devoid of values, and the Z
column refer to the seemingly unrelated weight attribute. The values in
the X, Y, Z, and Viz columns all help to specify a particular collection of
visualizations from a larger collection. 
When this collection is defined via the \zqname column, 
we no longer need to fill in the values for X, Y, Z, or Viz, except to select from
the collection---here, \zq only selects the items which satisfy the
constraint, {\tt weight < 10}.  

\techreport{Other set-like operators include
{\tt f1 - f2} for set minus and {\tt f1 \^{} f2} for intersection.}

\begin{table*}[t]
\vspace{-1pt}
  \centering\scriptsize
  \begin{adjustbox}{center}
    {\tt
      \begin{tabular}{|r|l|l|l|l|}
        \hline
        \zqname & X & Y & Z & \zqproc \\ \hline
        f1 & `year' & `profit' & v1 \IN `product'.* & \\ \hline
        f2 & `year' & `sales' & v1 & v2 \IN argmax$_{v1}[k=10] D(f1,f2)$ \\ \hline
        *f3  & `year' & `profit' & v2 & \\ \hline
      \end{tabular}
    }
   \end{adjustbox}
  \vspace{-10pt}  
  \caption{Query which returns the top 10 profit visualizations for products
    which are most different from their sales visualizations.}
  \label{tab:proc1}
\vspace{-5pt}
\end{table*}

\subsubsection{\zqproc}
%
The real power of \zq as a query language comes not from its ability to
effortlessly specify collections of visualizations, but rather from its ability
to operate on these collections somewhat declaratively. 
With \zq's processing
capabilities, users can filter visualizations based on trend, search for 
similar-looking visualizations, identify representative visualizations, and
determine outlier visualizations. 
Naturally, to operate on collections, \zq
must have a way to iterate over them; however, 
since different visual analysis tasks
might require different forms of traversals over the collections, 
we expose the iteration
interface to the user.

\paragraph{Iterations over collections.}
Since collections may be composed of varying values from multiple columns,
iterating over the collections is not straight-forward.
Consider
Table~\ref{tab:proc1}---the goal is to return profit
by year visualizations for the top-10 products whose profit by year visualizations 
look the most different
from the sales by year visualizations.
\techreport{This may indicate a product that deserves special attention. }%
While we will describe this query in 
detail below, at a high level the first row assembles
the visualizations for profit over year for all products ({\tt f1}),
the second row assembles the visualizations for sales over year for all products ({\tt f2}), 
followed by operating (via the \zqproc column) on these two collections 
by finding the top-10 products who
sales over year is most different from profit over year,
while the third row displays the profit over year for those top-10 products.
A array-based representation of the visualization collections {\tt f1} and {\tt f2},
would look like the following:
\vspace{-5pt}
\begin{align*}
\scriptsize
  {\tt f1} = 
  \begin{Bmatrix}
    \texttt{X: `year', Y: `profit'} \\
    \begin{array}{l}
      \hline
      \texttt{Z: `product.chair'} \\
      \texttt{Z: `product.table'} \\
      \texttt{Z: `product.stapler'}
    \end{array} \\
    \vdots
  \end{Bmatrix}
  {\tt f2} = 
  \begin{Bmatrix}
    \texttt{X: `year', Y: `sales'}  \\
    \begin{array}{l}
      \hline
      \texttt{Z: `product.chair'} \\
      \texttt{Z: `product.table'} \\
      \texttt{Z: `product.stapler'}
    \end{array} \\
    \vdots 
  \end{Bmatrix}
\end{align*}
We would like to iterate over the products, the Z dimension values, of both {\tt f1}
and {\tt f2} to make our comparisons. Furthermore, we must
iterate over the products in the \emph{same order} for both {\tt f1} and {\tt
f2} to ensure that a product's profit visualization correctly matches with its
sales visualization. 
Using a single index for this would be
complicated and need to take into account the sizes of each of the columns.
\techreport{While there may be other ways to architect this iteration for
a single attribute, it is virtually impossible to do when there are multiple
attributes that are varying. }%
Instead, \zq opts for a more powerful \emph{dimension-based} iteration, which
assigns each column (or dimension) a separate iterator called an
\emph{\axvar[]}.
This dimension-based iteration is a powerful idea that extends to any number of dimensions. 
%
As shown in Table~\ref{tab:proc1}, \axvar[s] are defined and assigned
using the syntax:
\mbox{$\langle variable \rangle $\IN$ \langle collection \rangle$};
\axvar[] {\tt v1} is assigned to
the Z dimension of {\tt f1} and iterates over all product values. 
For cases in which multiple collections must traverse
over a dimension in the same order, an \axvar[] must be shared across those
collections for that
dimension;
%
in Table~\ref{tab:proc1}, {\tt f1} and {\tt f2}
share {\tt v1} for their Z dimension, since we want
to iterate over the products in lockstep.

\paragraph{Operations on collections.}
With the \axvar[s] defined, the user can then formulate the high-level
operations on collections of visualizations as an optimization function which
maximizes/minimizes for their desired pattern. Given that {\tt
argmax$_{x}[k=10] \; g(x)$} returns the top-10 {\tt x} values which maximizes the
function $g(x)$, and $D(x,y)$ returns the ``{distance}'' between {\tt x} and
{\tt y}, now consider the expression in the \zqproc
column for Table~\ref{tab:proc1}.
Colloquially, the expression says to
find the top-10 {\tt v1} values whose $D(f1,f2)$ values are the
largest.
The $f1$ and $f2$ in $D(f1,f2)$ refer to the collections of
visualizations in the first and second row and are bound to the current value
of the iteration for {\tt v1}. In other words, for each product
{\tt v1'} in {\tt v1}, retrieve the
visualizations {\tt f1[z: v1']} from collection {\tt f1} and {\tt f2[z: v1']} 
from collection {\tt f2} and calculate the
``{distance}'' between these visualizations; then, retrieve the 10 {\tt
v1'} values for which this distance is the largest---these are the products, and assign {\tt v2} to this
collection. 
Subsequently, we can access this set of products, as we do in the Z column of 
the third line of Table~\ref{tab:proc1}.

\begin{table*}[t]
\vspace{0pt}
  \centering\scriptsize
  \begin{adjustbox}{center}
    {\tt
      \begin{tabular}{|r|l|l|l|l|}
        \hline
        \zqname & X & Y & Z & \zqproc \\ \hline
        f1 & `year' & `sales' & v1 \IN `product'.* & v2 \IN argmax$_{v1}[t<0]
        T(f1)$ \\ \hline
        *f2 & `year' & `sales' & v2 &  \\ \hline
      \end{tabular}
    }
  \end{adjustbox}
  \vspace{-8pt}
  \caption{Query which returns the sales visualizations for all products which
  have a negative trend.}
  \label{tab:trend}
\vspace{-15pt}
\end{table*}

\begin{table*}[t]
  \centering\scriptsize
  \begin{adjustbox}{center}
    {\tt
      \begin{tabular}{|r|l|l|l|l|}
        \hline
        \zqname & X & Y & Z & \zqproc \\ \hline
        f1 & `year' & `sales' & `product'.`chair' & \\ \hline
        f2 & `year' & `sales' & v1 \IN `product'.(* -  `chair') &
        v2 \IN argmin$_{v1}[k=10] D(f1,f2)$\\ \hline
        *f3 & `year' & `sales' & v2 & \\ \hline
      \end{tabular}
    }
  \end{adjustbox}
  \vspace{-10pt}
  \caption{Query which returns the sales visualizations for the 10 products
    whose sales visualizations are the most similar to the sales visualization
  for the chair.}
  \label{tab:similarity}
\vspace{-10pt}
\end{table*}

\begin{table*}[t]
  \centering\scriptsize
  \begin{adjustbox}{center}
    {\tt
      \begin{tabular}{|r|l|l|l|l|}
        \hline
        \zqname & X & Y & Z & \zqproc \\ \hline
        f1 & `year' & `sales' & v1 \IN `product'.* & \\ \hline
        f2 & `year' & `sales' & v2 \IN `product'.* & v3 \IN argmax$_{v1}[k=10]
        \sum_{v2} D(f1, f2)$ \\ \hline
        *f3 & `year' & `sales' & v3 & \\ \hline
      \end{tabular}
    }
  \end{adjustbox}
  \vspace{-10pt}
  \caption{Query which returns the sales visualizations for the 10 products
  whose sales visualizations are the most different from the others.}
  \label{tab:outlier}
  \vspace{-8pt}
\end{table*}

\paragraph{Formal structure.}
More generally, the basic structure of the \zqproc column is:
{\small
\vspace{-5pt}
\[
  \langle argopt \rangle_{\langle \axvar[a] \rangle}[\langle limiter \rangle] \langle expr \rangle \textrm{ \ \ \  where}
\]
\vspace{-2em}
\begin{align*}
  \langle expr \rangle \; & \rightarrow \; \left(\max | \min | \sum | \prod \right)_{\langle
  \axvar[a] \rangle} \langle expr \rangle \\
  & \rightarrow \; \langle expr \rangle \; (+|-|\times|\div) \; \langle expr \rangle \\
  & \rightarrow \; T(\langle \nmvar[a] \rangle) \\
  & \rightarrow \; D(\langle \nmvar[a] \rangle, \langle \nmvar[a] \rangle) \\
  \langle argopt \rangle \; & \rightarrow \; \left( \text{argmax} | \text{argmin} |
\text{argany} \right) \\
\langle limiter \rangle \; & \rightarrow  \; \left(k=\mathbb{N} \; | \; t>\mathbb{R}
\; | \; p=\mathbb{R} \right)
\end{align*}
}
\noindent where $\langle \axvar[a] \rangle$ refers to the \axvar[s], and $\langle \nmvar[a]
\rangle$ refers to collections of visualizations. $\langle argopt \rangle$ may
be one of argmax, argmin, or argany, which returns the values which have the largest, smallest, and
any expressions respectively. The $\langle limiter \rangle$ limits the number of results: 
$k = \mathbb{N}$ returns only the top-$k$ values;
$t>\mathbb{R}$ returns only values who are larger than a threshold value $t$
(may also be smaller, greater than equal, etc.); $p=\mathbb{R}$ returns
the top $p$-percentile values.
$T$ and $D$ are
two simple \emph{functional primitives} supported by \zq that can be applied to
visualizations to find desired patterns:
\begin{denselist}
    \item \ [$T(f) \rightarrow \mathbb{R}$]: $T$ is a function which takes a
    visualization $f$ and returns a real number measuring some visual property of the trend of
    $f$. One such property is ``growth'', which returns a positive number if the overall trend is
    ``{upwards}'' and a negative number otherwise; an
    example implementation might be to measure the slope of a linear fit to the
    given input visualization $f$.
    Other properties may measure the skewness, or the number of peaks, or noisiness
    of a visualization. 
  \item \ [$D(f, f') \rightarrow \mathbb{R}$]: $D$ is a function which takes two
    visualizations $f$ and $f'$ and measures the distance (or dissimilarity)
    between these visualizations. Examples of distance functions may include
    a pointwise distance function like 
    Euclidean distance, Earth Mover's Distance, or the Kullback-Leibler Divergence. 
    The distance $D$ could also be measured using 
    the difference in the number of peaks, or slopes, or some
    other property. 
\end{denselist}
\zq supports many different implementations for these two functional
primitives, and the user is free to choose any one. If
the user does not select one, \zv will
automatically detect the ``{best}'' primitive based on the data
characteristics. Furthermore, if \zq does not have an implementation of the
$T$ or $D$ function that the user is looking for, the user may write and use their own
function.

\paragraph{Concrete examples.}
With just dimension-based iteration, the optimization structure of the \zqproc
column, and the functional primitives $T$ and $D$, we found that we were able
to support the majority of the visual analysis tasks required by our users.
Common patterns include filtering based on overall trend
(Table~\ref{tab:trend}), searching for the most similar visualization
(Table~\ref{tab:similarity}), and determining outlier visualizations
(Table~\ref{tab:outlier}).
\techreport{
Table~\ref{tab:trend} describes a query where in the first row, the variable {\tt v2}
selects all products whose trend is decreasing, and the second row visualizes
these product's sales over year.
Table~\ref{tab:similarity} starts with the visualization 
sales over year for chair in the first row, then in the second row
computes the visualizations of sales over year for all products, and in the process
column computes the similarity with chair, assigning the top 10 to {\tt v2},
and the third row visualizes the sales over year for these products.
Table~\ref{tab:outlier} starts with the visualization collection of sales over year
for all products in the first row, followed by another collection of the same
in the second row, and in the process column computes the sum of pairwise distances,
assigning the 10 products whose visualizations are most distant to others to {\tt v3}, 
after which they are visualized.
}%
Table~\ref{tab:real2} features a realistic query inspired by one of our case
studies. The overall goal of the query is to find the
products which have positive sales and profits trends in locations and
categories which have overall negative trends; the user may want to look at
this set of products to see what makes them so special.
Rows 1 and 2 specify the sales and profit visualizations for all locations and
categories respectively, and the processes for these rows filter down to the
locations and categories which have negative trends. Then rows 3 and 4 specify
the sales and profit visualizations for products in these locations and
categories, and the processes filter the visualizations down to the ones that
have positive trends. Finally, row 5 takes the list of output products from the
processes in rows 3 and 4 and takes the intersection of the two returning the
sales and profits visualizations for these products.

\paragraph{Pluggable functions.}
While the general structure of the \zqproc column does cover the majority of
the use cases requested by our users, users may want to write their own
functions to run in a \zq query. To support this, \zq exposes a {\tt Java}-based API
for users to write their own functions. In fact, we use this interface to
implement the $k$-means algorithm for
\zq.
While the pluggable functions do allow virtually any capabilities to be
implemented, it is preferred that users write their queries using the syntax of
the \zqproc column; pluggable functions are considered black-boxes
and cannot be automatically optimized by the \zq compiler.




\begin{table*}
  \centering\scriptsize
  \begin{adjustbox}{center}
    {\tt
      \begin{tabular}{|r|l|l|l|l|l|l|}
        \hline
        \zqname & X & Y & Z & Z2 & Z3 & \zqproc \\ \hline
        f1 & `year' & `sales' & v1 \IN `location'.* & & & v2 \IN
        argany$_{v1}[t<0] T(f1)$\\ \hline
        f2 & `year' & `profit' & v3 \IN `category'.* & & & v4 \IN
        argany$_{v3}[t<0] T(f2)$\\ \hline
        f3 & `year' & `profit' & v5 \IN `product'.* & `location'.[? IN v2] &
        `category'.[? IN v4] & v6 \IN argany$_{v5}[t>0] T(f3)$ \\ \hline
        f4 & `year' & `sales' & v5 & `location'.[? IN v2] &
        `category'.[? IN v4] & v7 \IN argany$_{v5}[t>0] T(f4)$ \\ \hline
        *f5 & `year' & \{`profit', `sales'\} & v6 \^{} v7 & & & \\ \hline
      \end{tabular}
    }
  \end{adjustbox}
  \vspace{-10pt}
  \caption{Query which returns the profit and sales visualizations for products
    which have positive trends in profit and sales in locations and categories
  which have overall negative trends.}
  \label{tab:real2}
  \vspace{-15pt}
\end{table*}

 \vspace{-2pt}
\subsection{Discussion of Capabilities and Limitations}\label{subsec:cap-lim}
Although \zq can capture 
a wide range of visual exploration queries, it is not
limitless. Here, we give a brief description of what \zq can do.
A more formal quantification can be found in
\papertext{\cite{techreport}}\techreport{Section~\ref{sec:formal-expr}}.

\zq's primary goal is to support queries over visualizations---which are themselves
aggregate group-by queries on data. 
Using these queries, \zq can compose a collection of visualizations, 
filter them in various ways, compare them against benchmarks or against
each other, and sort the results. 
The functions $T$ and $D$, while intuitive, support the ability
to perform a range of computations on visualization collections---for example, 
any filter predicate on a single visualization, checking for a specific
visual property, can be captured under $T$.
\techreport{With the pluggable functions, the ability to perform sophisticated computation
on visualization collections is enhanced even further. }%
Then, via the dimension-based iterators, \zq supports the ability to chain
these queries with each other and compose new visualization collections.
These simple set of operations offer unprecedented power in being able to
sift through visualizations to identify desired trends.

Since \zq already operates one layer above the data---on the visualizations---it does
not support the {\em creation of new derived data}:
that is, \zq does not support the generation of derived attributes or values 
not already present in the data. 
The new data that is generated via \zq is limited to those from binning 
and aggregating via the Viz column.
This limits \zq's ability to perform prediction---since feature engineering is
an essential part of prediction; it also limits \zq's ability to 
compose visualizations on combinations of attributes at a time,
e.g., $\frac{A1}{A2}$ on the X axis. 
Among other drawbacks of \zq: \zq does not support 
\begin{inparaenum}[(i)] 
\item recursion; 
\item any data modification; 
\item non-foreign-key joins nor arbitrary nesting; 
\item dimensionality reduction or other changes to the attributes; 
\item other forms of processing visualization collections not expressible via $T$, $D$ or the
black box;
\techreport{\item multiple-dimensional visualizations;}
\techreport{\item intermediate variable definitions;}
\item merging of visualizations (e.g., by aggregating two visualizations);
and 
\item 
statistical tests.
\end{inparaenum}

}{
\vspace{-7pt}
\section{Query Language}
\label{sec:querylanguage}
\input{motiv-examples}

\subsection{Formalization}
\label{subsec:formalization}

%


We now formally describe the \zq syntax.
We assume that we are operating on a single relation
or a star schema where the attributes are unique
(barring key-foreign key joins).
In general, \zq could be applied to arbitrary collections
of relations by letting the user precede an attribute $A$
with the relation name $R$, e.g., $R.A$.
For ease of exposition, we focus on the single relation case.

\subsubsection{Overview}
As described earlier, a \zq query is composed using a table,
much like a QBE query.
Unlike QBE, the columns of a \zq query do not refer to attributes
of the table being operated on; instead, they are predefined, and have
fixed semantic meanings.
In particular, at a high level, the columns are:
\begin{inparaenum}[(i)]
\item {\em Name:} providing an identifier for a set or collection of visualizations,
and allowing us to indicate if a specific set of visualizations
are to be output
\item {\em X, Y:} specifying the X and Y axes of the collections of visualizations,
restricted to sets of attributes
\item {\em Z (optional):} specifying the ``slice'' (or subset) of data that we're varying
restricted to sets of attributes along with values for those attributes;
\item {\em Constraints (optional):} specifying optional constraints applied to
the data prior to any visualizations or collections of visualizations being generated
\item {\em Viz (optional):} specifying the mechanism of visualization,
e.g., a bar chart, scatterplot, as well as the associated
transformation, or aggregation, e.g., the X axis is binned in groups of 20,
while the Y axis attribute is aggregated using SUM.
If this is not specified, standard rules of thumb are used
to determine the appropriate visualization~\cite{wongvoyager2015,Mackinlay:1986:ADG:22949.22950}.
\item {\em Process (optional):} specifying the ``optimization'' operation performed
on a collection of visualizations, typically intended towards identifying
desired visualizations. 
\end{inparaenum}

A \zq query may have any number of rows, and conceptually each row represents a
set of visualizations of interest. 
The user can then process and
filtrate these rows until she is left 
with only the output visualizations she is
interested in.
The result of a \zq query is the \emph{data} used to generate visualizations.
The \zv front-end then generates the visualizations for the user to peruse.

\paragraph{The Merits of a Tabular Interface.} 
The reader may wonder why we chose a table as an interface
for entering \zq queries. 
Our choice is due to our end-users: primarily non-programmers
who are used to drop-down menus and spreadsheet tools like Microsoft
Excel, and feel more at home with a tabular interface.
Specifically, the tabular skeleton ensures that users would not 
``miss out'' on key columns, and can view the query on the web-client front-end 
as a collection
of correct steps, each of which corresponds to a row.
Indeed, our user study validates this point; even users
with minimal programming experience can still use \zq proficiently after a
short tutorial.
Additionally, this interface is much more suited for embedding into an
interactive web-client.
That said, a user with more experience in programming
may find the interface restrictive, and may prefer
issuing the query as a function call within a programming language.
Nothing in our underlying \zq backend is tied to the tabular interface:
specifically, we already support the issuing of \zq queries within our 
Java client library: users can easily embed \zq queries into other computation. 
\techreport{In the future, we plan to write wrappers for other libraries
so that users can embed \zq into computation in other settings.
We are exploring the use of Thrift~\cite{slee2007thrift} for this purpose.}

\subsubsection{X and Y Columns}
As mentioned, each row can be thought of as a set of visualizations, and the X
and Y columns represent the x- and y- axes for those visualizations. In
Table~\ref{tab:motiv-ex-1}, the first row's visualizations will have `year'
as their x-axis and `profit' as their y-axis. 

The only permissible entries for the X or Y column are either
a single attribute from the table, or 
a {\em set} of attributes, with a variable to iterate over them.
\techreport{The exact semantics of variables and sets are discussed later in
Section~\ref{sec:variables}, 
but essentially, a column is allowed to take on any
attribute from the set.}
For example, in Table~\ref{tab:y-set}, the y-axis is allowed to take on either
`profit' or `sales' as its attribute. 
Because the y-axis value can be
taken from a set of attributes, the resulting output of this query is actually
the \emph{set} of visualizations whose x-axis is the `time' and the y-axis is
one of `profit' and `sales' for the `stapler'. 
This becomes a powerful
notion later in the Process column where we try to iterate over a set of
visualizations and identify desired ones to return to the user
or select for down-stream processing.
\begin{table}[H]
\vspace{-10pt}
 \centering\scriptsize
  {\tt    
  \begin{tabular}{|r|l|l|l|} \hline
      \textbf{Name} & \textbf{X} & \textbf{Y} & \textbf{Constraints} \\ \hline
      *f1 & `year' & y1 \IN \{`profit', `sales'\} & product=`stapler' \\ \hline
    \end{tabular}
  }
  \vspace{-10pt}\caption{A query for a set of visualizations, one of which plots profit
  over year and one of which plots sales over time for the stapler.}
  \label{tab:y-set}
\vspace{-10pt}
\end{table}


\subsubsection{Z Column}
The Z column is used to either focus our attention on a specific slice 
(or subset) of the dataset or
iterate over a set of slices for one or more attributes. 
To specify a set of slices,
the Z column must specify (a) one or more attributes,
just like the X and Y column, and
(b) one or more attribute values for each of those attributes --- which 
allows us to slice the data in some way.
For both (a) and (b) the Z column could specify a single 
entry, or a variable associated with a set of entries.
Table~\ref{tab:z-1} gives an example of using the Z column to visualize the
sales over time data specifically with regards to the `chair' and `desk'
products, one per line. 
Note that the attribute name and the attribute value are separated
using a period.

Table~\ref{tab:z-2} depicts an example where we iterate over
a set of slices (attribute values)
for a given attribute. 
This query returns the set of sales over time
visualizations for each product. Here, {\tt v1} binds to
the values of the `product' category. Since \zq internally associates
attribute values with their respective attributes, there is no need to
specify the attribute explicitly
for {\tt v1}. 
\techreport{Another way to think about this is to think of v1 \IN `product'.* as
syntactic sugar for `product'.v1 \IN `product'.*. The * symbol denotes all
possible values; in this case, all possible values of the `product'
attribute.}

\begin{table}[H]
\vspace{-10pt}
  \centering\scriptsize
  {\tt  
    \begin{tabular}{|r|l|l|l|} \hline
      \textbf{Name} & \textbf{X} & \textbf{Y} & \textbf{Z} \\ \hline
      *f1 & `year' & `sales' & `product'.`chair' \\
      *f2 & `year' & `sales' & `product'.`desk' \\ \hline
    \end{tabular}
  }
  \vspace{-10pt}\caption{A \zq query that returns the sales over year visualization for
  chairs and the sales over time visualization for desks.}
  \label{tab:z-1}
\vspace{-10pt}\end{table}

\begin{table}[H]
\vspace{-10pt}
  \centering\scriptsize
  {\tt  
    \begin{tabular}{|r|l|l|l|} \hline
      \textbf{Name} & \textbf{X} & \textbf{Y} & \textbf{Z} \\ \hline
      *f1 & `year' & `sales' & v1 \IN `product'.* \\ \hline
    \end{tabular}
  }
  \vspace{-10pt}\caption{A \zq query that returns the set of sales over year visualizations
  for each product.}
  \label{tab:z-2}
\vspace{-10pt}\end{table}

\techreport{Furthermore, the Z column can be left blank if the user does not wish to slice the data in any way.}

\subsubsection{Constraints Column}
\label{sec:constraints}
The Constraints column is used to specify further constraints on the set of
data used to generate the set of visualizations. 
Conceptually, we can view the Constraints column as being applied
to the dataset first, following which a collection of visualizations
are generated on the constrained dataset.
While the Constraints column may appear to overlap in functionality with 
the Z column, the Constraints column does not admit iteration via
variables in the way the Z, X or Y column, admits.
It is instead intended to apply a fixed boolean predicate to each tuple of 
the dataset prior to visualization in much the same way the \texttt{WHERE} clause
is applied in \sql. 
\techreport{(We discuss this issue further in Section~\ref{sec:variables}.)}
A blank Constraints column means no constraints are applied\techreport{ to the
data prior to visualization}.



\begin{table*}[!htbp]
  \centering\scriptsize
  {\tt
      \begin{tabular}{|r|l|l|l|} \hline
        \textbf{Name} & \textbf{X} & \textbf{Y} & \textbf{Viz} \\ \hline
        *f1 & `weight' & `sales' & bar.(x=bin(20), y=agg(`sum')) \\ \hline
      \end{tabular}
  }
  \vspace{-10pt}\caption{A \zq query which returns the bar chart of overall sales for
  different weight classes.}
  \label{tab:viz-1}
\vspace{-10pt}\end{table*}

\subsubsection{Viz Column}
\label{sec:viz}
Given the data from the X, Y, Z, and Constraints columns, the Viz column 
determines how the data is shaped 
before returning it as a visualization. 
There are two aspects to the Viz column:
first, the visualization type 
(e.g., bar chart, scatter plot),
and the summarization type
(e.g., binning or grouping, aggregating in some way).
These two aspects are akin to the
geometric and statistical transformation layers
from the Grammar of Graphics, a visualization specification language, and the inspiration behind
\ggplot~\cite{ggplot}.

In \zq, the Viz column specifies the visualization type
and the summarization using a period delimiter.
Functions are used to represent both
the visualization type and the summarization.
Consider the query in Table~\ref{tab:viz-1}. 
Here, the user specifies that she wants a bar chart,
with {\tt bar}, and 
specifies the type of summarization in the accompanying tuple:
{\tt x=bin(20)} denotes that x-axis should be binned into bins
of size 20, and {\tt y=agg(`sum')} runs the {\tt SUM} aggregation on the
y-values when grouped by both the bins of the x-axis and the values in the z-axis,
(if any are specified). 

Often we can leave the Viz column blank.
In such cases, we would apply well-known rules of thumb for what types of
visualization and summarization would be appropriate given specific X and Y axes.
Work on recommending appropriate visualization types
dates back to the 80s~\cite{Mackinlay:1986:ADG:22949.22950},
that both Polaris/Tableau~\cite{DBLP:journals/cacm/StolteTH08}, and recent work builds on~\cite{wongvoyager2015,DBLP:conf/avi/KandelPPHH12}, determining
the best visualization type by examining the schema and statistical properties.
\techreport{In many of our examples of \zq queries, we omit the Viz column for this reason.}

\subsubsection{Name Column}
For any row of a \zq query, 
the combination of X, Y, Z, Constraints, and Viz
columns together represent the \emph{visual component} of that row. 
A visual component formally represents a set of visualizations. 
The Name column allows us to provide a name to this visual component 
by binding a variable to it.
These variables can be used in the Process column
to subselect the desired visualizations from the set of visualizations,
as we will see subsequently.

\begin{table*}[!htbp]
  \centering\scriptsize
  {\tt
    \begin{tabular}{|r|l|l|l|l|} \hline
      \textbf{Name} & \textbf{X} & \textbf{Y} & \textbf{Z} & \textbf{Process} \\ \hline
      *f1 & `year' & `sales' & `product'.`stapler' & \\
      f2 & `year' & `sales' & v1 \IN `product'.(* - `stapler') & v2 \IN
      $argmin_{v1}[k=10] D(f1,f2)$ \\
      *f3 & `year' & `sales' & v2 & \\ \hline
    \end{tabular}
  }
  \vspace{-10pt}\caption{A \zq query retrieving the sales over year visualization for the top
    10 products whose sales over year visualization looks the most similar to that
  of the stapler.}
  \label{tab:name-1}
\techreport{
\centering\scriptsize
  {\tt
    \begin{tabular}{|r|l|l|l|l|} \hline
      \textbf{Name} & \textbf{X} & \textbf{Y} & \textbf{Z} & \textbf{Process} \\ \hline
      -f1 & & & & \\
      f2 & x1 \IN \{`time', `location'\} & y1 \IN \{`sales', `profit'\} &
      `product'.`stapler' & x2, y2 \IN $argmin_{x1,y1}[k=10] D(f1,f2)$ \\
      *f3 & x2 & y2 & `product'.`stapler' & \\ \hline
    \end{tabular}
  }
  \vspace{-10pt}\caption{A \zq query retrieving two different visualisations (among different combinations of x and y) of stapler which are most similar to each other}
  \label{tab:name-2}}
\vspace{-10pt}\end{table*}

\papertext{
In Table~\ref{tab:name-1}, {\tt f1} is bound to the single visualization in
the first row, while {\tt f2} is bound to the variable which iterates over the
set of visualizations in the second row. Any rows prefaced with a * symbol are
output rows; visualizations corresponding to these rows are processed and
displayed by the \zv frontend.
}
\techreport{
In the \zq query given by Table~\ref{tab:name-1}, 
we see that the names for the visual components of
the rows are named {\tt f1}, {\tt f2}, and {\tt f3} in order of the rows. 
For the first row, the
visual component is a single visualization, since there are no sets, and {\tt f1}
binds to the single visualization. 
For the second row, we see that the visual
component is over the set of visualizations with varying Z column values, and
{\tt f2} binds to the variable which iterates over this set. 
Note that {\tt f1} and {\tt f3} in Table~\ref{tab:name-1} are prefaced
by a * symbol.
This symbol indicates that the visual component of the row
is designated to be part of the output.
As can be seen in the example, multiple rows can be part of the output,
and in fact, a row could correspond to multiple visualizations.
Visualizations corresponding to all the rows marked with * are 
processed and displayed by the \zv frontend.
}


\later{If multiple columns'
values are sets, the visual component becomes the set of visualizations that
would span over the Cartesian product of these column sets.
Table~\ref{tab:name-2} provides an example of this in the second row.
}

\later{Name cells may be blank, but then there would be no way to refer to the
visual component of that row, obviating the need for the row in the first place. If not blank,
Name cells may only have variable names possibly prefaced with either * or - as
it only possible values.}


\begin{table*}[!htbp]
  \centering\scriptsize
  {\tt
      \begin{tabular}{|r|l|l|l|l|} \hline
        \textbf{Name} & \textbf{X} & \textbf{Y} & \textbf{Z} & \textbf{Process}
        \\ \hline
        f1 & `year' & `sales' & v1 \IN `product'.* & \\
        f2 & `year' & `profit' & v1 & v2 \IN $argmax_{v1}[k=10]
        D(f1, f2)$ \\
        *f3 & `year' & `sales' & v2  & \\
        *f4 & `year' & `profit' & v2  & \\ \hline
      \end{tabular}
  }
  \vspace{-10pt}\caption{A \zq query which returns the set sales over years and the profit
    over years visualizations for the top 10 products for which these two
  visualizations are the most dissimilar.}
  \label{tab:vars-1}

  \centering\scriptsize
  {\tt
      \begin{tabular}{|r|l|l|l|l|l|} \hline
        \textbf{Name} & \textbf{X} & \textbf{Y} & \textbf{Z} &
        \textbf{Constraints} & \textbf{Process}
        \\ \hline
        f1 & `year' & `sales' & v1 \IN `product'.* & & v2 \IN $argmax_{v1}[k=10]
        T(f1)$ \\
        *f2 & `year' & `profit' & & product IN (v2.range) & \\ \hline
      \end{tabular}
  }
  \vspace{-10pt}\caption{A \zq query which plots the profit over years for the top 10
  products with the highest sloping trend lines for sales over the years.}
  \label{tab:vars-2}

  \centering\scriptsize
  {\tt
      \begin{tabular}{|r|l|l|l|l|} \hline
        \textbf{Name} & \textbf{X} & \textbf{Y} & \textbf{Z} &
        \textbf{Process}
        \\ \hline
        f1 & x1 \IN C & y1 \IN  M & `product'.`chair' & \\
        f2 & x1 & y1 & `product'.`desk' & x2,y2 \IN $argmax_{x1,y1}[k=10] D(f1,f2)$ \\
        *f3 & x2 & y2 & `product'.`chair' & \\
        *f4 & x2 & y2 & `product'.`desk' & \\ \hline
      \end{tabular}
  }
  \vspace{-10pt}\caption{A \zq query which finds the x- and y- axes which
  differentiate the chair and the desk most.}
  \label{tab:vars-3}
\vspace{-10pt}\end{table*}

\subsubsection{Sets and Variables}
\label{sec:variables}
As we described previously, sets in \zq must
always be accompanied by a variable which iterates over that set, to 
ensure that \zq traverses over sets of visualizations
in a consistent order when making comparisons. 
Consider Table~\ref{tab:vars-1}, which shows
a query that iterates over
the set of products, and for each, compares the sales over
years with the profits over years. Without a
variable enforcing a consistent iteration order over the set of products,
it is possible that the set of visualizations could be
traversed in unintended ways. For example, the sales vs.~year plot for chairs
from the first set could be compared with the profit vs.~year plot for desks in
the second set. By reusing {\tt
v1} in both the first and second rows, the user can force the \zv back-end to
step through the sets in sync.

\noindent {\bf Operations on Sets.} 
Constant sets in \zq must use \{\} to denote that the enclosed elements are
part of a set. As a special case, the user may also use * to represent the set of
all possible values. The union of
sets can be taken using the  sign {\tt |}, set difference can be taken using the
{\tt $\setminus$}
sign, and the intersection can be taken with {\tt \&}. 
\later{Users must take the
explicit difference between * and a set to get the complement.
Set operators have lower precedence than most other operators, so the use of
parentheses is encouraged:
{\tt `product'.(* $\setminus$ \{`stapler'\})}.
As a matter of convenience, all operands of a set operation
which are not sets are automatically converted into singleton sets which
contain that operand. This allows us to simplify the previous expression a
little
further to {\tt `product'.(* $\setminus$ `stapler')}.
The valid elements of sets are literals, the set expansion of variables ({\tt
.range}), and other valid sets. Variables in their raw form cannot be elements
in a set. The \zq, the only a Cartesian products of sets can be specified is
when a period is used (e.g., {\tt \{`product', 'color'\}.*}).}

\later{Labels for sets maybe used with the {\tt @} notation so that the user does not
need to specify the same set over and over again. For example, if the user has
decided that the measure attributes she is interested in are {\tt \{`profit',
`sales', 'revenue'\}}, she could use {\tt M @ \{`profit', `sales', `revenue'\}}
the first time when she defines the set, and {\tt M} every time afterwards.
Once a label has been set, it cannot be redefined.}

\noindent {\bf Ordering.} 
All sets in \zv are ordered, but when defined using the \{\}, the ordering is
defined arbitrarily. Only after a set has passed through a Process
column's ordering mechanism can the user depend on the ordering of the set.
Ordering mechanisms are discussed further in Section~\ref{sec:process}.
However, even for arbitrarily ordered sets, if a variable iterator is defined
for that set and reused, \zq guarantees at least a \emph{consistent} ordering
of traversal, allowing the sets in
Table~\ref{tab:vars-1} to be traversed in the intended order.
\later{
\alkim{
Variable names must start with a letter and can only contain letters, numbers,
and the underscore for the remaining characters. 
To be used, variables must either be defined in the current or a previous row.
}
}

\noindent {\bf Axis Variables.} 
In \zq, there are two types of variables: \emph{axis variables} and \emph{name
variables}.
Axis variables are the common variables used in any of the columns except the
Name column.  The declaration has the form:
\emph{$\langle$variable~name$\rangle$} \IN \emph{$\langle$set$\rangle$}.
The variable then can be used as an iterator in the Process column or reused in
a different table cell to denote that the set be traversed in the same way for
that cell. It is possible to declare multiple axis variables at once,
as we have seen from the Z and Process columns: (e.g., {\tt z.v \IN *.*}).

Sometimes, it is necessary to retrieve the set that an axis variable iterates
over. The {\tt .range} notation allows the user to expand variables to
their corresponding sets and apply arbitrary \zv set
operations on them. For example, {\tt v4 \IN (v2.range |
v3.range)} binds v4 to the union of the sets iterated by {\tt v2}
and {\tt v3}. 
\later{However, the expanded set for a variable may not be the final
value for any table cell in \zq; it must always be either a literal or a
raw variable.}

Axis
variables can be used freely in any column except the Constraints column. In
the Constraints column, only the expanded set form of a variable may be used.
Table~\ref{tab:vars-2} demonstrates how to use an axis variable in the
Constraints column. In this example, the user is trying to plot the overall
profits over years for the top 10 products which have had the most growth in
sales over the years. The user finds these top 10 products in the first row
and declares the variable {\tt v2} to iterate over that set. Afterwards, she
uses the constraint {\tt product \IN (v2.range)} to get the overall profits
across all 10 of these products in the second row.

\noindent {\bf Name Variables.}
Name variables (e.g., \texttt{f1, f2}) are declared only in the Name column and used only in the
Process column.  
Named variables are iterators
over the set of visualizations represented by the visual component. If the
visual component represents only a single visualization, the name variable is
set to that specific visualization. Name variables are \emph{bound} to the axis
variables present in their row. In Table~\ref{tab:vars-2}, {\tt f1} is bound
{\tt v1}, so if {\tt v1} progresses to the next product in the set {\tt f1}
also progresses to the next visualization in the set. This is what allows the
Process column to iterate over the axis variable ({\tt v1}), and still compare
using the name variable ({\tt f1}).
If multiple axis variables are present in the visual component, the name
variable iterates over set of visualizations produced by the Cartesian product
of the axis variables. 
\techreport{If the axis variables have been declared independently
of each other in the visual component, the ordering of the Cartesian product
  follows the ordered bag semantics described in
  Section~\ref{subsec:ordered-bag} in order of the columns laid out in \zq.
  However, the user may also superscript variables to control the order in
  which Cartesian product is done as well.
  However, if the variables were declared together in the Process 
column as is the case
with {\tt x2} and {\tt
y2} in Table~\ref{tab:vars-3}, the ordering is the same as the set in the
declaration.}


\begin{table*}[!htbp]
  \centering\scriptsize
 \vspace{-10pt}
  {\tt
      \begin{tabular}{|r|l|l|l|l|} \hline
        \textbf{Name} & \textbf{X} & \textbf{Y} & \textbf{Z} & \textbf{Process}
        \\ \hline
        f1 & `year' & `sales' & v1 \IN `product'.* & v2 \IN $R(10, v1, f1)$ \\
        f2 & `year' & `sales' & v2 & v3 \IN $argmax_{v1}[k=10] \; min_{v2}
        D(f1,f2)$ \\
        *f3 & `year' & `sales' & v3  & \\ \hline
      \end{tabular}
  }
  \vspace{-10pt}\caption{A \zq query which returns 10 sales over years visualizations for
  products which are outliers compared to the rest.}
  \label{tab:process-2}
\vspace{-10pt}\end{table*}

\subsubsection{Process Column}
\label{sec:process}
Once the visual component for a row has been named, the user may use the
Process column to sort, filter, and compare the visual component with
previously defined
visual components to isolate the set of visualizations she is interested in.
While the user is free to define her own functions for the Process column,
we have come up with a core set of primitives based on our case
studies which we believe can handle the vast majority of the common operations.
Each non-empty Process column entry is defined to be a {\em task}. 
Implicitly, a task may be impacted by one or more rows of a \zq query\techreport{ (which would be input to the process optimization),}
and may also impact one or more rows\techreport{ (which will use the outputs of the process optimization)}, 
as we will see below.

\noindent {\bf Functional Primitives.}
First, we introduce three simple functions that can be applied to
visualizations: $T$, $D$, and $R$.
\zv will use default settings for each of these functions, but
the user is free to specify their own variants for each of these functions
that are more suited to their application.

\noindent $\bullet \  T(f)$ measures the overall trend of visualization $f$. 
It is positive
if the overall trend indicates ``growth'' and negative if the overall trend goes
``down''.
There are obviously many ways that such a function can be
implemented, but one example implementation might be to measure the slope of a
linear fit to the given input visualization $f$.

\noindent $\bullet \ D(f,f')$ measures the distance between the two visualizations $f$ and
$f'$. For example, this might mean calculating the Earth Mover's Distance or the 
Kullback-Leibler Divergence between the induced probability distributions. 

\noindent $\bullet \ R(k,v,f)$ computes the set of $k$-representative visualizations given an axis
variable, $v$, and an iterator over the set of visualizations, $f$. Different
users might have different notions of what representative means, but one
example would be to run $k$-means clustering on the given set of visualizations
and return the $k$ centroids. In addition to taking in a single
axis variable, $v$, $R$ may also take in a tuple of axis variables to iterate
over. The return value of $R$ is the set of axis variable values which produced
the representative visualizations.

\noindent {\bf Processing.}
Given these functional primitives, 
\zq also provides some default sorting and filtering mechanisms: 
$argmin$, $argmax$, and $argany$. Although $argmin$ and $argmax$ are usually
used to find the best value for which the objective function is optimized, \zq
typically returns the top-$k$ values, sorted in order. $argany$ is used to
return any $k$ values. In addition to the top-$k$, the user might also like to
specify that she wants every value for which a certain threshold is met, and
\zq is able to support this as well.
Specifically, the expression {\tt v2 \IN $argmax_{v1}[k=10] D(f1,f2)$} returns
the top 10 {\tt v1} values for which $D(f1,f2)$ are maximized, sorts those in
decreasing order of the distance, and declares the variable {\tt v2} to iterate
over that set. 
This could be useful, for instance, to find the 10 visualizations with
the most variation on some attributes.
 The expression {\tt v2 \IN $argmin_{v1}[t<0] D(f1,f2)$} returns
the {\tt v1} values for which the objective function $D(f1,f2)$ is below the
threshold 0, sorts the values in increasing order of the objective function,
and declares {\tt v2} to iterate over that set. If a filtering option ($k$,$t$)
is not specified, the mechanism simply sorts the values, so {\tt v2} \IN
$argmin_{v1} T(f1)$ would bind {\tt v2} to iterate over the values of {\tt v1}
sorted in increasing order of $T(f1)$. {\tt v2 \IN $argany_{v1}[t>0] T(f1)$}
would set {\tt v2} to iterate over the values of {\tt v1} for which the $T(f1)$
is greater than 0. 
\later{Note that when using the $k$ filtering option for $argany$,
no objective function is required: {\tt v2 \IN $argany_{v1}[k=10]$}.}

Note that the mechanisms may also take in multiple axis variables, as we saw
from Table~\ref{tab:vars-3}.  The mechanism iterates over the Cartesian
product of its input variables. If a mechanism is given $k$ axis variables to
iterate over, the resulting set of values must also be bound to $k$ declared
variables. In the case of Table~\ref{tab:vars-3}, since there were two
variables to iterate over, {\tt x1} and {\tt y1}, there are two output
variables, {\tt x2} and {\tt y2}. The order of the variables is important as
the values of the $i$th input variable are set to the $i$th output variable.

\noindent 
{\bf User Exploration Tasks.}
By building on these mechanisms, the user can perform most
common tasks.
This includes the {\em similarity/dissimilarity search} 
performed in Table~\ref{tab:vars-1},
to find products that are dissimilar;
the {\em comparative search} performed in Table~\ref{tab:vars-3},
to identify attributes on which
two slices of data are dissimilar, or even the {\em outlier search}
query shown in Table~\ref{tab:process-2}, to identify the products
that are outliers on the sales over year visualizations. 
Note that in
Table~\ref{tab:process-2}, we use two levels of iteration. 
\later{Other mathematical
operations such as $\sum$ and $\prod$ are also available to the user. However,
other mathematical operations such as $max$ and $min$ retain their original
meaning.}

\later{
\subsubsection{Limitations}
\alkim{We probably want to move this section to another place}
\begin{itemize}
  \item \zq fundamentally assumes a 2-D visualization
\end{itemize}
}

}

\section{Expressiveness}
\label{sec:formal-expr}

In this section, we formally quantify the expressive power of \zq. To this end,
we formulate an algebra, called the \emph{\vzalg}.
Like relational algebra, \vzalg contains a basic
set of operators that we believe all \vzlans should be able to
express.
At a high level, the operators of
our \vzalg operate on sets of visualizations and are not
mired by the data representations of those visualizations, nor the details of how
the visualizations are rendered. Instead, the \vzalg is primarily
concerned with the different ways in which visualizations can be selected,
refined, and compared with each other.

Given a function $T$ that operates on a visualization at a time, 
and a function $D$ that operates on a pair of visualizations at a time, 
both returning real-valued numbers,
a \vzlan $L$ is defined to be
\emph{\vzcmp} $VEC_{T,D}(L)$ with respect to $T$ and $D$ if it supports
all the operators of the \vzalg.
These functions $T$ and $D$ (also defined previously) 
are ``functional primitives'' without which the resulting
algebra would have been unable to manipulate visualizations in the way
we need for data exploration. 
Unlike relational algebra, which does not have any ``black-box''
functions, \vzalg requires these functions for operating on visualizations
effectively. 
That said, these two functions are flexible and configurable and up to the user
to define (or left as system defaults).
Next, we formally define the \vzalg operators and prove
that \zq is  \vzcmp. 


\subsection{Ordered Bag Semantics}
\label{subsec:ordered-bag}
In \vzalg, relations have bag semantics. However, since users want to
see the most relevant visualizations first, ordering is critical. So, we
adapt the operators from relational algebra to preserve ordering
information.

Thus, we operate on {\em ordered bags} (i.e., a bag that has an inherent
order). 
We describe the details of how to operate on ordered bags below.
We use the variables $R,S$ to denote ordered bags. 
We also use the notation $R = [t_1, \ldots, t_n]$ to refer to an ordered bag,
where $t_i$ are the tuples. 

The first operator that we define is an indexing operator, much like
indexing in arrays.
The notation $R[i]$ refers to the $i$th tuple within $R$, and $R[i:j]$ refers
to the ordered bag corresponding to the list of tuples from the $i$th
to the $j$th tuple, both inclusive. In the notation $[i:j]$ if either one of
$i$ or $j$ is omitted, then it is assumed to be $1$ for $i$, and $n$ for $j$,
where $n$ is the total number of tuples.

Next, we define a union operator $\cup$: $R \cup S$ refers to the concatenation
of the two ordered bags $R$ and $S$.  
If one of $R$ or $S$ is empty, then the result of the union is simply the other
relation. We define the union operation first 
because it will come in handy for subsequent operations.
 
We define the $\sigma$ operator like in relational algebra, via a recursive definition:
$$\sigma_\theta (R) = \sigma_\theta ([R[1]]) \cup \sigma_\theta (R[2:])$$
where $\sigma_\theta$ when applied to an ordered bag with a single tuple ([t])
behaves exactly like in the relational algebra case, returning the same
ordered bag ([t]) if the condition is satisfied, and the empty ordered bag
([]) if the condition is not satisfied.
The $\pi$ operator for projection is defined similarly to $\sigma$ in the equation above,
with the $\pi$ operator on an ordered bag with a 
single tuple simply removing the irrelevant attributes from that tuple,
like in relational algebra. 

Then, we define the $\setminus$ operator, for ordered bag difference. 
Here, the set difference operator operates on every tuple in the first ordered
bag and removes it if it finds it in the second ordered bag.
Thus:
$$R \setminus S = ([R[1]] \setminus S) \cup (R[2:] \setminus S)$$
where $[t] \setminus S$ is defined like in relational algebra, 
returning $[t]$ if $[t]$ is not in $S$, and $[]$ otherwise. The intersection
operator $\cap$ is defined similarly to $\cup$ and $\setminus$.

Now, we can define the duplicate elimination operator as follows:
$$\delta(R) = [R[1]] \cup (R[2:] \setminus [R[1]])$$
Thus, the duplication elimination operator preserves ordering, while maintaining the first
copy of each tuple at the first position that it was found in the ordered bag.

Lastly, we have the cross product operator, as follows:
$$R \times S = ([R[1]] \times S) \cup (R[2:] \times S)$$
where further we have
$$[t] \times S = ([t] \times [S[1]]) \cup ([t] \times S[2:])$$
where $[t] \times [u]$ creates an ordered bag with the result of the cross product 
as defined in relational algebra.

Given these semantics for ordered bags, we can develop the visual exploration algebra.



\subsection{Basic Notation}
Assume we are given a $k$-ary relation $\calR$ with attributes
$(A_1,A_2,\dots,A_k)$.
Let $\calX$ be the unary relation with attribute X whose values are the
names of the attributes in $\calR$ that can appear on the x-axis. If the
x-axis attributes are not specified by the user for relation $\calR$, the
default behavior is to include all attributes in $\calR$: $\{A_1,\dots,A_k\}$. Let
$\calY$ be defined similarly with Y for attributes that can appear on the y-axis.
Given $\calR$, $\calX$, and $\calY$, we define $\calV$, the \emph{\vzuni},
as follows: $
  \calV = \nu(\calR) = \calX \times \calY \times
  \left(
    \mathlarger{\mathlarger{\times}}_{i=1}^k \pi_{A_i}(\calR) \cup \{*\}
  \right)$
where $\pi$ is the projection operator from relational algebra and \wild
is a special wildcard symbol, used to denote all values of an attribute.
Table~\ref{tab:setup} shows an example of what a sample $\calR$ and
corresponding $\calX$, $\calY$, and $\calV$ would look like.
\techreport{
    At a high level, the visual universe specifies all subsets of data that may
    be of interest, along with the intended attributes to be visualized. Unlike
    relational algebra, \vzalg mixes schema and data elements, but in a special
    way in order to operate on a collection of visualizations.
}



\begin{table*}[!t]
  \vspace{10pt}
  \begin{subtable}{0.4\linewidth}
    \scriptsize
    \begin{tabular}{|c|c|c|c|c|c|}
      \hline
      year & month & product & location & sales & profit \\ \hline
      2016 & 4 & chair & US & 623,000 & 314,000 \\ \hline
      2016 & 3 & chair & US & 789,000 & 410,000 \\ \hline
      2016 & 4 & table & US & 258,000 & 169,000 \\ \hline
      2016 & 4 & chair & UK & 130,000 & 63,000 \\
      \multicolumn{6}{|c|}{{\bf \vdots}} \\
    \end{tabular}
    \caption{Example $\calR$}
    \label{subtab:ex-R}
     \end{subtable}
       \hspace{-10pt}
  \begin{subtable}{.1\linewidth}
    \scriptsize
    \begin{tabular}{|c|}
      \hline
      X \\ \hline
      year \\ \hline
      month \\ \hline
    \end{tabular}
    \caption{$\calX$\hphantom{ for $\calR$}}
    \label{subtab:ex-X}
  \end{subtable}
  \hspace{-10pt}
  \begin{subtable}{.1\linewidth}
        \scriptsize
    \begin{tabular}{|c|}
      \hline
      Y \\ \hline
      sales \\ \hline
      profit \\ \hline
    \end{tabular}
    \caption{$\calY$\hphantom{  for $\calR$}}
    \label{subtab:ex-Y}
  \end{subtable}
  \hspace{-10pt}
  \begin{subtable}{0.3\linewidth}
    \scriptsize
    \begin{tabular}{|c|c|c|c|c|c|c|c|}
      \hline
      X & Y & year & month & product & location & sales & profit \\ \hline
      year & sales & \wild & \wild & \wild & \wild & \wild & \wild \\ \hline
      year & profit & \wild & \wild & \wild & \wild & \wild & \wild \\ \hline
      year & sales & \wild & \wild & chair & \wild & \wild & \wild \\ \hline
      year & sales & \wild & \wild & chair & US & \wild & \wild \\
      \multicolumn{8}{|c|}{{\bf \vdots}} \\
    \end{tabular}
    \caption{$\calV$ for $\calR$}
    \label{subtab:ex-V}
  \end{subtable}
  \vspace{-5pt}
\caption{An example relation $\calR$ and its resultant $\calX$, $\calY$,
  and $\calV$.}
  \label{tab:setup}
  \vspace{-15pt}
\end{table*}

Any subset relation $V \subseteq \calV$ is called a \emph{\vzgrp}, and any
$k+2$-tuple from $\calV$ is called a \emph{\vzchn}.
The last $k$ portions (or attributes) of a tuple from $\calV$
comprise the {\em data source} of the \vzchn.
Overall, a \vzchn
represents a visualization that can be rendered from a selected data source, and a
set of \vzchns is a \vzgrp. The X
and Y attributes of the \vzchn determine the x- and y- axes, and the
selection on the data source is determined by attributes $A_1,\dots,A_k$. If
an attribute has the wildcard symbol \wild as its value, no subselection is
performed on that attribute for the data source.
For example, the third row of Table~\ref{subtab:ex-V} is a \vzchn that
represents the visualization with year as the x-axis and sales as the y-axis for
chair products. Since the value of location is \wild, all locations are
considered valid or pertinent for the data source.
In relational algebra, the data source
for the third row can be written 
as $\sigma_{\text{product=chair}}(\calR)$.
The
\wild symbol therefore attempts to emulate the lack of presence 
of a selection condition on that attribute in the $\sigma$
operator of the relational algebra. 
Readers familiar with OLAP will notice the similarity between the use of the
symbol \wild here 
and the GROUPING SETS functionality in SQL.

Note that infinitely many
visualizations can be produced from a single visual source, due to different
granularities of binning, aggregation functions, and types
of visualizations that can be constructed, 
since a visualization generation engine can use a
visualization rendering grammar like ggplot~\cite{ggplot} that provides that functionality.
Our focus in defining the \vzalg is to specify the inputs to a visualization
and attributes of interest as opposed to the aesthetic aspects and encodings.
Thus, for our discussion, we assume that each visual source maps 
to a singular visualization.
Even if the details of the encoding and aesthetics are not provided,
standard rules may be applied for this mapping as alluded
earlier~\cite{wongvoyager2015,DBLP:journals/cacm/StolteTH08} in
Section~\ref{subsec:formalization}.
Furthermore,
a \vzchn does not specify the data representation of the underlying data source;
therefore the expressive power of \vzalg is not tied to any specific backend data storage
model.
The astute reader will have noticed that the format for a \vzchn looks fairly
similar to a collections of visualizations in \zq; this is no accident. In fact, we
will use the visualization collections of \zq as a proxy to \vzchns when proving that
\zq is \vzcmp.

\subsection{Functional Primitives}

Earlier, we mentioned that a \vzalg is \vzexp complete 
with respect to two functional primitives: $T$ and $D$.
Here we define the formal types for these functional primitives with respect to
\vzalg.

The function $T: \calV \rightarrow \mathbb{R}$ returns a real number given
a \vzchn. This function can be used to assess whether a trend:
defined by the visualization corresponding to a specific \vzchn,
is ``increasing'', or ``decreasing'', or satisfies some other fixed property. Many such $T$ can be defined and used within the \vzalg.

The function $D: \calV \times \calV \rightarrow \mathbb{R}$ returns a real
number given a pair of \vzchns. This function can be used
to compare pairs of visualizations (corresponding to the \vzchns)
with respect to each other. The most natural way to define $D$ is
via some notion of distance, e.g., Earth Mover's or Euclidian distance, 
but once again, the definition can be provided by the user or assumed as a fixed black box.


\subsection{\VzAlg Operators}
\label{sec:viz-exp-oper}

Similar to how operators in
ordered bag algebra operate on and result in ordered bags, 
operators in \vzalg operate on and result in \vzgrps. 
Many of the symbols for operators in \vzalg are also derived from
relational algebra, with some differences. 
To differentiate, operators in \vzalg are superscripted
with a $v$ (e.g., $\vzsel$, $\vzsort$).
The unary operators for \vzalg include 
\begin{inparaenum}[\itshape (i)\upshape]
\item $\vzsel$ for selection, \item $\vzsort$ for
sorting a \vzgrp based on the trend-estimating function $T$, \item $\vzlimit$ for
limiting the number of \vzchns in a \vzgrp, and \item $\vzdedup$ for duplicate \vzchn removal.
\end{inparaenum}. 
The binary operators include 
\begin{inparaenum}[\itshape (i)\upshape]
\item  $\vzunion$ for union, \item $\vzdiff$ for
difference, \item $\vzswap$ for replacing the attribute values of the \vzchns in one \vzgrp's with
another's, \item $\vzdist$ to reorder the first \vzgrp based on the \vzchns'
distances to the \vzchns of another \vzgrp based on metric $D$, and \item $\vzfind$
to reorder the \vzchns in a \vzgrp based on their distance to a reference
\vzchn from a singleton \vzgrp based on $D$. 
\end{inparaenum}
These operators are described below, and listed
in Table~\ref{tab:vizalgebopt}.

\begin{table*}[!t]
  \vspace{0pt}
\scriptsize
    \centering
\begin{tabular}{|c|l|l|p{5cm}|l|}      \hline
       Operator & Name & Derived from Bag Algebra  & Meaning & Unary/Binary \\ \hline \hline
       $\vzsel$ & Selection & Yes  & Subselects \vzchns & Unary \\ \hline
       $\vzsort$ & Sort & No & Sorts \vzchns in increasing order & Unary \\ \hline
       $\vzlimit$ & Limit & Yes & Returns first k \vzchns & Unary \\ \hline 
       $\vzdedup$ & Dedup & Yes  & Removes duplicate \vzchns & Unary \\ \hline 
       $\vzunion/\vzdiff/\vzint$ & Union/Diff/Int & Yes & Returns the union
  of/differences between/intersection of two \vzgrps & Binary \\ \hline 
       $\vzswap$ & Swap & No & Returns a \vzgrp in which values of an attribute in  one \vzgrp is  replaced  with values of the same attribute in another \vzgrp & Binary \\ \hline 
       $\vzdist$ & Dist & No & Sorts a \vzgrp based on pairwise distance to another \vzgrp  & Binary \\ \hline 	
       $\vzfind$ & Find & No & Sorts a \vzgrp in increasing order based on their distances to a single reference \vzchn & Binary \\ \hline  
   \end{tabular}
   \caption{Visual Exploration Algebra Operators}
  \label{tab:vizalgebopt}
\end{table*}

\subsubsection{Unary Operators.}
\paragraph{$\vzsel_{\theta}(V)$:}
$\vzsel$ selects a \vzgrp from $V$ based on selection criteria
$\theta$, like ordered bag algebra. 
However, $\vzsel$ has a
more restricted $\theta$; while $\vee$ and $\wedge$ may still be used, 
only the binary comparison operators $=$ and
$\ne$ are allowed. 
As an example, \\
$\vzsel_\theta (\calV)$ where $\theta = ($X=`year' $\land$ Y=`sales' $\land$ year=\wild $\land$ month=\wild $\land$ product $\ne$ \wild $\land$ location=`US' $\land$ sales=\wild $\land$ profit=\wild$)$ from Table~\ref{subtab:vzsel} 
on $\calV$ from
Table~\ref{tab:setup} would result in the \vzgrp of time 
vs. sales visualizations for different products in the US.

In this example, note that the product is
specifically set to not equal \wild so that the resulting \vzgrp will include
all products.
On the other hand, the location is explicitly set to be equal to US. 
The other attributes, e.g., sales, profit, year, month are set to equal \wild: 
this implies that the visual groups are not employing any additional constraints
on those attributes. 
(This may be useful, for example when those attributes are not
relevant for the current visualization or set of visualizations.)
As mentioned before, \vzgrps have the semantics of ordered bags. Thus, $\vzsel$
operates on one tuple at a time in the order they appear in
$V$, and the result is in the same order the tuples are operated on. 

\paragraph{$\vzsort_{F(T)}(V)$:}
$\vzsort$ returns the \vzgrp sorted in an increasing order based on applying $F(T)$ on
each \vzchn in $V$, where $F(T)$ is a procedure that uses  
function $T$.  For example, $\vzsort_{-T}(V)$ might return the
visualizations in $V$ sorted in decreasing order of estimated slope.
This operator is not present in the ordered bag semantics,
but may be relevant when we want to reorder the ordered bag using a different criterion.
The function $F$ may be any higher-order function with no side effects. For a language to
\vzcmp, the language must be able to support any arbitrary $F$.

\paragraph{$\vzlimit_{k}(V)$:}
$\vzlimit$ returns the first $k$ \vzchns of $V$ ordered in the same way they
were in $V$. $\vzlimit$ is equivalent to the \texttt{LIMIT} statement in \sql.
$\vzlimit$ is often used in conjunction with $\vzsort$ to retrieve the top-$k$
visualizations with greatest increasing trends (e.g. $\vzlimit_{k}(\vzsort_{-T}(V))$).
When instead of a number $k$, the subscript to $\vzlimit$ is actually $[a: b]$, 
then the items of $V$ that are between
positions $a$ and $b$ in $V$ are returned.
Thus $\vzlimit$ offers identical functionality to the $[a:b]$ in ordered bag algebra,
with the convenient functionality of getting the top $k$ results by just having
one number as the subscript.
Instead of using $\vzlimit$, \vzalg also supports the use
of the syntax $V[i]$ to refer to the $i$th \vzchn in $V$, and 
$V[a:b]$ to refer to the ordered bag of \vzchns from positions $a$ to $b$.

\paragraph{$\vzdedup(V):$}
$\vzdedup$ returns the \vzchns in $V$ with the duplicates removed, in the order
of their first appearance.
Thus, $\vzdedup$ is defined identically to ordered bag algebra.

\subsubsection{Binary Operators.}
\paragraph{$V \vzunion U$ | $V \vzdiff U$ | $V \vzint U$:}
Returns the union / difference / intersection of $V$ and $U$.
These operations are just like the corresponding operations in
ordered bag algebra.

\paragraph{$\vzswap_{A}(V, U)$:}
$\vzswap$ returns a \vzgrp in which values of attribute
$A$ in $V$ are replaced with
the values of $A$ in $U$. 
Formally, assuming $A_i$ is the $i$th attribute of $V$
and $V$ has $n$ total attributes:
$\vzswap_{A_i}(V, U) = \pi_{A_1,\dots, A_{i-1}, A_{i+1}, \dots, A_{n}}(V) \times \pi_{A_i}(U)$. This can be useful for when the user would like to change
an axis: $\beta_{\text{X}}(V, \vzsel_{\text{X=year}}(\calV))$ will change the \vzchns in $V$
to have year as their x-axis. $\vzswap$ can also be used to combine multiple
dimensions as well. If we assume that $V$ has multiple Y values, we can do
$\vzswap_{\text{X}}(V, \vzsel_{\text{X}\ne*}(\calV))$ to have the \vzchns in
$V$ vary over both X and Y.
\techreport{This operator allows us to start with a set of visualizations and then ``pivot''
to focus on a different attribute, e.g., start with sales over time
visualizations and pivot to look at profit. 
Thus, the operator allows us to transform the space of \vzchns.}

\paragraph{$\vzdist_{F(D),A_1,...,A_j}(V, U)$:}
$\vzdist$ sorts the \vzchns in $V$ in increasing order based on their
distances to the corresponding \vzchns in $U$. More specifically, $\vzdist$
computes $F(D)(\vzsel_{A_1=a_1 \land ... \land A_j=a_j}(V),\\ \vzsel_{A_1=a_1
  \land ... \land A_j=a_j}(U)) \forall a_1,...,a_j \in
  \pi_{A_1,...,A_j}(V)$ and returns an increasingly sorted $V$ based on the
  results. If $\vzsel_{A_1=a_1 \land ... \land A_j=a_j}$ for either $V$ or $U$
  ever returns a non-singleton \vzgrp for any tuple
  $(a_1,...,a_j)$, the result of the operator is undefined.

\paragraph{$\vzfind_{F(D)}(V, U)$:}
$\vzfind$ sorts the \vzchns in $V$ in increasing order based on their distances
to a single reference \vzchn in singleton \vzgrp $U$. 
Thus, $U = [t]$.
$\vzfind$ computes $F(D)(V[i],
U[1]) \quad$ $\forall i \in \{1,\dots,|V|\}$, and returns a reordered $V$ based
on these values, where $F(D)$ is a procedure that uses $D$. 
If $U$ has
more than one \vzchn, the operation is undefined. 
$\vzfind$ is useful for
queries in which the user would like to find the top-$k$ most similar
visualizations to a reference: $\vzlimit_{k}(\vzfind_{D}(V,U))$,
where $V$ is the set of candidates and $U$ contains the reference.
Once again, this operator is similar to $\vzsort$, except that
it operates on the results of the comparison of individual \vzchns to a specific \vzchn.

\begin{table}[!]
  \scriptsize
    \centering
    \begin{tabular}{|c|c|c|c|c|c|c|c|}
      \hline
      X & Y & year & month & product & location & sales & profit \\ \hline
      year & sales & \wild & \wild & chair & US & \wild & \wild \\ \hline
      year & sales & \wild & \wild & table & US & \wild & \wild \\ \hline
      year & sales & \wild & \wild & stapler & US & \wild & \wild \\ \hline
      year & sales & \wild & \wild & printer & US & \wild & \wild \\
      \multicolumn{8}{|c|}{\vdots} \\
    \end{tabular}
  \caption{Results of performing unary operators on $\calV$ from Table~\ref{tab:setup}: $\vzsel_\theta (\calV)$ where $\theta = ($X=`year' $\land$ Y=`sales' $\land$ year=\wild $\land$ month=\wild $\land$ product $\ne$ \wild $\land$location=`US' $\land$ sales=\wild $\land$ profit=\wild$)$}.
  \label{tab:viz-un-opers}    \label{subtab:vzsel}
  \vspace{-10pt}
\end{table}


\subsection{Proof of \VzCmpness}
\label{subsec:viz-exp-proof}
We now attempt to quantify the expressiveness of \zq within the context of
\vzalg and the two functional primtives $T$ and $D$.
More
formally, we prove the following theorem:

\vspace{-7pt}
\begin{framed}
  \vspace{-12pt}
  \begin{theorem}
    Given well-defined functional primitives $T$ and $D$, \zq is \vzcmp
    with respect to $T$ and $D$: $VEC_{T,D}(\text{\zq})$ is true.
    \label{thm:main}
  \end{theorem}
  \vspace{-12pt}
\end{framed}
\vspace{-7pt}


Our proof for this theorem involves two major steps:
\begin{description}
  \item[Step 1.] We show that a visualization collection in \zq has as much
    expressive power as a \vzgrp of \vzalg, and therefore a visualization
    collection in
    \zq serves as an appropriate proxy of a \vzgrp in \vzalg.
  \item[Step 2.] For each operator in \vzalg, we show that there exists a \zq
    query which takes in visualization collection semantically equivalent to the
    \vzgrp operands and produces visualization collection semantically equivalent to
    the resultant \vzgrp.
\end{description}

\begin{lemma}
  A visualization collection of \zq has at least as much expressive power as a
  \vzgrp in \vzalg.
  \label{lemma:1}
\end{lemma}
\begin{proof}
  A \vzgrp $V$, with $n$ \vzchns, is a relation with $k+2$ columns and $n$
  rows, where $k$ is the number of attributes in the original relation. We show
  that for any \vzgrp $V$, we can come up with a \zq query $q$ which can
  produce a visualization collection that represents the same set of visualizations as
  $V$.
  
  \begin{table}[H]
    \centering\scriptsize
    {\tt   
      \begin{tabular}{|r|l|l|l|l|l|} \hline
        \textbf{Name} & \textbf{X} & \textbf{Y} & \textbf{Z1} & \textbf{...}
        & \textbf{Zk} \\ \hline
        f1 & $\pi_{\text{X}}(V[1])$ & $\pi_{\text{Y}}(V[1])$ & $E_{1,1}$ & ... &
        $E_{1,k}$ \\
        \vdots & \vdots & \vdots & \vdots & \vdots & \vdots \\
        fn & $\pi_{\text{X}}(V[n])$ & $\pi_{\text{Y}}(V[n])$ & $E_{n,1}$ & ... &
        $E_{n,k}$ \\
        *fn+1 \IN f1+...+fn & & & & & \\
        \hline
      \end{tabular}
    }
    \vspace{-10pt}\caption{\zq query $q$ which produces a visualization collection
    equal in expressiveness to \vzgrp $V$.}
    \label{tab:lemma-1}
  \vspace{-10pt}\end{table}

  Query $q$ has the format given by Table~\ref{tab:lemma-1}, where $V[i]$
  denotes the $i$th tuple of relation $V$ and:
  \[
    E_{i,j} = 
    \begin{cases}
      \text{`` ''} & \text{if } \pi_{A_j}(V[i]) = * \\
      A_j.\pi_{A_j}(V[i]) & \text{otherwise}
    \end{cases}
  \]
  Here, $A_j$ refers to the $j$th attribute of the original relation. The
  $i$th \vzchn of $V$ is represented with the {\tt fi} from $q$. The X and Y
  values come directly from the \vzchn using projection. For the {\tt Zj}
  column, if the $A_j$ attribute of \vzchn has any value than other
  than \wild, we must filter the data based on that value, so $E_{i,j} =
  A_j.\pi_{A_j}{V[i]}$. However, if the $A_j$ attribute is equal to
  \wild, then the corresponding element in {\tt fi} is left blank, signaling no
  filtering based on that attribute.

  After, we have defined a visualization collection {\tt fi} for each $i$th \vzchn in
  $V$, we take the sum (or concatenation) across all these visualization collections as
  defined in Appendix \ref{name-appex}, and the resulting
  {\tt fn+1} becomes equal to the \vzgrp $V$.
\end{proof}

\begin{lemma}
  $\vzsel_{\theta}(V)$ is expressible in \zq for all valid constraints $\theta$
  and \vzgrps $V$.
\end{lemma}
\begin{proof}
  We prove this by induction.
  
  The full context-free grammar (CFG) for $\theta$ in
  $\vzsel_{\theta}$ can be given by:
  \begin{flalign}
    \theta &\rightarrow E \;|\; E \land E \;|\; E \lor E \;|\; \epsilon &\\
    E &\rightarrow C \;|\; (E) \;|\; E \land C \;|\; E \lor C & \label{cfg:e}\\
    C &\rightarrow T_1 = B_1 \;|\; T_1 \ne B_1 \;|\; T_2 = B_2 \;|\; T_2 \ne
    B_2 &\\
    T_1 &\rightarrow X \;|\; Y &\\
    B_1 &\rightarrow A_1 \;|\; ... \;|\; A_k &\\
    T_2 &\rightarrow A_1 \;|\; ... \;|\; A_k &\\
    B_2 &\rightarrow \text{string} \;|\; \text{number} \;|\; *
  \end{flalign}
  where $\epsilon$ represents an empty string (no selection), and $X$, $Y$,
  and $A_1$, ..., $A_k$ refer to the attributes of $\calV$.

  To begin the proof by induction,
  we first show that \zq is capable of expressing the base expressions
  $\vzsel_{C}(V)$:
  $\vzsel_{T_1 = B_1}(V)$,
  $\vzsel_{T_1 \ne B_1}(V)$,
  $\vzsel_{T_2 = B_2}(V)$, and
  $\vzsel_{T_2 \ne B_2}(V)$.
  The high level idea for each of these proofs is to be come up with a
  \emph{filtering \vzgrp} $U$ which we take the intersection with to arrive at
  our desired result: $\exists U, \vzsel_{C}(V) = V \vzint U$.

  In the first two expressions, $T_1$ and $B_1$ refer to filters on the $X$ and
  $Y$ attributes of $V$; we
  have the option of either selecting a specific attribute ($T_1=B_1$) or
  excluding a specific attribute ($T_1 \ne B_1$). Tables \ref{tab:vzsel-c1} and
  \ref{tab:vzsel-c2} show \zq queries which express $\vzsel_{T_1=B_1}(V)$ for
  $T_1 \rightarrow X$ and $T_1 \rightarrow Y$ respectively. The \zq queries
  do the approximate equivalent of $\vzsel_{T_1=B_1}(V) = V \vzint
  \vzsel_{T_1=B_1}(\calV)$.

  \begin{table}[H]
    \centering\scriptsize
    {\tt   
      \begin{tabular}{|r|l|l|l|l|l|} \hline
        \textbf{Name} & \textbf{X} & \textbf{Y} & \textbf{Z1} & \textbf{...}
        & \textbf{Zk} \\ \hline
        f1 & - & - & - & ... & - \\
        f2 \IN f1 & & y1 \IN \_ & v1 \IN $A_1$.\_ & ... & vk \IN $A_k$.\_ \\
        f3 & $B_1$ & y1 & v1 & ... & vk \\
        *f4 \IN f1\^{}f3 & & & & & \\
        \hline
      \end{tabular}
    }
    \vspace{-10pt}\caption{\zq query which expresses $\vzsel_{X=B_1}(V)$.}
    \label{tab:vzsel-c1}
  \vspace{-10pt}\end{table}

  \begin{table}[H]
    \centering\scriptsize
    {\tt   
      \begin{tabular}{|r|l|l|l|l|l|} \hline
        \textbf{Name} & \textbf{X} & \textbf{Y} & \textbf{Z1} & \textbf{...}
        & \textbf{Zk} \\ \hline
        f1 & - & - & - & ... & - \\
        f2 \IN f1 & x1 \IN \_ & & v1 \IN $A_1$.\_ & ... & vk \IN $A_k$.\_ \\
        f3 & x1 & $B_1$ & v1 & ... & vk \\
        *f4 \IN f1\^{}f3 & & & & & \\
        \hline
      \end{tabular}
    }
    \vspace{-10pt}\caption{\zq query which expresses $\vzsel_{Y=B_1}(V)$.}
    \label{tab:vzsel-c2}
  \vspace{-10pt}\end{table}

  We have
  shown with Lemma~\ref{lemma:1} that a visualization collection is capable of
  expressing a \vzgrp, so we assume that {\tt f1}, the
  visualization collection which represents the operand $V$, is given to us for both of
  these tables. Since we
  do not know how {\tt f1} was derived, we use - for its axis variable columns.
  The second rows of these tables derive {\tt f2} from {\tt f1} and bind axis
  variables to the values of the non-filtered attributes. Here, although the
  set of visualizations present in {\tt f2} is exactly the same as {\tt f1},
  we now have a convenient way to iterate over the non-filtered attributes of
  {\tt f1} (for more information on derived visualization collections, please refer to
  Appendix~\ref{name-appex}).
  The third row
  combines the specified attribute $B_1$ with the non-filtered attributes of
  {\tt f2} to form the \emph{filtering visualization collection} {\tt f3}, which
  expresses the filtering \vzgrp $U$ from above. We then take the
  intersection between {\tt f1} and the filtering visualization collection {\tt f3} to
  arrive at our desired visualization collection {\tt f4}, which represents
  the resultant \vzgrp $\vzsel_{T_1=B_1}(V)$. Although, we earlier said that we
  would come up with {\tt f3} $=\vzsel_{T_1=B_1}(\calV)$, in truth, we come up
  with {\tt f3} $=B_1 \times \pi_{Y,A_1,...,A_k}(V)$ for $T_1 \rightarrow X$
  and {\tt f3} $=\pi_{X,A_1,...,A_k}(V) \times B_1$ for $T_1 \rightarrow Y$
  because they are easier to express in \zq; regardless we still end up with
  the correct resulting set of visualizations.

  Tables \ref{tab:vzsel-c3} and \ref{tab:vzsel-c4} show \zq queries which
  express $\vzsel_{T_1 \ne B_1}(V)$ for $T_1 \rightarrow X$ and $T_1
  \rightarrow Y$ respectively. Similar to the queries above, these queries
  perform the approximate
  equivalent of $\vzsel_{T_1 \ne B_1}(V) = V \vzint \vzsel_{T_1 \ne
  B_1}(\calV)$. We once again assume {\tt f1} is a given visualization collection which
  represents the operand $V$, and we come up with a filtering visualization collection
  {\tt f3} which mimics the effects of (though is not completely equivalent to) $\vzsel_{T_1
    \ne B_1}(\calV)$. We then take the intersection between {\tt f1} and {\tt
    f3} to arrive at {\tt f4} which represents the resulting $\vzsel_{T_1 \ne
    B_1}(V)$.

  \begin{table*}[!htbp]
    \centering\scriptsize
    {\tt   
      \begin{tabular}{|r|l|l|l|l|l|} \hline
        \textbf{Name} & \textbf{X} & \textbf{Y} & \textbf{Z1} & \textbf{...}
        & \textbf{Zk} \\ \hline
        f1 & - & - & - & ... & - \\
        f2 \IN f1 & x1 \IN \_ & y1 \IN \_ & v1 \IN $A_1$.\_ & ... & vk \IN $A_k$.\_ \\
        f3 & x2 \IN x1 - \{$B_1$\} & y1 & v1 & ... & vk \\
        *f4 \IN f1\^{}f3 & & & & & \\
        \hline
      \end{tabular}
    }
    \vspace{-10pt}\caption{\zq query which expresses $\vzsel_{X \ne B_1}(V)$.}
    \label{tab:vzsel-c3}
  \vspace{-10pt}\end{table*}

  \begin{table*}[!htbp]
    \centering\scriptsize
    {\tt   
      \begin{tabular}{|r|l|l|l|l|l|} \hline
        \textbf{Name} & \textbf{X} & \textbf{Y} & \textbf{Z1} & \textbf{...}
        & \textbf{Zk} \\ \hline
        f1 & - & - & - & ... & - \\
        f2 \IN f1 & x1 \IN \_ & y1 \IN \_ & v1 \IN $A_1$.\_ & ... & vk \IN $A_k$.\_ \\
        f3 & x1 & y2 \IN y1 - \{$B_1$\} & v1 & ... & vk \\
        *f4 \IN f1\^{}f3 & & & & & \\
        \hline
      \end{tabular}
    }
    \vspace{-10pt}\caption{\zq query which expresses $\vzsel_{Y \ne B_1}(V)$.}
    \label{tab:vzsel-c4}
  \vspace{-10pt}\end{table*}

  The expressions $\vzsel_{T_2=B_2}$ and $\vzsel_{T_2 \ne B_2}$ refer to filters on the
  $A_1,...,A_k$ attributes of $V$. Specifically, $T_2$ is some attribute $A_j
  \in \{A_1,...,A_k\}$ and $B_2$ is the attribute value which is selected or
  excluded.
  Here, we have an additional complication to the proof since any attribute
  $A_j$ can also filter for or exclude \wild. First, we
  show \zq is capable of expressing $\vzsel_{T_2 = B_2'}$ and $\vzsel_{T_2
  \ne B_2'}$ for which $B_2' \neq *$; that is $B_2'$ is any attribute
  value which is not \wild. Tables \ref{tab:vzsel-c5} and \ref{tab:vzsel-c6}
  show the \zq queries which express $\vzsel_{T_2 = B_2'}(V)$ and $\vzsel_{T_2
  \ne B_2'}(V)$ respectively. Note the similarity between these queries and the
  queries for $\vzsel_{T_1=B_1}(V)$ and $\vzsel_{T_1 \ne B_1}(V)$.

  \begin{table*}[!htbp]
    \centering\scriptsize
    {\tt   
      \begin{tabular}{|r|l|l|l|l|l|l|l|} \hline
        \textbf{Name} & \textbf{X} & \textbf{Y} & \textbf{Z1} & \textbf{...} &
        \textbf{Zj} & \textbf{...} & \textbf{Zk} \\ \hline
        f1 & - & - & - & ... & - & ... & - \\
        f2 \IN f1 & x1 \IN \_ & y1 \IN \_ & v1 \IN $A_1$.\_ & ... & & ... & vk \IN $A_k$.\_ \\
        f3 & x1 & y1 & v1 & ... & $B_2'$ & ... & vk \\
        *f4 \IN f1\^{}f3 & & & & & & & \\
        \hline
      \end{tabular}
    }
    \vspace{-10pt}\caption{\zq query which expresses $\vzsel_{A_j =  B_2'}(V)$
    when $B_2' \ne *$.}
    \label{tab:vzsel-c5}
  \vspace{-10pt}\end{table*}

  \begin{table*}[!htbp]
    \centering\scriptsize
    {\tt   
      \begin{tabular}{|r|l|l|l|l|l|l|l|} \hline
        \textbf{Name} & \textbf{X} & \textbf{Y} & \textbf{Z1} & \textbf{...} &
        \textbf{Zj} & \textbf{...} & \textbf{Zk} \\ \hline
        f1 & - & - & - & ... & - & ... & - \\
        f2 \IN f1 & x1 \IN \_ & y1 \IN \_ & v1 \IN $A_1$.\_ & ... & vj \IN $A_j$.\_ & ... & vk \IN $A_k$.\_ \\
        f3 & x1 & y1 & v1 & ... & uj \IN vj - \{$B_2'$\} & ... & vk \\
        *f4 \IN f1\^{}f3 & & & & & & & \\
        \hline
      \end{tabular}
    }
    \vspace{-10pt}\caption{\zq query which expresses $\vzsel_{A_j \ne
    B_2'}(V)$ when $B_2' \ne *$.}
    \label{tab:vzsel-c6}
  \vspace{-10pt}\end{table*}

  For $\vzsel_{T_2=*}(V)$ and $\vzsel_{T_2 \ne *}(V)$, Tables
  \ref{tab:vzsel-c7} and \ref{tab:vzsel-c8} show the corresponding queries. In
  Table~\ref{tab:vzsel-c7}, we explicitly avoid setting a value for {\tt Zj}
  for {\tt f3} to emulate $A_j=*$ for the filtering visualization collection. In
  Table~\ref{tab:vzsel-c8}, {\tt f3}'s {\tt Zj} takes on all possible values
  from $A_j${\tt.*}, but that means that a value is set for {\tt Zj} (i.e.,
  $T_2 \ne *$).

  \begin{table*}[!htbp]
    \centering\scriptsize
    {\tt   
      \begin{tabular}{|r|l|l|l|l|l|l|l|} \hline
        \textbf{Name} & \textbf{X} & \textbf{Y} & \textbf{Z1} & \textbf{...} &
        \textbf{Zj} & \textbf{...} & \textbf{Zk} \\ \hline
        f1 & - & - & - & ... & - & ... & - \\
        f2 \IN f1 & x1 \IN \_ & y1 \IN \_ & v1 \IN $A_1$.\_ & ... & & ... & vk \IN $A_k$.\_ \\
        f3 & x1 & y1 & v1 & ... & & ... & vk \\
        *f4 \IN f1\^{}f3 & & & & & & & \\
        \hline
      \end{tabular}
    }
    \vspace{-10pt}\caption{\zq query which expresses $\vzsel_{A_j = *}(V)$.}
    \label{tab:vzsel-c7}
  \vspace{-10pt}\end{table*}

  \begin{table*}[!htbp]
    \centering\scriptsize
    {\tt   
      \begin{tabular}{|r|l|l|l|l|l|l|l|} \hline
        \textbf{Name} & \textbf{X} & \textbf{Y} & \textbf{Z1} & \textbf{...} &
        \textbf{Zj} & \textbf{...} & \textbf{Zk} \\ \hline
        f1 & - & - & - & ... & - & ... & - \\
        f2 \IN f1 & x1 \IN \_ & y1 \IN \_ & v1 \IN $A_1$.\_ & ... & & ... & vk \IN $A_k$.\_ \\
        f3 & x1 & y1 & v1 & ... & vj \IN $A_j$.* & ... & vk \\
        *f4 \IN f1\^{}f3 & & & & & & & \\
        \hline
      \end{tabular}
    }
    \vspace{-10pt}\caption{\zq query which expresses $\vzsel_{A_j \ne *}(V)$.}
    \label{tab:vzsel-c8}
  \vspace{-10pt}\end{table*}

  Now that we have shown how to express the base operations, we next assume \zq
  is capable of expressing any arbitrary complex filtering operations
  $\vzsel_{E'}$ where $E'$ comes from Line~\ref{cfg:e} of the CFG.
  Specifically, we assume that given a visualization collection {\tt f1} which
  expresses $V$, there exists a filtering visualization collection {\tt f2} for which
  $\vzsel_{E'}(V)=$ {\tt f1\^{}f2}. Given this assumption, we now must take the
  inductive step, apply Line~\ref{cfg:e}, and prove that $\vzsel_{E \rightarrow (E')}(V)$,
  $\vzsel_{E \rightarrow E' \land C}(V)$, and $\vzsel_{E \rightarrow E' \lor
  C}(V)$ are all expressible in
  \zq for any base constraint $C$.

  \paragraph{$\vzsel_{E \rightarrow (E')}(V)$:} This case is trivial. Given
  {\tt f1} which represents $V$ and {\tt f2} which is the filtering
  visualization
  collection for $E'$, we simply the intersect the two to get {\tt f3 \IN f1\^{}f2}
  which represents $\vzsel_{E \rightarrow (E')}$.

  \paragraph{$\vzsel_{E \rightarrow E' \land C}$:} Once again assume we are
  given {\tt f1} which represents $V$ and {\tt f2} which is the filtering
  visualization collection of $E'$. Based on the base expression proofs above, we know
  that given any base constraint $C$, we can find a filtering visualization collection
  for it; call this filtering visualization collection {\tt f3}. We can then see that
  {\tt f2\^{}f3} is the filtering visualization collection of $E \rightarrow E' \land C$, and {\tt
  f4 \IN f1\^{}(f2\^{}f3)} represents $\vzsel_{E \rightarrow E \land C}(V)$.

  \paragraph{$\vzsel_{E \rightarrow E' \lor C}$:} Once again assume we are
  given {\tt f1} which represents $V$, {\tt f2} which is the filtering
  visualization collection of $E'$, and we can find a filtering visualization collection {\tt
  f3} for $C$.  We can then see that {\tt f2+f3} is the filtering visualization
  collection of $E \rightarrow E' \lor C$, and {\tt f4 \IN f1\^{}(f2+f3)} represents $\vzsel_{E
  \rightarrow E \lor C}(V)$.

  With this inductive step, we have shown that for all complex constraints $E$
  of the form given by Line~\ref{cfg:e} of the CFG, we can find a \zq query
  which expresses $\vzsel_{E}(V)$. Given this, we can finally show that \zq is capable
  of expressing $\vzsel_{\theta}(V)$ for all $\theta$: $\vzsel_{\theta \rightarrow E}(V)$,
  $\vzsel_{\theta \rightarrow E \land E')}(V)$,
  $\vzsel_{\theta \rightarrow E \lor E'}(V)$, and
  $\vzsel_{\theta \rightarrow \epsilon}(V)$.

  \paragraph{$\vzsel_{\theta \rightarrow E}(V)$:} This case is once again
  trivial. Assume, we are given {\tt f1} which represents $V$, and {\tt f2},
  which is the filtering visualization collection of $E$, {\tt f3 \IN f1\^{}f2} represents
  $\vzsel_{\theta \rightarrow E}(V)$.

  \paragraph{$\vzsel_{\theta \rightarrow E \land E'}(V)$:} Assume, we are given
  {\tt f1} which represents $V$, {\tt f2}, which is the filtering visualization collection of
  $E$, and {\tt f3}, which is the filtering visualization collection of $E'$.
  {\tt f2\^{}f3} is the filtering visualization collection of $\theta \rightarrow E
  \land E'$, and {\tt f4 \IN f1\^{}(f2\^{}f3)} represents $\vzsel_{\theta
  \rightarrow E \land E'}(V)$.

  \paragraph{$\vzsel_{\theta \rightarrow E \lor E'}(V)$:} Assume, we are given
  {\tt f1} which represents $V$, {\tt f2} which is the filtering visualization collection of
  $E$, and {\tt f3} which is the filtering visualization collection of $E'$.
  {\tt f2+f3} is the filtering visualization collection of $\theta \rightarrow E
  \lor E'$, and {\tt f4 \IN f1\^{}(f2+f3)} represents $\vzsel_{\theta
  \rightarrow E \lor E'}(V)$.

  \paragraph{$\vzsel_{\theta \rightarrow \epsilon}(V)$:} This is the case in
  which no filtering is done. Therefore, given {\tt f1} which represents $V$,
  we can simply return {\tt f1}.
\end{proof}

\begin{lemma}
  $\vzsort_{F(T)}(V)$ is expressible in \zq for all valid
  functionals $F$ of $T$ and \vzgrps $V$.
\end{lemma}
\begin{proof}
  \begin{table*}[!htbp]
    \centering\scriptsize
    {\tt   
      \begin{tabular}{|r|l|l|l|l|l|l|} \hline
        \textbf{Name} & \textbf{X} & \textbf{Y} & \textbf{Z1} & \textbf{...}
        & \textbf{Zk} & \textbf{Process} \\ \hline
        f1 & - & - & - & ... & - & \\
        f2 \IN f1 & x1 \IN \_ & y1 \IN \_ & v1 \IN $A_1$.\_ & .. & vk \IN $A_k$.\_
        &
        x2, y2, u1, ..., uk \IN $argmin_{x1,y1,v1,...,vk}[k=\infty] F(T)(f2)$ \\
        *f3 & x2 & y2 & u1 & ... & uk & \\
        \hline
      \end{tabular}
    }
    \vspace{-10pt}\caption{\zq query $q$ which expresses $\vzsort_{F(T)}(V)$.
    }
    \label{tab:proof-vzsort}
  \vspace{-10pt}\end{table*}

  Assume {\tt f1} is the visualization collection which represents $V$. Query $q$
  given by Table~\ref{tab:proof-vzsort} produces visualization collection {\tt f3}
  which expresses $\vzsort_{F(T)}(V)$.
\end{proof}

\begin{lemma}
  $\vzlimit_{[a:b]}(V)$ is expressible in \zq for all valid intervals $a:b$
  and \vzgrps $V$.
\end{lemma}
\begin{proof}
  \begin{table*}[!htbp]
    \centering\scriptsize
    {\tt   
      \begin{tabular}{|r|l|l|l|l|l|l|} \hline
        \textbf{Name} & \textbf{X} & \textbf{Y} & \textbf{Z1} & \textbf{...}
        & \textbf{Zk} & \textbf{Process} \\ \hline
        f1 & - & - & - & ... & - & \\
        *f2 \IN f1[a:b] &  & & & & & \\
        \hline
      \end{tabular}
    }
    \vspace{-10pt}\caption{\zq query $q$ which expresses $\vzlimit_{[a:b]}(V)$.
    }
    \label{tab:proof-vzlimit}
  \vspace{-10pt}\end{table*}

  Assume {\tt f1} is the visualization collection which represents $V$. Query $q$
  given by Table~\ref{tab:proof-vzlimit} produces visualization collection {\tt f2}
  which expresses $\vzlimit_{[a:b]}(V)$.
\end{proof}

%

\begin{lemma}
  $\vzdedup(V)$ is expressible in \zq for all valid \vzgrps $V$.
\end{lemma}
\begin{proof}
  \begin{table*}[!htbp]
    \centering\scriptsize
    {\tt   
      \begin{tabular}{|r|l|l|l|l|l|l|} \hline
        \textbf{Name} & \textbf{X} & \textbf{Y} & \textbf{Z1} & \textbf{...}
        & \textbf{Zk} & \textbf{Process} \\ \hline
        f1 & - & - & - & ... & - & \\
        *f2 \IN f1 &  & & & & & \\
        \hline
      \end{tabular}
    }
    \vspace{-10pt}\caption{\zq query $q$ which expresses $\vzdedup(V)$.
    }
    \label{tab:proof-vzdedup}
  \vspace{-10pt}\end{table*}

  Assume {\tt f1} is the visualization collection which represents $V$. Query $q$
  given by Table~\ref{tab:proof-vzdedup} produces visualization collection {\tt f2}
  which expresses $\vzdedup(V)$.
\end{proof}

\begin{lemma}
  $V \vzunion U$ is expressible in \zq for all valid \vzgrps $V$ and $U$.
\end{lemma}
\begin{proof}
  \begin{table*}[!htbp]
    \centering\scriptsize
    {\tt   
      \begin{tabular}{|r|l|l|l|l|l|l|} \hline
        \textbf{Name} & \textbf{X} & \textbf{Y} & \textbf{Z1} & \textbf{...}
        & \textbf{Zk} & \textbf{Process} \\ \hline
        f1 & - & - & - & ... & - & \\
        f2 & - & - & - & ... & - & \\
        *f3 \IN f1+f2 & & & & ... & & \\
        \hline
      \end{tabular}
    }
    \vspace{-10pt}\caption{\zq query $q$ which expresses $V \vzunion U$.
    }
    \label{tab:proof-vzunion}
  \vspace{-10pt}\end{table*}

  Assume {\tt f1} is the visualization collection which represents $V$ and {\tt f2}
  represents $U$.  Query $q$ given by Table~\ref{tab:proof-vzunion} produces
  visualization collection {\tt f3} which expresses $V \vzunion U$.
\end{proof}

\begin{lemma}
  $V \vzdiff U$ is expressible in \zq for all valid \vzgrps $V$ and $U$.
\end{lemma}
\begin{proof}
  \begin{table*}[!htbp]
    \centering\scriptsize
    {\tt   
      \begin{tabular}{|r|l|l|l|l|l|l|} \hline
        \textbf{Name} & \textbf{X} & \textbf{Y} & \textbf{Z1} & \textbf{...}
        & \textbf{Zk} & \textbf{Process} \\ \hline
        f1 & - & - & - & ... & - & \\
        f2 & - & - & - & ... & - & \\
        *f3 \IN f1-f2 & & & & ... & & \\
        \hline
      \end{tabular}
    }
    \vspace{-10pt}\caption{\zq query $q$ which expresses $V \vzdiff U$.
    }
    \label{tab:proof-vzdiff}
  \vspace{-10pt}\end{table*}

  Assume {\tt f1} is the visualization collection which represents $V$ and {\tt f2}
  represents $U$.  Query $q$ given by Table~\ref{tab:proof-vzdiff} produces
  visualization collection {\tt f3} which expresses $V \vzdiff U$. The proof for
  $\vzint$ can be shown similarly.
\end{proof}

\begin{lemma}
  $\vzswap_{A}(V, U)$ is expressible in \zq for all valid attributes $A$ in
  $\calV$ and \vzgrps $V$ and $U$.
\end{lemma}
\begin{proof}

  \begin{table*}[!htbp]
    \centering\scriptsize
    {\tt   
      \begin{tabular}{|r|l|l|l|l|l|} \hline
        \textbf{Name} & \textbf{X} & \textbf{Y} & \textbf{Z1} & \textbf{...}
        & \textbf{Zk} \\ \hline
        f1 & - & - & - & ... & - \\
        f2 & - & - & - & ... & - \\
        f3 \IN f1 & & y1 \IN \_ & v1 \IN $A_1$.\_ & ... & vk \IN $A_k$.\_ \\
        f4 \IN f2 & x1 \IN \_ & & & & \\
        *f5 & x1$^2$ & y1$^1$ & v1$^1$ & ... & vk$^1$ \\
        \hline
      \end{tabular}
    }
    \vspace{-10pt}\caption{\zq query $q$ which expresses $\vzswap_{A}(V, U)$
      where $A=X$.
    }
    \label{tab:proof-vzswap1}
  \vspace{-10pt}\end{table*}

  \begin{table*}[!htbp]
    \centering\scriptsize
    {\tt   
      \begin{tabular}{|r|l|l|l|l|l|} \hline
        \textbf{Name} & \textbf{X} & \textbf{Y} & \textbf{Z1} & \textbf{...}
        & \textbf{Zk} \\ \hline
        f1 & - & - & - & ... & - \\
        f2 & - & - & - & ... & - \\
        f3 \IN f1 & x1 \IN \_ & & v1 \IN $A_1$.\_ & ... & vk \IN $A_k$.\_ \\
        f4 \IN f2 & & y1 \IN \_ & & & \\
        *f5 & x1$^1$ & y1$^2$ & v1$^1$ & ... & vk$^1$ \\
        \hline
      \end{tabular}
    }
    \vspace{-10pt}\caption{\zq query $q$ which expresses $\vzswap_{A}(V, U)$
      where $A=Y$.
    }
    \label{tab:proof-vzswap2}
  \vspace{-10pt}\end{table*}

  \begin{table*}[!htbp]
    \centering\scriptsize
    {\tt   
      \begin{tabular}{|r|l|l|l|l|l|l|l|l|l|} \hline
        \textbf{Name} & \textbf{X} & \textbf{Y} & \textbf{Z1} & \textbf{...}
        & \textbf{Zj-1} & \textbf{Zj} & \textbf{Zj+1} & \textbf{...}
        & \textbf{Zk} \\ \hline
        f1 & - & - & - & ... & - & - & - &... & - \\
        f2 & - & - & - & ... & - & - & - &... & - \\
        f3 \IN f1 & x1 \IN \_ & y1 \IN \_ & v1 \IN $A_1$.\_ & ... & vj-1 \IN
        $A_{j-1}$.\_ & & vj+1 \IN $A_{j+1}$.\_ & ... & vk \IN $A_k$.\_ \\
        f4 \IN f2 & & & & & & uj \IN $A_{j}$.\_ & & & \\
        *f5 & x1$^1$ & y1$^1$ & v1$^1$ & ... & vj-1$^1$ & uj$^2$ & vj+1$^1$ & ... & vk$^1$ \\
        \hline
      \end{tabular}
    }
    \vspace{-10pt}\caption{\zq query $q$ which expresses $\vzswap_{A}(V, U)$
      where $A=A_j$ and $A_j$ is an attribute from $\calR$
    }
    \label{tab:proof-vzswap3}
  \vspace{-10pt}\end{table*}

  Assume {\tt f1} is the visualization collection which represents $V$ and {\tt f2}
  represents $U$. There are three cases we must handle depending on the value
  of $A$ due to the structure of columns in \zq: \begin{inparaenum}[\itshape
    (i)\upshape] \item $A=X$ \item $A=Y$ \item $A=A_j$ where $A_j$ is an
  attribute from the original relation $\calR$ \end{inparaenum}. For each
  of the three cases, we produce a separate query which expresses $\vzswap$.
  For $A=X$, the query given by Table~\ref{tab:proof-vzswap1} produces {\tt f5}
  which is equivalent to $\vzswap_{X}(V,U)$ We use the superscripts 
  in the last row so that cross product conforms to the ordering defined in
  Section~\ref{sec:viz-exp-oper}. For more information about the superscripts,
  please refer to Appendix~\ref{name-appex}. For $A=Y$, the query given by
  Table~\ref{tab:proof-vzswap2} produces {\tt f5} which is equivalent to
  $\vzswap_{Y}(V,U)$, and for $A=A_j$, the query given by
  Table~\ref{tab:proof-vzswap3} produces {\tt f5} which is equivalent to
  $\vzswap_{A_j}(V,U)$.
\end{proof}

\begin{lemma}
  $\vzdist_{F(D),A_1,...,A_j}(V, U)$ is expressible in \zq for all valid
  attributes $A_1,...,A_j$ and \vzgrps $V$ and $U$.
\end{lemma}
\begin{proof}

  \begin{table*}[!htbp]
    \centering\scriptsize
    {\tt   
      \begin{tabular}{|r|l|l|l|l|l|l|} \hline
        \textbf{Name} & \textbf{X} & \textbf{Y} & \textbf{Z1} & \textbf{...} &
        \textbf{Zj} &
        \textbf{Process} \\ \hline
        f1 & - & - & - & ... & - & \\
        f2 & - & - & - & ... & - & \\
        f3 \IN f1 & & & v1 \IN $A_1$.\_ & ...  & vj \IN
        $A_j$.\_ & \\
        f4 \IN f2.order & & & v1 \TO & ... & vj \TO & u1, ..., uj \IN
        $argmin_{v1,...,vj}[k=\infty] F(D)(f3,f4)$\\
        *f5 \IN f1.order & & & u1 \TO & ... & uj \TO & \\
        \hline
      \end{tabular}
    }
    \vspace{-10pt}\caption{\zq query $q$ which expresses
      $\vzdist_{F(D),A_1,...,A_j}(V, U)$.
    }
    \label{tab:proof-vzdist}
    \vspace{-10pt}\end{table*}

  Assume {\tt f1} is the visualization collection which represents $V$, and {\tt f2}
  represents $U$. Without loss of generality, assume the attributes we want to
  match on ($A_1,...,A_j$) are the first $j$ attributes of $\calR$.
  Query $q$ given by Table~\ref{tab:proof-vzdist} produces
  visualization collection {\tt f5} which expresses $\vzdist_{F(D),A_1,...,A_j}(V,U)$.
  In the table, we first retrieve the values for $(A_1,...,A_j)$ using {\tt f3} and reorder
  {\tt f2} based on these values to get {\tt f4}.
  We then compare the visualizations in {\tt f3}
  and {\tt f4} with respect to $(A_1,...,A_j)$ using the distance function
  $F(D)$ and retrieve the increasingly sorted $(A_1,...,A_j)$ values from the
  $argmin$. We are guaranteed that visualizations in {\tt f3} and {\tt f4}
  match up perfectly with respect to $(A_1,...,A_j)$ since the definition in
  Section~\ref{sec:viz-exp-oper} allows exactly one \vzchn to result from any
  $\vzsel_{A_1=a_1 \land ... \land A_j=a_j}$. Finally, we
  reorder {\tt f1} according to these values to retrieve {\tt f5}.
  For more information on the {\tt .order} operation, please refer to
  Appendix~\ref{name-appex}.
\end{proof}

\begin{lemma}
  $\vzfind_{F(D)}(V, U)$ is expressible in \zq for all valid functionals $F$ of
  $D$ and \vzgrps $V$ and singleton \vzgrps $U$.
\end{lemma}
\begin{proof}
  \begin{table*}[!htbp]
    \centering\scriptsize
    {\tt   
      \begin{tabular}{|r|l|l|l|l|l|l|} \hline
        \textbf{Name} & \textbf{X} & \textbf{Y} & \textbf{Z1} & \textbf{...}
        & \textbf{Zk} & \textbf{Process} \\ \hline
        f1 & - & - & - & ... & - & \\
        f2 & - & - & - & ... & - & \\
        f3 \IN f1 & x1 \IN \_ & y1 \IN \_ & v1 \IN $A_1$.\_ & ... & vk \IN $A_k$.\_
        &
        x2, y2, u1, ..., uk \IN $argmin_{x1,y1,v1,...,vk}[k=\infty] {F(D)}(f3,f2)$
        \\
        *f4 & x2 & y2 & u1 & ... & uk & \\
        \hline
      \end{tabular}
    }
    \vspace{-10pt}\caption{\zq query $q$ which expresses $\vzfind_{F(D)}(V, U)$.
    }
    \label{tab:proof-vzfind}
  \vspace{-10pt}\end{table*}

  Assume {\tt f1} is the visualization collection which represents $V$ and {\tt f2}
  represents $U$.  Query $q$ given by Table~\ref{tab:proof-vzfind} produces
  visualization collection {\tt f4} which expresses $\vzfind_{F(D)}(V, U)$.
\end{proof}

Although we have come up with a formalized algebra to measure the
expressiveness of \zq, \zq is actually more expressive than \vzalg. For
example, \zq allows the user to nest multiple levels of iteration in the
Process column as in Table~\ref{tab:outlier}.
Nevertheless,
\vzalg serves as a useful minimum metric for determining the expressiveness of
\vzlans.
Other visual analytics tools like Tableau are capable of expressing the
selection operator $\vzsel$ in \vzalg, but they are incapable of expressing
the other operators which compare and filter visualizations based on
functional primitives $T$ and $D$. 
General purpose programming
languages with analytics libraries such as Python and
Scikit-learn~\cite{pedregosa2011scikit} are \vzcmp since they are
Turing-complete, but \zq's declarative syntax strikes a novel balance between
simplicity and expressiveness which allows even non-programmers to become data
analysts as we see in Section~\ref{sec:userstudy}.


\vspace{-3pt}

\vspace{-4pt}
\section{Query Execution}
\label{subsec:query_execution}
\vspace{-2pt}

In \zv, \zq queries are automatically parsed and executed by the back-end.
The \zq compiler translates \zq queries into a combination of \sql
queries to fetch the visualization collections and processing tasks to
operate on them.
We present a basic graph-based translation for \zq
 and then provide 
several optimizations to the graph which reduce the
overall runtime considerably.
%
\vspace{4pt}
\subsection{Basic Translation}
\begin{figure}
\vspace{-5pt}
  \centering
  \includegraphics[width=.6\linewidth]{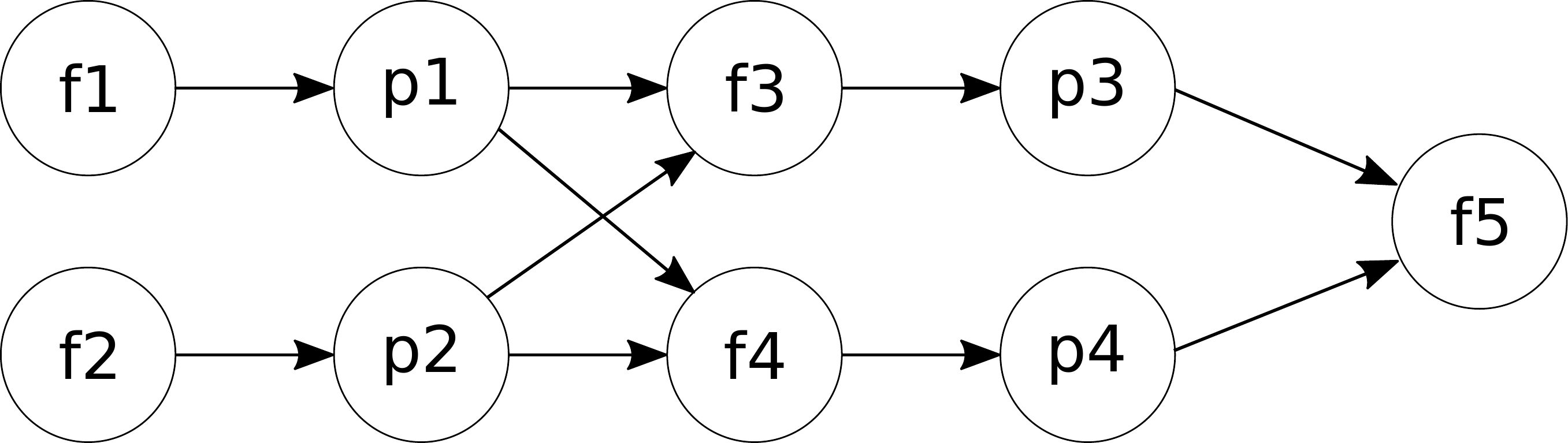}
  \caption{The query plan for the query presented in Table~\ref{tab:real2}.}
  \label{fig:real2-plan}
  \vspace{-15pt}
\end{figure}

Every valid \zq query can be transformed into a query plan 
in the form of a directed
acyclic graph (DAG). 
The DAG contains \cnode[s] (or \emph{collection nodes})
to represent the collections of visualizations in the \zq query and \pnode[s]
(or \emph{process nodes}) to represent the optimizations (or
\emph{processes}) in the \zqproc column. 
Directed edges are drawn between nodes
that have a dependency relationship. 
Using this query plan,
the \zq engine can determine at each step which visualization collection to
fetch from the database or which process to execute.
The full steps to build a query plan for any \zq query is as follows: 
\begin{inparaenum}[(i)]
\item
  Create a \cnode[] or \emph{collection node} for every collection of
  visualizations (including singleton collections).
\item
  Create a \pnode[] or \emph{processor node} for every optimization (or
  \emph{process}) in the \zqproc column.
\item
  For each \cnode[], if any of its \axvar[s] are derived as a result of a
  process, connect a directed edge from the corresponding \pnode[].
\item
  For each \pnode[], connect a directed edge from the \cnode[] of each
  collection which appears in the process.
\end{inparaenum}
\later{
\begingroup
\plitemsep=.2em
\pltopsep=.2em
\begin{compactenum}
\item
  Create a \cnode[] or \emph{collection node} for every collection of
  visualizations (including singleton collections).
\item
  Create a \pnode[] or \emph{processor node} for every optimization (or
  \emph{process}) in the \zqproc column.
\item
  For each \cnode[], if any of its \axvar[s] are derived as a result of a
  process, connect a directed edge from the corresponding \pnode[].
\item
  For each \pnode[], connect a directed edge from the \cnode[] of each
  collection which appears in the process.
\end{compactenum}
\endgroup
}
%
Following these steps, we can translate our realistic query example in
Table~\ref{tab:real2} to its query plan presented in
Figure~\ref{fig:real2-plan}. The \cnode[s] are annotated with {\tt f\#},
and the \pnode[s] are annotated with {\tt p\#} (the $i$th \pnode[] refers to the
process in the $i$th row of the table). Here, {\tt f1} is a root node
with no dependencies since it does not depend on any process, whereas {\tt f5}
depends on the results of both {\tt p3} and {\tt p4} and have edges coming from
both of them.
Once the query plan has been constructed, 
the \zq engine can execute it using the simple algorithm presented in
in Algorithm~\ref{simplealg:basic}.

  \vspace{-0.75em}
\begin{framed}
  \vspace{-1.5em}
\begin{simplealg}
\small
Algorithm to execute \zq query plan:
\begin{compactenum}
\item
  Search for a node with either no parents or one whose parents have all been
  marked as done.
\item 
  Run the corresponding task for that node and mark the node as done.
\item
  Repeat steps 1 and 2 until all nodes have been marked as done.
\end{compactenum}
\label{simplealg:basic}
\end{simplealg}
  \vspace{-1.5em}
\end{framed}
  \vspace{-1em}
For \cnode[s], the corresponding task is to retrieve the data for visualization
collection, while for \pnode[s], the corresponding task is to execute the
process.

\paragraph{\cnode[] translation:}
At a high level, for \cnode[s], the appropriate \sql group-by queries
are issued to the database to compose the data for multiple visualizations
at once. 
Specifically,
for the simplest setting where there are no collections specified
for X or Y, a \sql query in the form of:
\begin{quoting}\small
\begin{verbatim}
      SELECT X, A(Y), Z, Z2, ...  WHERE C(X, Y, Z, Z2, ...)
        GROUP BY X, Z, Z2, ...  ORDER BY X, Z, Z2, ...
\end{verbatim}
\end{quoting}
is issued to the database, where {\tt X} is the X column attribute, {\tt Y} is
the Y column attribute, {\tt A(Y)} is the aggregation function on Y (specified
in the Viz column), {\tt Z, Z2, ...} are the attributes/dimensions we are
iterating over in the Z columns, 
while {\tt C(X, Y, Z, Z2, ...)} refers to any additional constraints
specified in the {\tt Z} columns. The {\tt ORDER BY} is inserted to ensure
that all rows corresponding to a visualization are grouped together, in order. 
As an example, the \sql
query for the \cnode[] for {\tt f1} in Table~\ref{tab:outlier} would have the form:
\begin{quoting}\small
\begin{verbatim}
      SELECT year, SUM(sales), product
        GROUP BY year, product  ORDER BY year, product
\end{verbatim}
\end{quoting}
If a collection is specified for the y-axis, each attribute in the collection
is appended to the {\tt SELECT} clause. If a collection is specified for the
x-axis, a separate query must be issued for every X attribute in the collection.
The results of the \sql query are then packed into a $m$-dimensional array
(each dimension in the array corresponding to a dimension in the collection)
and labeled with its {\tt f\#} tag. 

\paragraph{\pnode[] translation:} 
At a high level, for \pnode[s], depending on the structure
of the expression within the process, 
the appropriate pseudocode is generated
to operate on the visualizations.
To illustrate, say our process is trying to find the
top-10 values for which a trend is maximized/minimized 
with respect to various
dimensions (using $T$), and the process has the form:
\vspace{-5pt}
\begin{equation}
  \langle argopt \rangle_{v0}[k=k']
  \; \bigg[ \langle op_1 \rangle_{v1}
    \bigg[
      \langle op_2 \rangle_{v2}
      \cdots
      \bigg[
        \langle op_m \rangle_{vm}
        T(f1)
      \bigg]
    \bigg]
  \bigg]
  \label{eq:trend}
  \vspace{-5pt}
\end{equation}
where $\langle argopt \rangle$ is one of {\tt argmax} or {\tt argmin}, and
$\langle op \rangle$ refers to one of 
$(\max|\min|\sum|\prod)$. 
Given this, the pseudocode which optimizes this
process can automatically be generated based on the actual values of $\langle
argopt \rangle$, $\langle op \rangle$, and the number of operations. 
In short, for each $\langle op \rangle$ or
dimension traversed over, the \zq engine generates a new nested for loop.
Within each for loop, we iterate over all values of that dimension,
evaluate the inner expression, and then eventually apply the overall 
operation (e.g., max, $\sum$). 
\iftechreportcode
\vspace{10pt}
For Equation~\ref{eq:trend}, 
the generated pseudocode would look like the one given by
Listing~\ref{lst:pseudocode}. Here, {\tt f} refers to the visualization
collection being operated on by the \pnode[], which the parent \cnode[] should
have already retrieved.  

\vspace{-5pt}
\begin{lstlisting}[columns=fullflexible,basicstyle=\ttfamily\scriptsize,
    caption={Pseudocode for a process in the form of
  Equation~\ref{eq:trend}.}, label={lst:pseudocode}, captionpos=b]
  f = make_ndarray(SQL(...))
  tmp0 = make_array(size=len(v0))
  for i0 in [1 .. len(v0)]:
    tmp1 = make_array(size=len(v1))
    for i1 in [1 .. len(v1)]:
      tmp2 = make_array(size=len(v2))
      for i2 in [1 .. len(v2)]:
        ...
          tmpm = make_array(size=len(vn))
          for im in [1 .. len(vn)]:
            tmpm[im] = T(f[0, i1, i2, ..., im])
          tmpm-1[im-1] = opm(tmpm)
        ...
      tmp1[i1] = op1(tmp2)
    tmp0[i0] = op0(tmp1)
  return argopt(tmp0)[:k']
\end{lstlisting}
\vspace{-5pt}

\fi


\techreport{Although this is the translation for one specific type of process, it is easy
to see how the code generation would generalize to other patterns.}

\vspace{-3pt}
\subsection{Optimizations}
We now present several optimizations to the previously introduced basic
translator. 
In preliminary experiments, we found that the \sql queries for the \cnode[s] 
took the majority of the runtime for \zq queries, 
so we concentrate our efforts on
reducing the cost of computing \cnode[s]. 
However, we do present one \pnode[]-based
optimization for process-intensive \zq queries.
We start with the simplest optimization schemes, and add more sophisticated
variations later.

\vspace{-4pt}
\subsubsection{Parallelization}\label{sec:para}
\vspace{-3pt}
One natural way to optimize the graph-based query plan is to take advantage of
the multi-query optimization (MQO)~\cite{DBLP:journals/tods/Sellis88} present
in databases and issue in parallel the \sql queries for independent \cnode[s]---the \cnode[s]
for which there is no dependency between them.
With MQO, the database can receive multiple \sql queries at the same time and
share the scans for those queries, thereby reducing the number of
times the data needs to be read from disk.

To integrate this optimization, we make two simple
modifications to Algorithm~\ref{simplealg:basic}. 
In the first step, instead of searching for a
single node whose parents have all been marked done, search for \emph{all}
nodes whose parents have been marked as done. Then in step 2, issue the
\sql queries for all \cnode[s] which were found in step 1 in parallel at the
same time. For example, the \sql queries for {\tt f1} and {\tt f2}
could be issued at the same time in Figure~\ref{fig:real2-plan}, and once {\tt
p1} and {\tt p2} are executed, \sql queries for {\tt f3} and {\tt f4} can
be issued in parallel.

\vspace{-6pt}
\subsubsection{Speculation}\label{sec:spec}
\vspace{-3pt}

While parallelization gives the \zq engine a substantial increase in
performance, we found that many realistic \zq queries intrinsically 
have a high level
of interdependence between the nodes in their query plans. 
To further optimize the performance, we use \emph{speculation}:
the \zq engine pre-emptively issues \sql
queries to retrieve the superset of visualizations for each \cnode[],
considering all possible outcomes for the \axvar[s]. 
Specifically, by tracing the provenance of each \axvar[] 
back to the root, we can determine the superset of all values for each \axvar[];
then, by considering the Cartesian products of these sets, 
we can determine a superset of the relevant visualization collection for a \cnode[].
After the \sql queries have returned, the \zq
engine proceeds through the graph as before, and
once all parent \pnode[s] for a \cnode[] have been
evaluated, the \zq engine isolates the correct subset of data for that
\cnode[] from the pre-fetched data.

For example, in the query in Table~\ref{tab:real2}, {\tt f3} depends on
the results of {\tt p1} and {\tt p2} since it has constraints based on {\tt v2}
and {\tt v4}; specifically {\tt v2} and {\tt v4} should be locations and
categories for which {\tt f1} and {\tt f2} have a negative trend. However, we
note that {\tt v2} and {\tt v4} are derived as a result of {\tt v1} and {\tt
v3}, specified to take on all locations and categories in rows 1 and
2. So, a
superset of {\tt f3}, the set of profit over year
visualizations for various products for all locations and categories 
(as opposed to just those that satisfy {\tt p1} and {\tt p2}), could be
retrieved pre-emptively.
Later, when the \zq engine executes {\tt p1} and {\tt p2}, this superset
can be filtered down correctly.

One downside of speculation is that a lot more data must be
retrieved from the database, but we found that blocking on the retrieval of
data was more expensive in runtime than retrieving extra data. Thus,
speculation ends up being a powerful optimization which compounds the positive
effects of parallelization.
\vspace{-6pt}
\subsubsection{Query Combination}\label{sec:comb}
\vspace{-3pt}
From extensive modeling of relational databases,
we found that the overall runtime of concurrently 
running issuing \sql queries 
is heavily dependent on the number of queries being run in parallel.
Each additional query constituted a $T_q$ increase in
the overall runtime (e.g., for our settings of PostgreSQL, we found $T_q$ =
\texttildelow900ms). 
To reduce the total number of running queries, we use \emph{query combination}; that is,
given two \sql queries $Q_1$ and $Q_2$, we
combine these two queries into a new $Q_3$ 
which returns the data for both
$Q_1$ and $Q_2$. 
In general, if we have $Q_1$ (and $Q_2$) in the form of:
\begin{quoting}\small
\begin{verbatim}
SELECT X1, A(Y1), Z1    WHERE C1(X1, Y1, Z1)
       GROUP BY X, Z1    ORDER BY X, Z1
\end{verbatim}
\end{quoting}
we can produce a combined $Q_3$ which has the
form:
\begin{quoting}\small
\begin{verbatim}
SELECT X1, A(Y1), Z1, C1, X2, A(Y2), Z2, C2
    WHERE C1 or C2
    GROUP BY X1, Z1, C1, X2, Z2, C2
    ORDER BY X1, Z1, C1, X2, Z2, C2
\end{verbatim}
\end{quoting}
where {\tt C1 = C1(X1, Y1, Z1)} and {\tt C2} is defined similarly.
From the combined query $Q_3$, it is possible to regenerate the data which would
have been retrieved using queries $Q_1$ and $Q_2$ by aggregating over the
non-related groups for each query. For $Q_1$, we would select the data for which
{\tt C1} holds, and for each {\tt (X1, Z1)} pair, we would aggregate over the
{\tt X2}, {\tt Z2}, and {\tt C2} groups.

While query combining is an effective optimization, there are limitations. We found
that the overall runtime also depends on the number of unique group-by values
per query, and the number of unique group-by values for a combined query is the
product of the number of unique group-by values of the constituent queries.
Thus, the number of average group-by values per query grows super-linearly with
respect to the number of combinations. However, we found that as long as the
combined query had less than $M_G$ unique group-by values, it was more
advantageous to combine than not (e.g., for our settings of PostgreSQL, we
found $M_G$ = 100k).

\paragraph{Formulation.} Given the above findings, we can now formulate the problem of deciding which
queries to combine as an optimization problem:
{\em Find the best combination of \sql queries that 
minimizes: $\alpha \times$(total number
of combined queries) + $\sum_i$ (number of unique group-by values in combined query $i$),
such that no single combination has more than $M_G$ unique group-by
values.} 

\papertext{
As we show in the technical report~\cite{techreport}, this optimization
problem is {\sc NP-Hard} via a reduction from the {\sc Partition Problem}.
}
\techreport{
The cost of adding a thread, $\alpha$, is generally more than $M_G$---for instance, in PostgreSQL we found $\alpha$ > 100k ($M_G$) for different experimental settings. By further 
assuming that the cost of processing all group by values < $M_G$ is same, we can simplify the problem to finding the minimum number of combined queries such that the maximum number of group by values per 
combined query is less than $M_G$. We prove that the solution to this problem is {\sc NP-Hard} by reduction from the {\sc Partition Problem}. 
\begin{proof} 
Let $g_1,g_2,...g_n$ be the group by values for the queries $Q_1,Q_2,...,Q_n$ we want to combine.
We want to find minimum number $m$ of combined queries, such that each combined query $G_i$ 
has at most $M_G$ maximum group by values. 
Recall that in the Partition problem, we are given an instance of n numbers $a_1,a_2,...,a_n$, and we are asked to decide if
there is a set $S$ such that $\sum_{a_i \subset S}a_i=\sum_{a_i \not \subset S}a_i$. 
Let $A=\sum a_i$ and consider an instance of Query Combination problem with $g_i=M_G^{\frac{2\times a_i}{A}}$. 
With this setting, it is easy to see that the answer to the Partition instance is YES 
if and only if the minimum number of combined queries is 2.
\end{proof} 
}

\paragraph{Wrinkle and Solution.} However, a wrinkle to the above formulation is that it assumes no two \sql
queries share a group-by attribute.
If two queries have a shared group-by attribute, it may be
more beneficial to combine those two, since the number of group-by values does
not increase upon combining them. 
Overall, we developed the metric $EFGV$
or the effective increase in the number of group-by values to determine the
utility of combining query $Q'$ to query $Q$:
$EFGV_{Q}(Q') = \prod_{g \in G(Q')} \#(g)^{[[g \notin G(Q)]]}$
where $G(Q)$ is the set of group-by values in $Q$, $\#(g)$ calculates the
number of unique group-by values in $g$,  and $[[g \notin G(Q)]]$
returns 1 if $g \notin G(Q)$ and 0 otherwise. In other words, this calculates
the product of group-by values of the attributes which are in $Q'$ but not in $Q$.
Using the $EFGV$ metric, we then apply a variant of agglomerative
clustering~\cite{anderberg2014cluster} to decide the best choice of
queries to combine. 
As we show in the experiments section, this technique 
leads to very good performance. 


\vspace{-4pt}
\subsubsection{Cache-Aware Execution}\label{sec:cache}
\vspace{-2pt}
Although the previous optimizations were all I/O-based optimizations for \zq,
there are cases in which optimizing the execution of \pnode[s] is important as
well. In particular, when a process has multiple nested for loops, the cost of
the \pnode[] may start to dominate the overall runtime. To address this
problem, we adapt techniques developed in high-performance computing---specifically,
cache-based optimizations similar to those used in matrix
multiplication~\cite{goto2008anatomy}. With cache-aware execution, the \zq
engine partitions the iterated values in the for loops into blocks of data
which fit into the L3 cache. Then, the \zq engine reorders the order of
iteration in the for loops to maximize the time that each block of data remains
in the L3 cache. This allows the system to reduce the amount of data transfer between
the cache and main memory, minimizing the time taken by the
\pnode[s].


\vspace{-5pt}
\vspace{-2pt}
\section{{\Large \zv} System Description}
\label{sec:architecture}
\vspace{-2pt}

\noindent We now give a brief description of the \zv system.


\begin{figure}
\vspace{-10pt}
\centering
        \includegraphics[width=7cm]{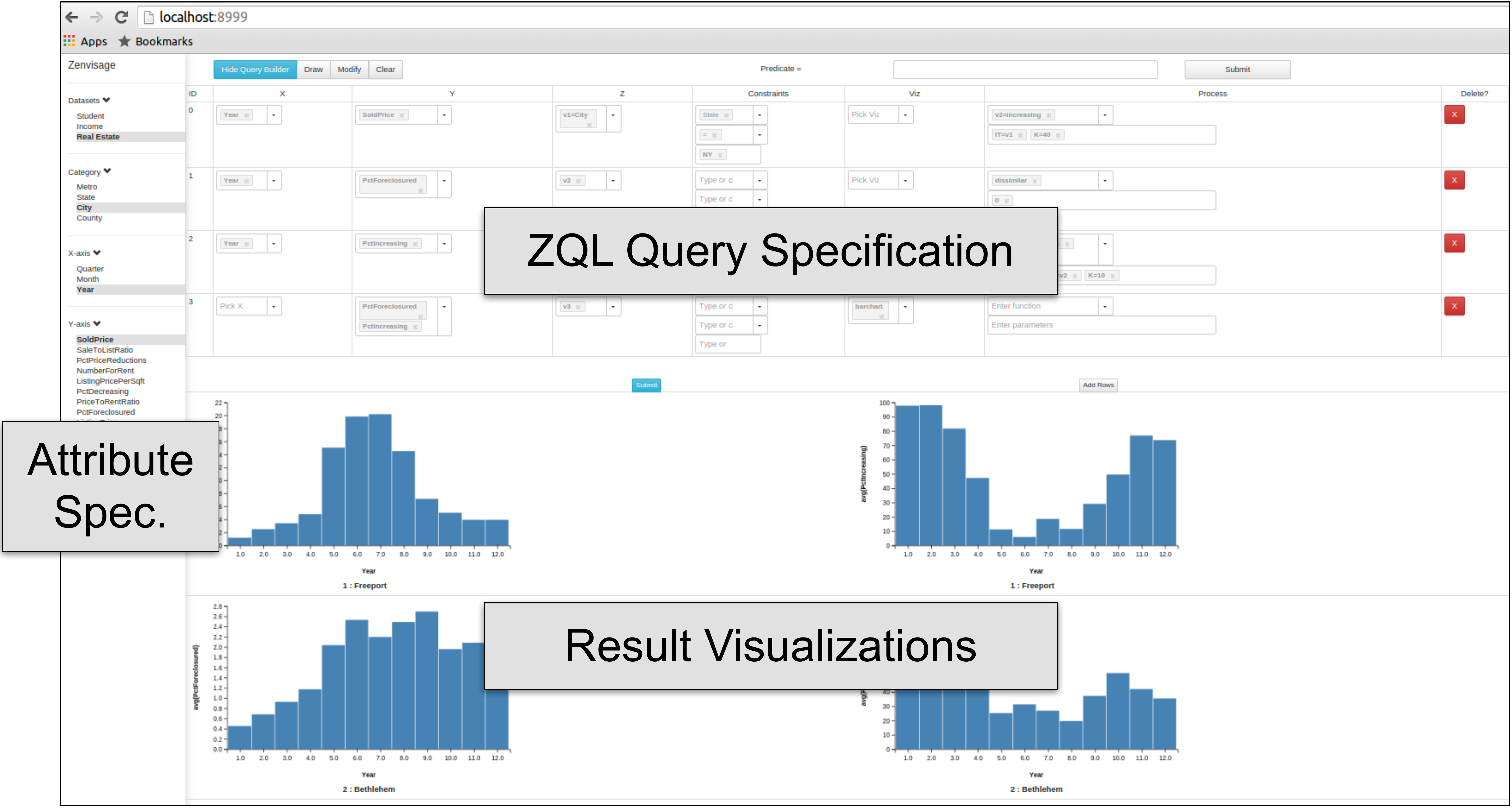}
    \vspace{-5pt}
        \caption{\zv basic functionalities}
        \label{fig:querytable}
\vspace{-20pt}
    
\end{figure}

\paragraph{Front-end.}
The \zv front-end is designed as a 
lightweight web-based client application.
It provides a GUI to compose \zq queries, 
and displays the resulting visualizations
using Vega-lite~\cite{wongvoyager2015}. 
A screenshot of \zv in action is shown in Figure~\ref{fig:querytable}.
A list of attributes, divided into qualitative and quantitative, is provided on
the left; a table to enter \zq queries, with auto-completion, 
is on top, and the resulting
visualizations are rendered at the bottom.
Users also have the option of hiding the \zq specification table and
instead using a simpler drop-down menu-based interface complemented by a
sketching canvas.
The sketching canvas allows users to draw their desired trend that can then be
used to search for similar trends. The menu-based interface makes it easy for
users to perform some of the more common visual exploration queries, such as
searching for representative or outlier visualizations. Furthermore, the user
may drag-and-drop visualizations from the results onto the sketching canvas,
enabling further interaction with the results.

\paragraph{Back-end.}
The \zv front-end issues \zq queries to the back-end over a {\sf REST}
protocol. The back-end (written in {\tt node.js}) receives the queries and
forwards them to the \zq engine (written in {\tt Java}), which is
responsible for parsing, compiling, and optimizing the queries as in Section~\ref{subsec:query_execution}. \sql queries issued by the \zq
engine are submitted to one of our back-end databases (which
currently include PostgreSQL and Vertica), and the resultant visualization data
is returned back to the front-end encoded in {\tt JSON}.

\vspace{-13pt}

\vspace{-4pt}
\section{Experimental Study}
\label{sec:experiments}
\vspace{0pt}

In this section, we evaluate the runtime performance of the \zq engine.
We present the runtimes for executing both synthetic and realistic \zq queries
and show that we gain speedups of up to 3$\times$
with the optimizations from Section~\ref{subsec:query_execution}.
We also varied the characteristics of a synthetic \zq query to observe
their impact on our optimizations. Finally, we show that disk I/O
was a major bottleneck for the \zq engine, and if we switched our back-end
database to a column-oriented database and cache the dataset in memory, we
can
achieve interactive run times for datasets as large as 1.5GB.


\paragraph{Setup.}
All experiments were conducted on a 64-bit Linux server
with 8 3.40GHz Intel Xeon E3-1240 4-core processors and 8GB of
1600 MHz DDR3 main memory. We used 
PostgreSQL with working memory size set to 512 MB and shared buffer
size set to 256MB for the majority of the experiments; the last set of 
experiments demonstrating interactive run times additionally 
used Vertica Community Edition with a
working memory size of 7.5GB.

\techreport{
\paragraph{PostgreSQL Modeling.}
For modeling the performance
on issuing multiple parallel queries with varying number of group by values, 
we varied the 
number of parallel queries issued (\#Q) from 1 to 100,
and the group by values per query (\#V) from 10 to 100000, 
and recorded the response times (T).
We observed that the time taken for a batch of queries was 
practically linearly dependent 
to both the number of queries as 
well as the group by values. 
Fitting a linear equation by performing 
multiple regression over the observed data, 
we derived the following cost-model,
$$ T(ms)= 908\times(\#Q) + 1.22\times\frac{(\#V)}{100} + 1635$$
As per the above model, adding a thread leads to the same rise in response time as increasing the number 
of group by values by 75000 over the existing threads in the batch. In other words, it is better to merge queries 
with small number of group by values. Moreover, since there is a fixed cost (1635 ms) associated with
every batch of queries, we tried to minimize the number of batches by packing as many queries as possible 
within the memory constraints.
}

\paragraph{Optimizations.} The four versions of the \zq engine we use are:
\begingroup
\plitemsep=.2em
\pltopsep=.2em
\begin{inparaenum}[(i)]
\item \noopt: The basic translation from Section~\ref{subsec:query_execution}.
\item \para: Concurrent \sql queries for independent nodes from Section~\ref{sec:para}.
\item \spec: Speculating and pre-emptively issuing \sql queries from Section~\ref{sec:spec}.
\item \fuse: Query combination with speculation from Section~\ref{sec:comb}. 
\end{inparaenum}
\endgroup
\noindent In our experiments, we consider \noopt and the MQO-dependent \para to be our
baselines, while \spec and \fuse were considered to be completely novel optimizations.
For certain experiments later on, we also evaluate the performance of the caching optimizations 
from Section~\ref{sec:cache} on \fuse.
 
\begin{table*}
\vspace{-5pt}
\small
  \scalebox{0.90}{
    \begin{tabular}{|C{1pt}|L{220pt}|C{33pt}|C{33pt}|C{30pt}|C{30pt}|C{30pt}|C{40pt}|C{48pt}|}
      \hline
      & Query Description & \# \cnode[s] & \# \pnode[s] & \# $T$ & \# $D$ & \#
      Visualizations & \# \sql Queries: \noopt & \# \sql Queries: \fuse \\
      \hline
      1 & Plot the related visualizations for airports which have a correlation
      between arrival delay and traveled distances for flights arriving there.
     %
      & 6 & 3 & 670 & 93,000 & 18,642 & 6 & 1 \\ \hline
      2 & Plot the delays for carriers whose delays have gone up at airports
      whose average delays have gone down over the years.
      & 5 & 4 & 1,000 & 0 & 11,608 & 4 & 1 \\ \hline
      3 & Plot the delays for the outlier years, outlier airports, and outlier
      carriers with respect to delays.
      & 12 & 3 & 0 & 94,025 & 4,358 & 8 & 2 \\ \hline
    \end{tabular}
}
%
%
   \vspace{-5pt}
   \caption{Realistic queries for the airline dataset with the \# of
   \cnode[s], \# of \pnode[s], \# of $T$ functions calculated, \# of $D$ functions
 calculated, \# of visualizations explored, \# of \sql queries issued with
 \noopt, and \# of \sql queries issued with \fuse per query.}
  \label{tab:realworldqueries}
  \vspace{-15pt}
\end{table*}

\noindent

\begin{figure}[!h]
\vspace{-10pt}
\begin{center}
  \begin{subfigure}[c]{0.48\columnwidth}
    \includegraphics[width=\textwidth]{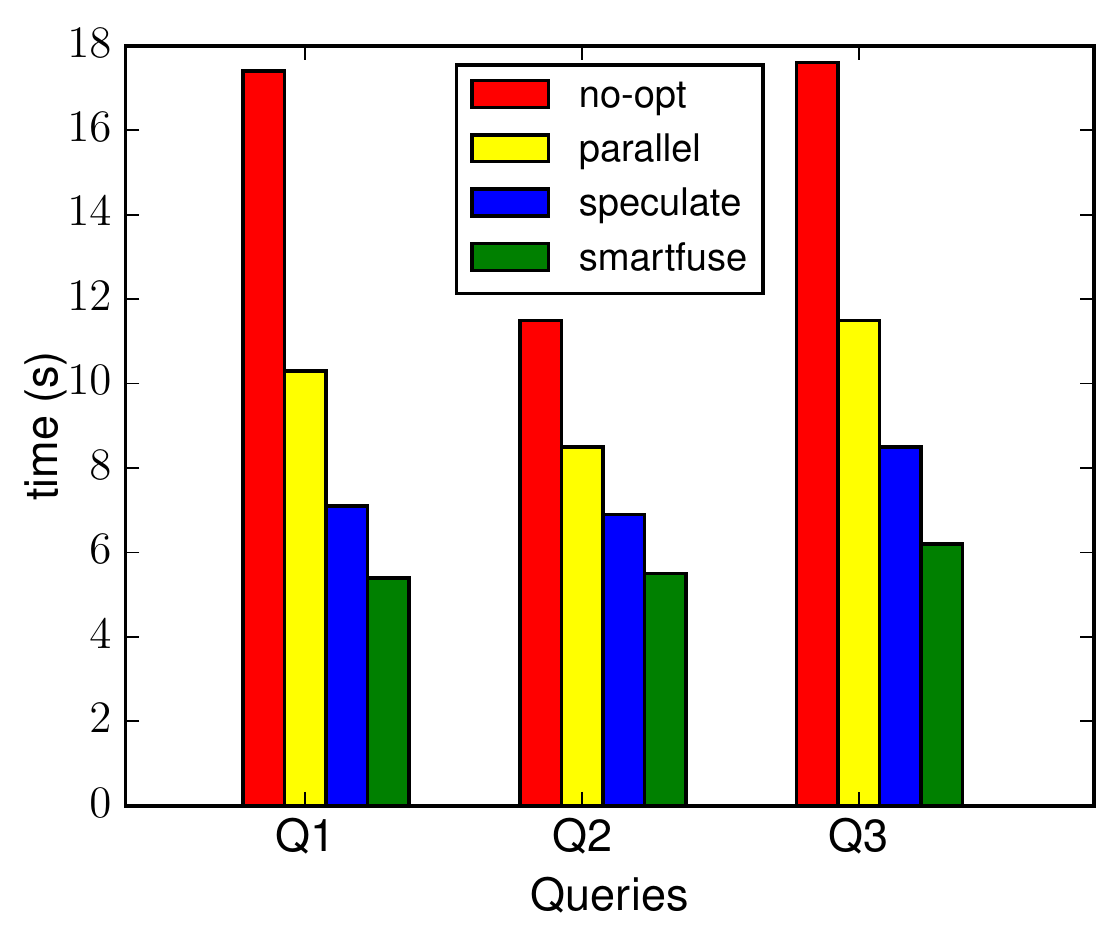}
  \end{subfigure}
  \begin{subfigure}[c]{0.48\columnwidth}
    \includegraphics[width=\textwidth]{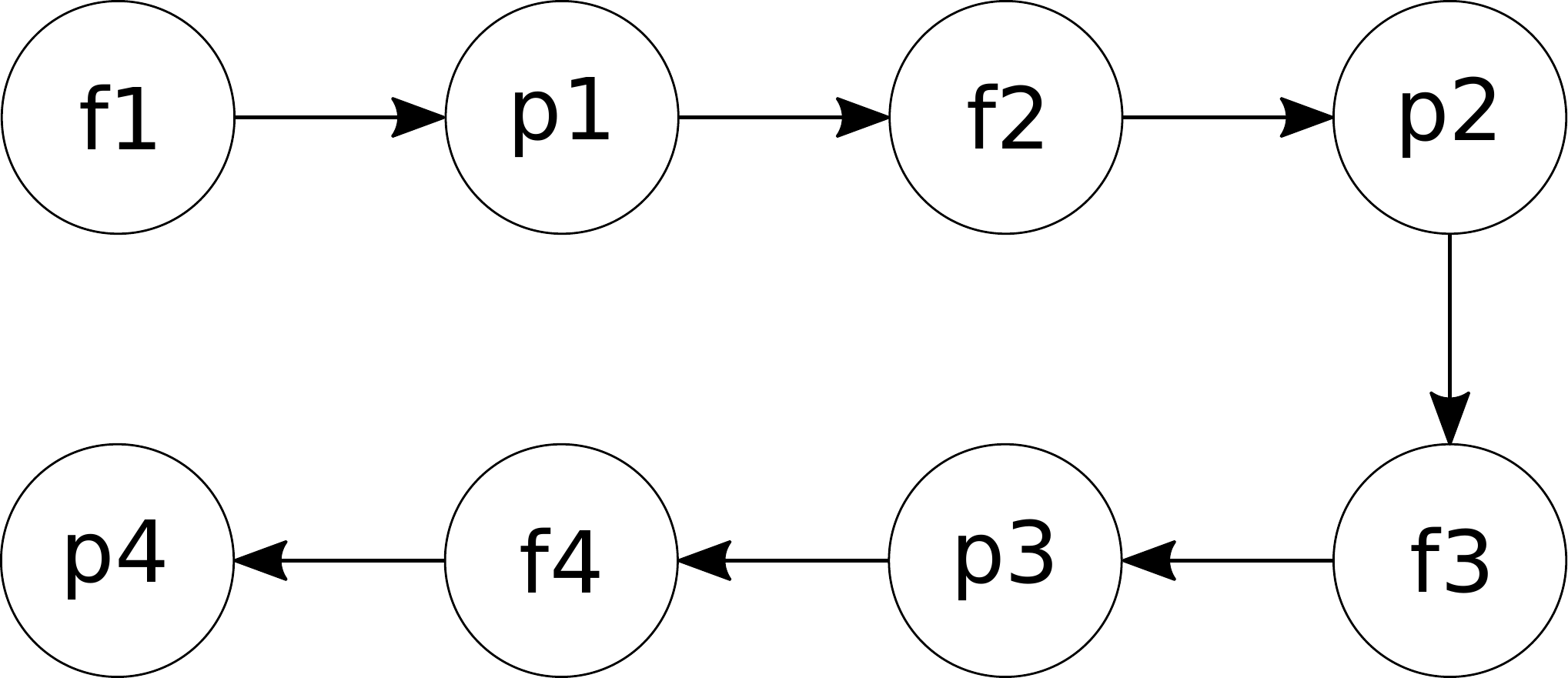}
  \end{subfigure}
\end{center}
\vspace{-20pt}
\caption{Runtimes for queries on real dataset (left) and single chain synthetic query (right)}
\vspace{-8pt}
\label{fig:optrealnviz}
\end{figure}

\begin{figure}[!h]
\vspace{-5pt}
\begin{center}
\includegraphics[width=0.48\columnwidth]{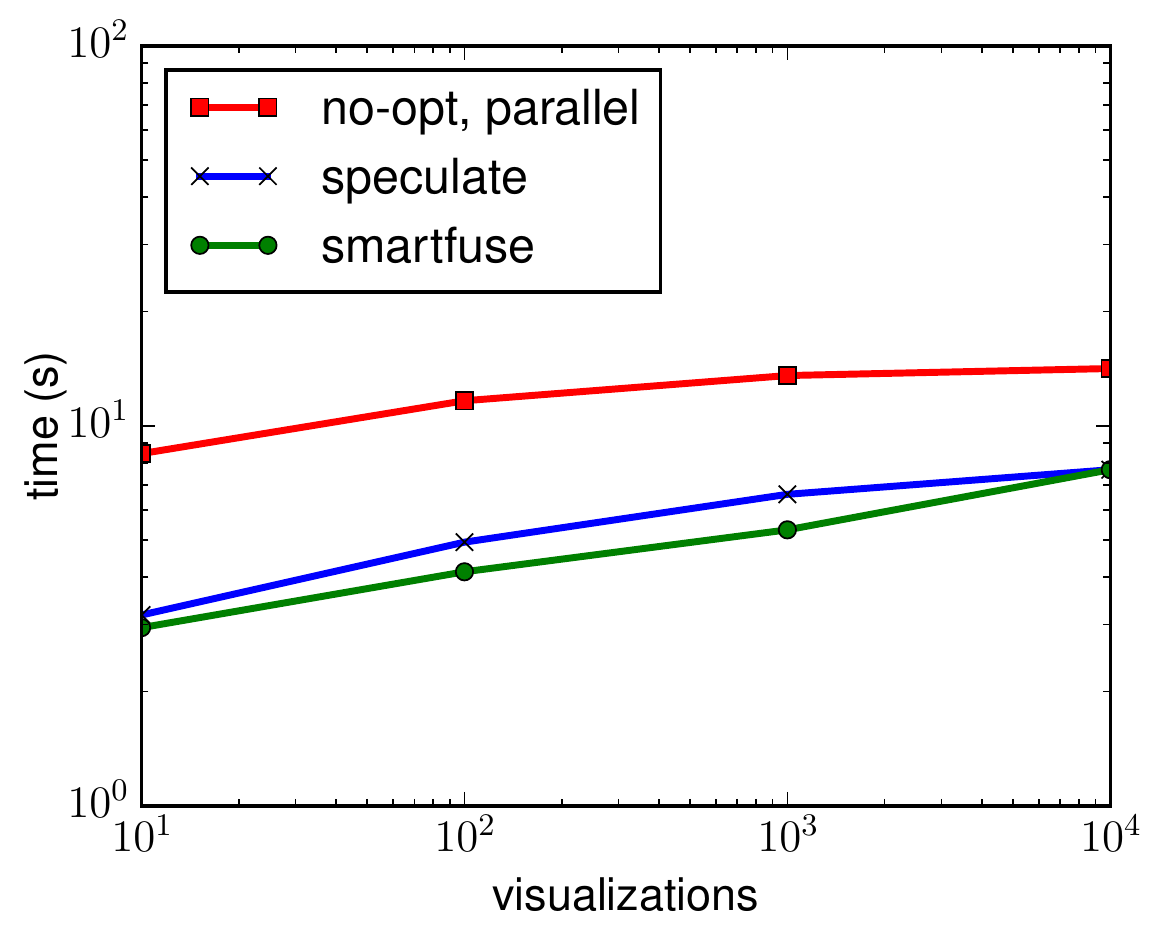}
\includegraphics[width=0.48\columnwidth]{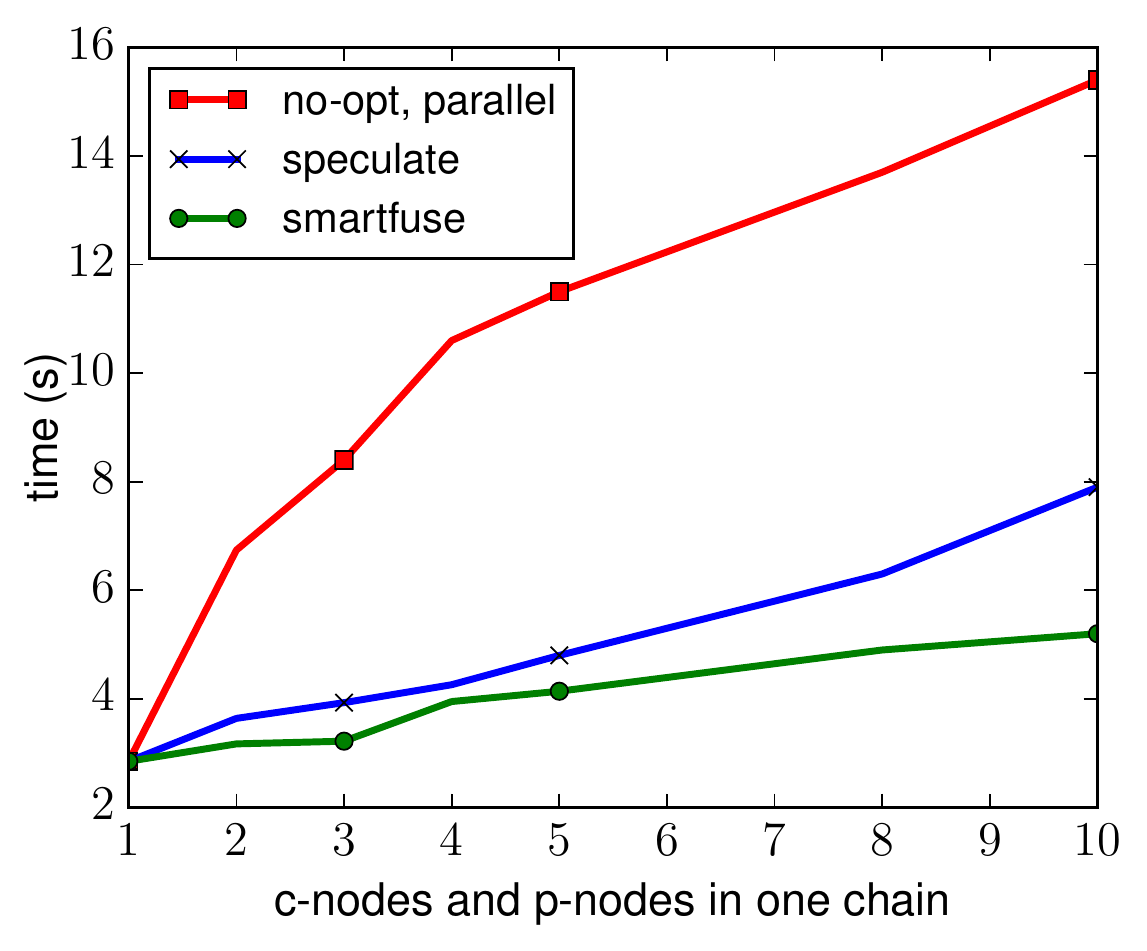}
\end{center}
\vspace{-20pt}
\caption{Effect of number of  visualizations (left) and length
of the chain (right) on the overall runtimes.}
\vspace{-15pt}
\label{fig:optsynviz}
\end{figure}

\vspace{2pt}
\subsection{Realistic Queries}
\vspace{-3pt}

For our realistic queries, we used 20M rows of a real 1.5GB airline
dataset~\cite{airlinedata} which contained the details of flights within the
USA from 1987-2008, with 11 attributes. On
this dataset, we performed 3 realistic \zq queries inspired by the case
studies in our introduction. Descriptions of the queries can be found in Table~\ref{tab:realworldqueries}.

Figure~\ref{fig:optrealnviz} (left) depicts the runtime performance of the
three realistic \zq queries, for each of the optimizations. 
For all queries, each level of optimization provided a substantial speedup
in execution time compared to the previous level.
{\em Simply by going from \noopt to \para, we saw a 45\% reduction in runtime. From
\para to \spec and \spec to \fuse, we saw 15-20\% reductions in
runtime}.
A large reason for why the optimizations were so effective was because \zq
runtimes are heavily dominated by the execution time of the issued \sql
queries. In fact, we found that for these three queries, 94-98\% of the overall
runtime could be contributed to the \sql execution time. We can see from
Table~\ref{tab:realworldqueries}, \fuse always managed to lower the number of
\sql queries to 1 or 2 after our optimizations, thereby heavily impacting the
overall runtime performance of these queries.

\vspace{-2pt}
\subsection{Varying Characteristics of {\large \zq} Queries}
\vspace{-2pt}

We were interested in evaluating the efficacy of our optimizations with respect
to four different characteristics of a \zq query:
\begin{inparaenum}[(i)]
\item the number of visualizations to explore,
\item the complexity of the \zq query,
\item the level of interconnectivity within the \zq query, and
\item the complexity of the processes.
\end{inparaenum} 
To control for all variables except these characteristics, we used a synthetic
chain-based \zq query to conduct these experiments. Every row of the
chain-based \zq query specified a collection of visualizations based on the
results of the process from the previous row, and every process was applied on
the collection of visualizations from the same row. Therefore, when we created
the query plan for this \zq query, it had the chain-like structure depicted by
Figure~\ref{fig:optrealnviz} (right).
Using the chain-based \zq query, we could then
\begin{inparaenum}[(i)]
\item vary the number of visualizations explored,
\item use the length of the chain as a measure of complexity,
\item introduce additional independent chains to decrease
  interconnectivity, and
\item increase the number of loops in a \pnode[] to control the complexity of
  processes.
\end{inparaenum}

\smallskip
\noindent To study these characteristics, we used a 
synthetic dataset with 10M rows and 15 
attributes (10 dimensional and 5 measure) 
with cardinalities of dimensional attributes varying from 10 to 10,000. 
By default, we set the input number of visualizations per chain to be 100,
with 10 values for the X attribute, 
number of \cnode[s] per chain as 5, the process as $T$ (with a single for loop)
 with a selectivity of .50, and  number of chains as 1.

\paragraph{Impact of number of visualizations.} 
Figure \ref{fig:optsynviz} (left) shows the performance of \noopt, \spec, and
\fuse on our chain-based \zq query as we increased the number of visualizations
that the query operated on. The number of visualizations was increased by
specifying larger collections of Z column values in the first \cnode[]. We
chose to omit \para here since 
it performs identically to \noopt.
With the increase in visualizations, the overall response time increased for
all versions
because the amount of processing per \sql query increased. \fuse showed better
performance than \spec up to
10k visualizations due to reduction in the total number of \sql queries issued.
However, at 10k visualization, we reached the threshold of the number of unique
group-by values per combined query (100k for PostgreSQL), so it was less optimal
to merge queries. At that point, \fuse behaved similarly to \spec.

\begin{figure}[!h]
\vspace{-10pt}
\begin{center}
\includegraphics[width=0.48\columnwidth]{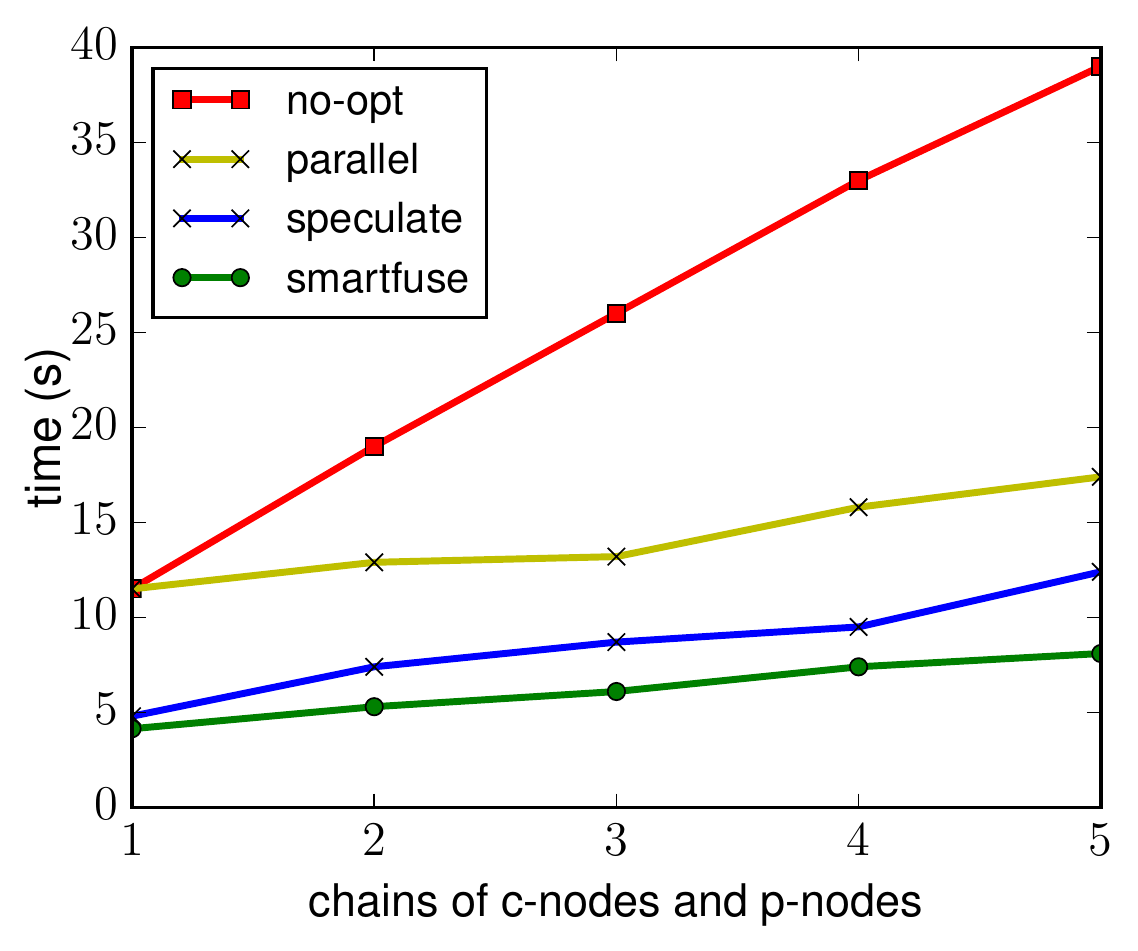}
\includegraphics[width=0.48\columnwidth]{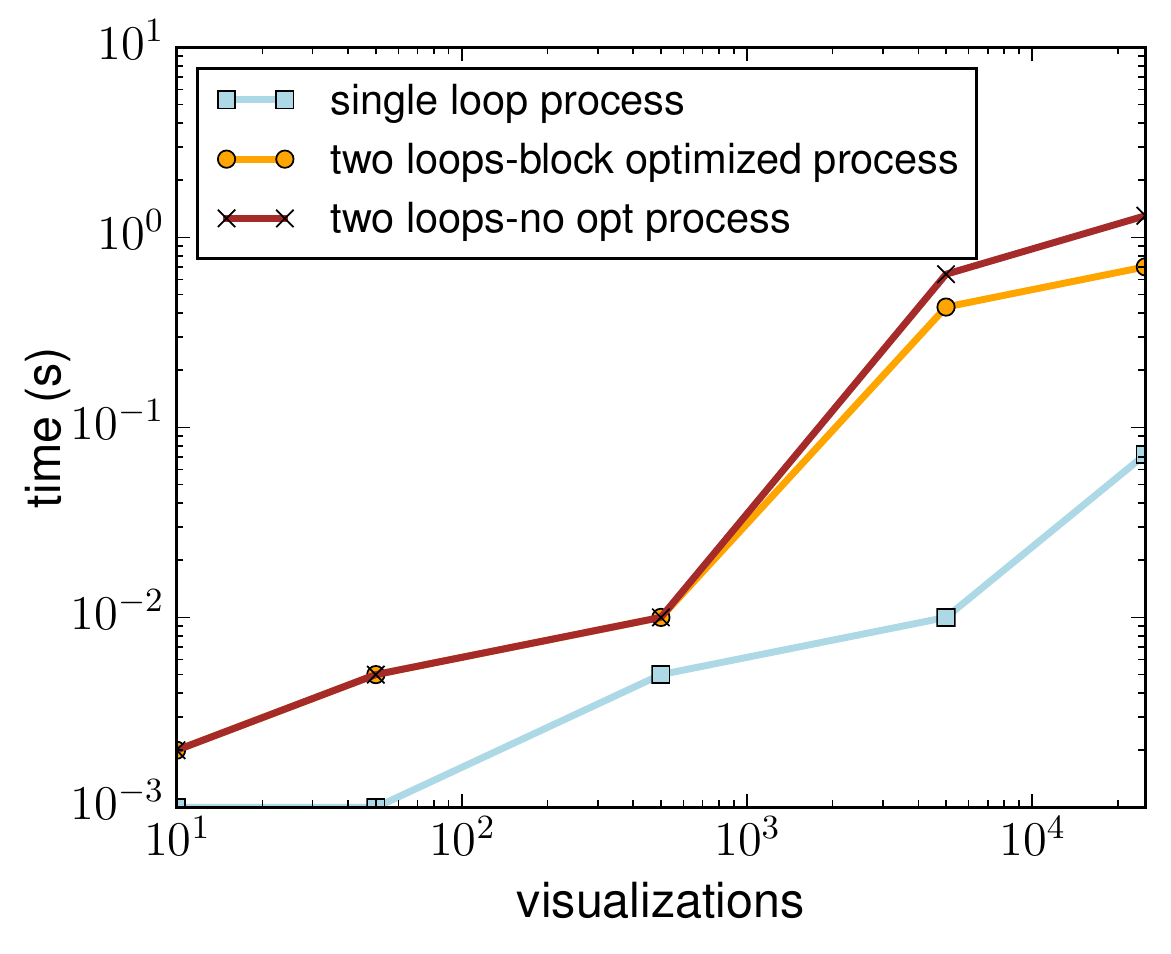}
\end{center}
\vspace{-15pt}
\caption{Effect of number of independent chains (left) and the number of loops
in a \pnode[] (right) on the overall runtimes.}
\vspace{-10pt}
\label{fig:opt_syn_nodes}
\end{figure}

\paragraph{Impact of the length of the chain.}
We varied the length of the chain in the query plan (or the number of rows in
the \zq query) to simulate a change in the complexity of the \zq query and
plotted the results in Figure~\ref{fig:optsynviz} (right). As the number of
nodes in the query plan grew, the overall runtimes for the different
optimizations also grew. However, while the runtimes for both \noopt and \spec
grew at least linearly, the runtime for \fuse grew sublinearly due to its query
combining optimization. While the runtime for \noopt was much greater than for \spec,
since the overall runtime is linearly dependent on the number of
\sql queries run in parallel, we see a linear growth for \spec.

\paragraph{Impact of the number of chains.}
We increased the number of independent chains from 1 to 5 to observe the effect
on runtimes of our optimizations; the results are presented in
Figure~\ref{fig:opt_syn_nodes} (left). While \noopt grew linearly as expected, all
\para, \spec, and \fuse were close to constant with respect to the number of
independent chains. We found that while the overall runtime for concurrent \sql
queries did grow linearly with the number of \sql queries issued, they grew much
slower compared to issuing those queries sequentially, thus leading to an
almost flat line in comparison to \noopt. 

\paragraph{Impact of process complexity.}
We increased the complexity of processes by increasing the number of loops in
the first \pnode[] from 1 to 2. For the single loop, the \pnode[] filtered
based on a positive trend via $T$, while for the double loop, the \pnode[] found
the outlier visualizations.
Then, we varied the number of visualizations to
see how that affected the overall runtimes. Figure~\ref{fig:opt_syn_nodes}
(right) shows the results. For this experiment, we compared regular \fuse with
cache-aware \fuse to see how much of a cache-aware execution made.
We observed that there was not much difference
between cache-aware \fuse and regular \fuse below 5k visualizations when all
data could fit in cache.
After 5k visualizations, not all the visualizations could be
fit into the cache the same time, and thus the cache-aware execution of the
\pnode[] had
an improvement of 30-50\% as the number of visualizations increased from 5k to
25k. However, this improvement, while substantial, is only a minor change in 
the overall runtime. 

\vspace{-2pt}
\subsection{Interactivity}
\vspace{-2pt}

\begin{figure}[!h]
\vspace{-10pt}
\begin{center}
\includegraphics[width=0.48\columnwidth]{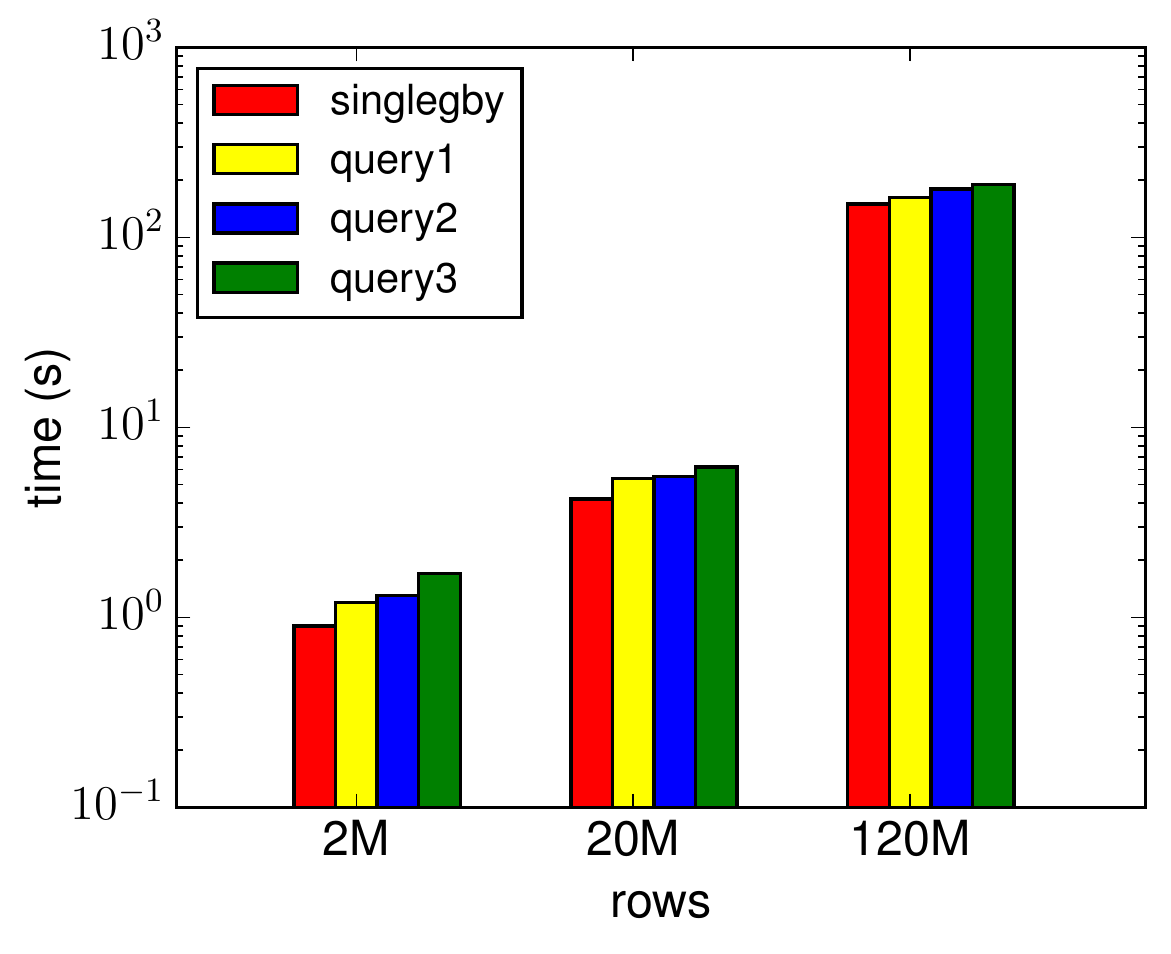}
\includegraphics[width=0.48\columnwidth]{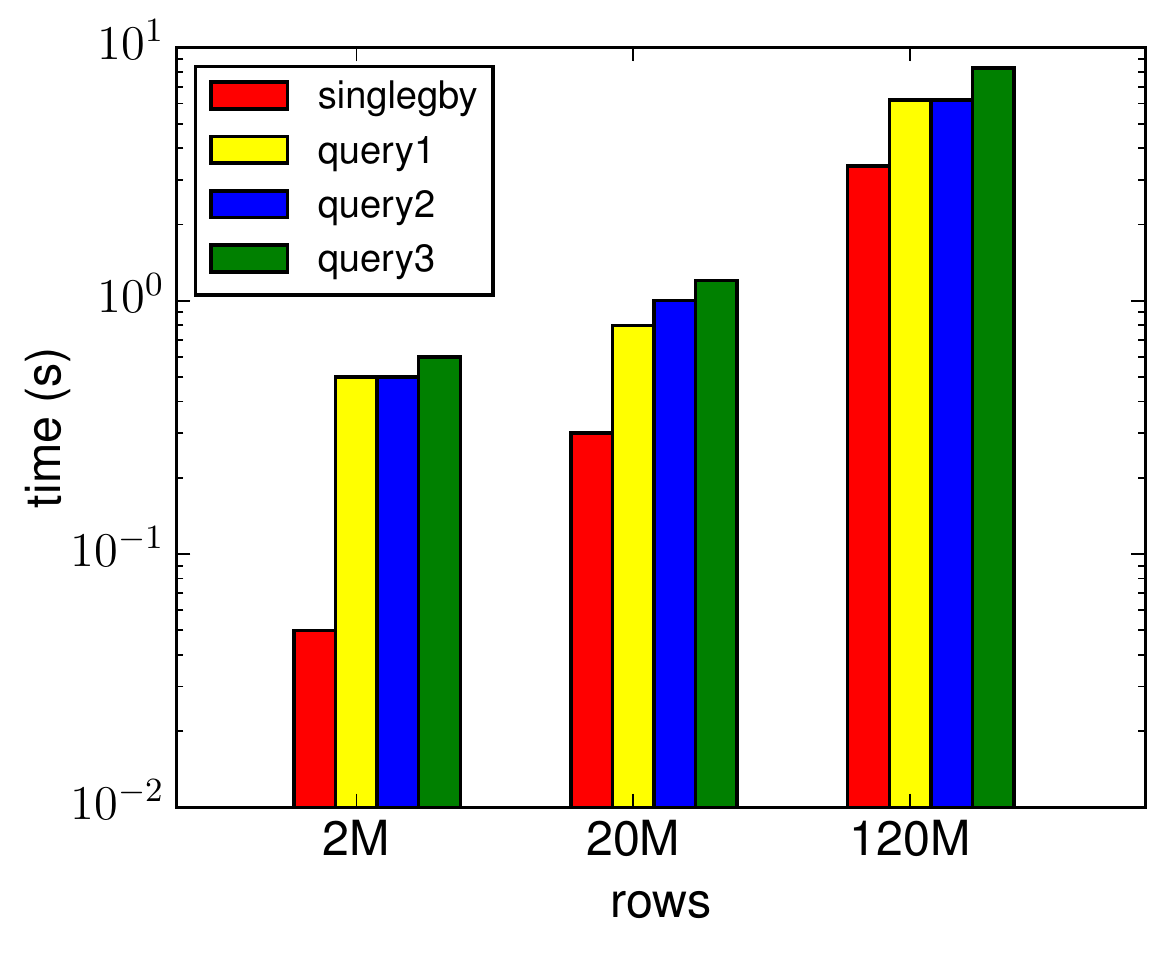}
\end{center}
\vspace{-20pt}
\caption{\fuse  on PostgreSQL (left) and Vertica (right)}
\vspace{-15pt}
\label{fig:optrowcolumstores}
\end{figure}

The previous figures showed that the overall execution times of \zq queries took
several seconds, even with \fuse, thus perhaps indicating
\zq is not fit for interactive use with large datasets.
However, 
we found that this was primarily due
to the disk-based I/O bottleneck of \sql queries. In
Figure~\ref{fig:optrowcolumstores} (left), we show the \fuse runtimes of the
3 realistic queries from before on varying size subsets of the airline dataset,
with the time that it takes to do a single group-by scan of the dataset. As we
can see, the runtimes of the queries and scan time are \emph{virtually the
same}, indicating that \fuse comes very close to the optimal I/O runtime
(i.e., a ``fundamental limit'' for the system). 

To further test our hypothesis, we ran our \zq engine with Vertica with a large
working memory size to cache the data in memory to avoid expensive disk I/O. The
results, presented in Figure~\ref{fig:optrowcolumstores} (right), showed that
there was a 50$\times$ speedup in using Vertica over PostgreSQL with these
settings. Even with a large dataset of 1.5GB, we were able to achieve
sub-second response times for many queries. Furthermore, for the dataset
with 120M records (11GB, so only 70\% could be cached), we were able to
reduce the overall response times from 100s of seconds to less than 10 seconds.
Thus, once again, \zv returned results in a small multiple of the time 
it took to execute a single group-by query.
\noindent
Overall, \fuse is interactive on moderate sized datasets on PostgreSQL, or on 
large datasets that can be cached in memory and operated on using a columnar database---which 
is standard practice adopted by visual analytics tools~\cite{tableau-inmem}. 
Improving on interactivity is impossible due to fundamental limits to the system;
in the future, we plan to explore returning approximate answers using samples,
since even reading the entire dataset is prohibitively slow.

%


\vspace{-5pt}
\section{User Study}
\label{sec:userstudy}
\vspace{-3pt}

We conducted a user study to evaluate the utility of \zv for data exploration versus two types of systems---first, visualization tools, similar to Tableau, and second, general database and data mining tools, which also support interactive analytics to a certain extent.
In preparation for the user study, we conducted interviews with data analysts to identify the typical exploration tasks and tools used in their present workflow. Using these interviews, we identified a set of tasks to be used in the user study for \zv. We describe these interviews first, followed by the user study details.


\vspace{-5pt}
\subsection{Analyst Interviews and Task Selection}
\label{sec:interviews}
\vspace{-2pt}

We hired seven data analysts via Upwork~\cite{upwork},
a freelancing plat\-form---we 
found these analysts by searching for freelancers
who had the keywords \texttt{analyst} or \texttt{tableau}
in their profile. 
We conducted one hour interviews with them
to understand how they perform data exploration tasks.
The interviewees had 3---10 years of prior experience
and explained every step of their workflow;
from receiving the dataset to presenting the
analysis to clients.
The rough workflow of all interviewees identified 
was the following: first, data cleaning is performed;
subsequently, the analysts perform data exploration;
then, the analysts develop presentations using their findings.
We then drilled down onto the data exploration step.

We first asked the analysts what types of tools 
they use for data exploration.
Analysts reported nine different tools---the most popular ones 
included Excel (5), Tableau (3), and SPSS (2). 
The rest of the tools were reported by just one analyst:
Python, SQL, Alteryx, Microsoft Visio, Microsoft BI, SAS. 
Perhaps not surprisingly, analysts use both visualization
tools (Tableau, Excel, BI), programming languages (Python),
statistical tools (SAS, SPSS), and relational databases (SQL)
for data exploration.

Then, to identify the common tasks used in data exploration, 
we used a taxonomy of
abstract exploration tasks proposed by Amar et al.~\cite{amar2005low}.
Amar et al. developed their taxonomy through summarizing
the analytical questions that arose
during the analysis of five different datasets, 
independent of the capabilities of existing tools or interfaces.
The exploration tasks in Amar et al.~include:
filtering ({\sf f}), sorting ({\sf s}), determining range ({\sf r}), 
characterizing distribution ({\sf d}),
finding anomalies ({\sf a}), clustering ({\sf c}), 
correlating attributes ({\sf co}), retrieving value ({\sf v}),  
computing derived value ({\sf dv}), and finding extrema ({\sf e}).
When we asked the data analysts which tasks they use in their workflow, 
the responses were consistent in that all of them use all of these
tasks, except for three exceptions---{\textsf c}, reported by four participants,
and {\sf e, d}, reported by six participants. 


Given these insights, we selected a small number of appropriate tasks for our user study
encompassing eight of the ten exploration tasks described above:
{\sf f}, {\sf s}, {\sf r}, {\sf d}, {\sf a}, {\sf c}, {\sf co}, {\sf v}.
The other two---{\sf dv} and {\sf e}---finding derived values
and computing extrema, are important tasks in data analysis,
but existing tools (e.g., Excel) already provide adequate
capabilities for these tasks, and we did not expect \zv to 
provide additional benefits.

\vspace{-3pt}
\subsection{User Study Methodology}
\vspace{-2pt}

The goal of our user study was to evaluate \zv with other tools,
on its ability to effectively support data exploration.

\paragraph{Participants.} 
We recruited 12 graduate students as participants with 
varying degrees of expertise in data analytics.
\papertext{In short, half of them used databases; eight of them 
used Matlab, R, Python or Java; eight of them used
spreadsheet software; and four of them used Tableau.
Data for other not as popular tools are not reported. 
}
\techreport{Table \ref{tab:toolsrep} depicts the participants' experience with different categories of tools.
\begin{table}[H]
\vspace{-10pt}
 \centering\scriptsize
  {\tt    
  \begin{tabular}{|l|l|} \hline
      \textbf{Tools} & \textbf{Count} \\ \hline
      Excel, Google spreadsheet, Google Charts  & 8 \\ \hline
      Tableau & 4 \\ \hline
      SQL, Databases & 6 \\ \hline
      Matlab,R,Python,Java & 8 \\ \hline
      Data mining tools such as weka, JNP & 2 \\ \hline
      Other tools like D3 & 2 \\ \hline
     \end{tabular}
  }
  \vspace{-10pt}\caption{Participants' prior experience with data analytic tools}
  \label{tab:toolsrep}
\vspace{-10pt}\end{table}
}

\paragraph{Baselines.}
For the purposes of our study, we explicitly wanted to
do a head-to-head qualitative and quantitative comparison with visual analytics tools,
and thus we developed a baseline tool to compare \zv against directly.
Further, via qualitative interviews, we compared \zv versus
against other types of tools, such as databases, data mining, and programming tools.
Our baseline tool was developed by replicating the visualization 
selection capabilities of visual analytics tools with a styling scheme identical to \zv to control for external factors.
The tool allowed users to specify the x-axis, y-axis, dimensions, and filters.
The tool would then populate all visualizations meeting the specifications.

\techreport{
\paragraph{Comparison Points.} 
There are no tools that offer the same functionalities as
\zv. Visual analytics tools do not offer the ability
to search for specific patterns, or issue complex visual exploration queries; 
data mining toolkits do not offer the ability to search for visual patterns 
and are instead tailored for general machine learning and prediction. 
\papertext{We decided to implement our baseline tool by replicating
the basic visualization specification capabilities of tools like Tableau,
augmented with additional filtering capabilities. 
The styling scheme was similar to \zv to control for external factors;
details can be found in \cite{techreport}.}\techreport{Since visual analytics tools are closer in spirit and functionality 
to \zv, we decided to implement a visual analytics tool as our baseline.
Thus, our baseline tool replicated the basic query specification and output visualization
capabilities of existing tools such as Tableau.  
We augmented the baseline tool with the ability to specify an arbitrary number of filters, allowing
users to use filters to drill-down on specific visualizations.
This baseline visualization tool was implemented with a styling scheme similar to \zv to control for external factors. 
As depicted in Figure \ref{fig:baseline}, the baseline allowed users to visualize data by allowing them to specify the x-axis, y-axis, category, and filters. The baseline tool would populate all the visualizations, which fit the user specifications, using an alpha-numeric sort order.} In addition to task-based comparisons with this baseline, we
also explicitly asked participants to compare \zv with existing data mining and visual analytics 
tools that they use in their workflow.
}

\paragraph{Dataset.}
We used a housing dataset from Zillow.com \cite{housing_data}, 
consisting of housing sales data for different cities, counties, and states from 2004-15, 
with over 245K rows, and 15 attributes. We selected this dataset since participants could relate to the dataset and understand the usefulness of the tasks.


\paragraph{Tasks.}
We designed the user study tasks with the case studies 
from Section~\ref{sec:intro} in mind,
and translated them into the housing dataset. 
Further, we ensured that these tasks together evaluate 
eight of the exploration tasks described above---{\sf f}, {\sf s}, {\sf r}, {\sf d}, {\sf a}, {\sf c}, {\sf co}, 
and {\sf v}.
One task used in the user study is as follows: ``Find three cities in the state
of NY where the sold price vs year trend is very different from the
state\textquotesingle s overall trend.'' This query required the participants to
first retrieve the trend of NY ({\sf v}) and characterize its distribution ({\sf
d}), then separately filter to retrieve the cities of NY ({\sf f}), compare the
values to find a negative correlation ({\sf co}), sort the results ({\sf s}),
and report the top three cities on the list.

\paragraph{Study Protocol.} 
The user study was conducted using a within-subjects study design~\cite{bordens2002research},
forming three phases.
First, participants described their previous experience with data 
analytics tools. 
Next, participants performed exploration tasks using 
\zv (Tool A) and the baseline tool (Tool B), with the orders 
randomized to reduce order effects.
Participants were provided a 15-minute tutorial-cum-practice session
per tool to get familiarized before performing the tasks.
Finally, participants completed a survey
that both measured their satisfaction levels and preferences,
along with open-ended questions on the strengths
and weaknesses of \zv and the baseline, when compared to 
other analytics tools they may have used.\techreport{ The average study session lasted for 75 minutes on average. 
Participants were paid ten dollars per hour for their participation.}
After the study, we reached out to participants with backgrounds
in data mining and programming, and asked if they could complete
a follow-up interview where they use their favorite analytics tool for performing
one of the tasks. 

\paragraph{Metrics.} 
Using data that we recorded, we collected the following metrics: 
completion time, accuracy, and the usability ratings 
and satisfaction level from the survey results. 
In addition, we also explicitly 
asked participants to compare \zv with tools that they use in their workflow.
For comparisons between \zv and general database and data mining tools
via follow-up interviews, 
we used the number of lines of code to evaluate the differences. 

\paragraph{Ground Truth.} 
Two expert data analysts prepared the ground truth 
for each the tasks in the form of ranked answers,
along with score cut-offs on a 0 to 5 scale (5 highest). 
Their inter-rater agreement, measured using Kendall's Tau coefficient, 
was 0.854. 
We took the average of the two scores to rate the participants' answers.

\techreport{
\begin{figure}
\vspace{-5pt}
\begin{center}
\fbox{\includegraphics[width=0.6\columnwidth]{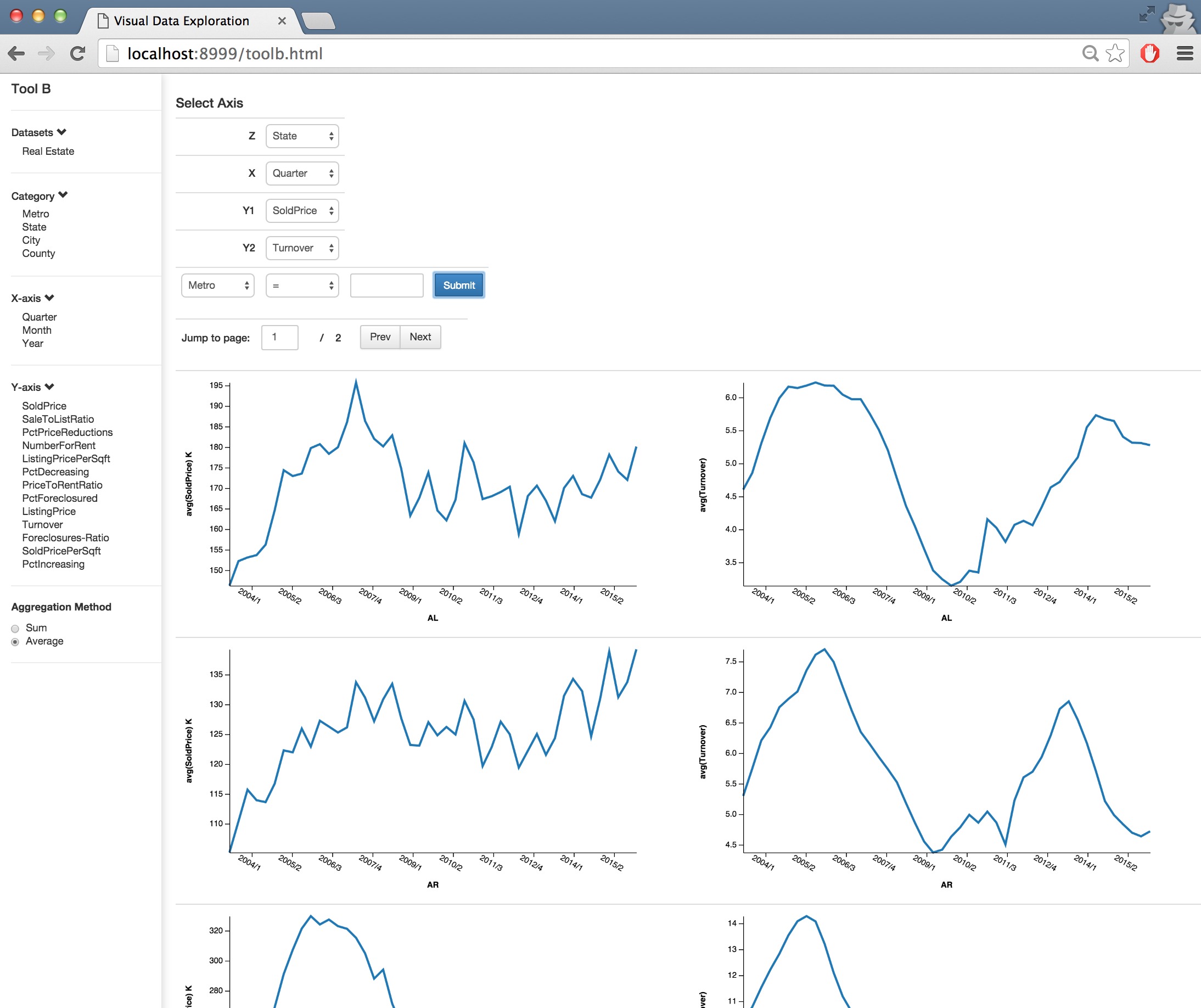}}
\end{center}
\vspace{-20pt}
\caption{The baseline interface implemented for the user study.}
\label{fig:baseline}
\vspace{-10pt}
\end{figure}
}

\vspace{-8pt}
\subsection{Key Findings}
\vspace{-2pt}

Three key findings emerged from the study and are described below.
We use $\mu$, $\sigma$, $\chi^2$ to denote average, standard deviation, 
and Chi-square test scores, respectively.

\paraem{Finding 1: \zv enables faster and more accurate exploration than existing visualization tools.}
Since all of our tasks involved generating multiple visualizations
and comparing them to find desired ones, participants were not only able to complete the tasks
 {\em faster}---$\mu$=115s, $\sigma$=51.6 for 
\zv vs.~$\mu$=172.5s, $\sigma$=50.5 for the baseline---but also more 
{\em accurately}---$\mu$=96.3\%, $\sigma$=5.82 for \zv vs.~$\mu$=69.9\%, $\sigma$=13.3 for the baseline.
\techreport{A one-way between-subjects ANOVA,
followed by a post-hoc Tukey's test~\cite{tukey1949comparing},
we found that \zv
had statistically significant
faster task completion times compared to the baseline interface,
with $p$ value of 0.0069.}
The baseline required considerable manual exploration to complete
the same task, explaining the high task completion times. In addition, participants frequently compromised
by selecting suboptimal answers before browsing
the entire list of results for better ones, explaining the 
low accuracy.
On the other hand, \zv was able to 
automate the task of finding desired visualizations, considerably reducing manual
effort.
Also of note is the fact that the accuracy with \zv was close to 100\%---indicating
that a short 15 minute tutorial on \zq was enough to equip users with the knowledge they
needed to address the tasks---and that too, within 2 minutes (on average). 

When asked about using \zv vs.~the baseline in their current workflow, 
9 of the 12 participants stated that they would use \zv in their workflow, 
whereas only two participants stated that they would use our baseline tool ($\chi^2=8.22$, p<0.01). 
When the participants were asked how, 
one participant provided a specific scenario: \textit{``If I am doing my social science study, 
and I want to see some specific behavior among users, then I can use tool A [\zv] 
since I can find the trend I am looking for and easily see what users fit into the pattern.''} (P7). 
In response to the survey question \emph{``I found the tool to be effective in visualizing the data I want to see''}, 
the participants rated \zv higher ($\mu$=4.27, $\sigma$=0.452) than the baseline ($\mu$=2.67, $\sigma$=0.890) 
on a five-point Likert scale. A participant experienced in Tableau 
commented: \textit{``In Tableau, there is no pattern searching. If I see some pattern in Tableau, 
such as a decreasing pattern, and I want to see if any other variable is decreasing in that month, 
I have to go one by one to find this trend. But here I can find this through the query table.''} (P10). 


\paraem{Finding 2: \zv complements existing database and data mining systems, and programming languages.}
When explicitly asked about comparing \zv with the
tools they use on a regular basis for data analysis, 
all participants acknowledged that \zv adds 
value in data exploration not encompassed by their tools. 
\zq augmented with inputs from the sketching canvas proved to be extremely effective.
For example P8 stated: \emph{``you can just [edit] and draw to find out similar patterns. You\textquotesingle ll 
need to do a lot more through Matlab to do the same thing.''}
Another experienced participant mentioned the benefits of not needing to know much programming to accomplish certain tasks: \emph{``The obvious good thing is that you can do complicated queries, and you don\textquotesingle t have to write SQL queries... I can imagine a non-cs student [doing] this.''} (P9).
When asked about the specific tools they would use to solve the user study tasks,
all participants reported a programming language like Matlab or Python. 
This is despite half of the participants reporting using a relational database regularly,
and a smaller number of participants (2) reporting using a data mining tool regularly.
Additionally, multiple participants, even those with extensive programming experience,
reported that \zv would take less time and fewer lines of code for certain data exploration tasks.
(Indeed, we found that all participants were able to complete the user study tasks in under 2 minutes.)
In follow-up email interviews, we asked a few participants to respond with code from their
favorite data analytics tool for the user study tasks.
Two participants responded --- one with Matlab code, one with Python code.
Both these code snippets were much longer than \zq:
as a concrete example, the participant accomplished the same task with 
38 lines of Python code compared to 4 lines of \zq.
While comparing code may not be fair, the roughly order of magnitude
difference demonstrates the power of \zv over existing systems.

\paraem{Finding 3: \zv can be improved.}  
\papertext{Participants outlined some areas for improvement: some requested drag-and-drop interactions 
to support additional operations, such as outlier finding; others wanted a more polished interface;
and some desired bookmarking and search history capabilities.}
\techreport{
 While the participants looked forward to using custom query builder in their own workflow, a few of them were interested in directly exposing the commonly-used trends/patterns such as outliers, through the drag and drop interface. Some were interested in knowing how they could integrate custom functional primitives (we could not cover it in the tutorial due to time constraints). In order to improve the user experience, participants suggested adding instructions and guidance for new users as part of the interface. Participants also commented on the unrefined look and feel of the tool, as well as the lack of a diverse set of usability related features, such as bookmarking and search history, that are offered in existing systems.
}

\later{
\paraem{Finding 1: \zv enabled over 50\% and 100\% faster task completion times than the baseline for the custom query builder and drag and drop interfaces, respectively.} 
We found that the drag and drop interface
had the smallest, and most predictable completion time ($\mu=74s$, $\sigma=15.1$),
the custom query builder had a higher completion time and a higher variance ($\mu=115s$, $\sigma=51.6$),
while the baseline had a significantly higher completion time ($\mu=172.5s$, $\sigma=50.5$), more than twice the completion time of the drag and drop interface,
and almost 50\% higher than the custom query builder interface.
This is not surprising, given that the baseline tool requires more
manual exploration compared to the other two interfaces.
Using 
a one-way between-subjects ANOVA,
followed by a post-hoc Tukey's test~\cite{tukey1949comparing},
we found that the
drag and drop interface and the custom query builder interface
had statistically significant
faster task completion times compared to the baseline interface,
with $p$ values of 0.0010 and 0.0069, respectively.
The difference between the drag and drop interface
and the custom query builder interface was not statistically significant,
with a $p$ value of $0.06$. Detailed statistics can be found in Table\ref{tab:turkeytest}.

\techreport{
\begin{table}[H]
 \centering \scriptsize
  {\tt    
  \scalebox{0.75}{
  \begin{tabular}{|l|l|l|l|} \hline
      \textbf{Treatments} & \textbf{Q statistic} & \textbf{p-value} & \textbf{inference} \\ \hline
	Drag and drop interface vs. Custom query builder & 3.3463 & 0.0605331 & insignificant \\ \hline
	Drag and drop interface vs. Baseline tool & 7.9701 & 0.0010053 & significant (p<0.01) \\ \hline
	Custom query builder vs. Baseline tool & 4.6238 & 0.0069276 & significant (p<0.01) \\ \hline    
     \end{tabular}
  }
  }
\vspace{-10pt}
\caption{Tukey's test on task completion time}
\label{tab:turkeytest}
\vspace{-10pt}
\end{table}
}

\begin{figure}
\begin{center}
\includegraphics[width=0.6\columnwidth]{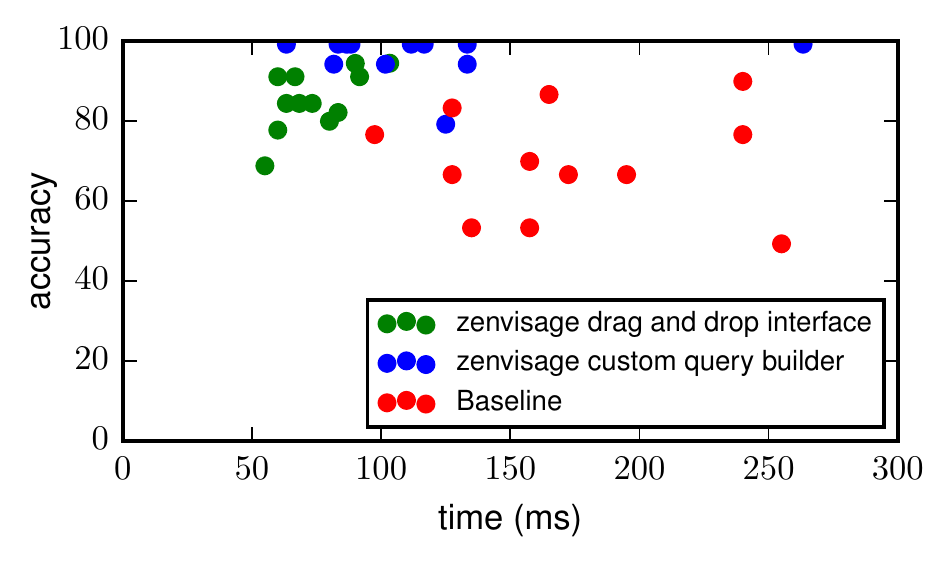}
\end{center}
\vspace{-25pt}
\caption{Accuracy over time results for zenvisage and baseline}
\label{fig:accuracyvstime}
\vspace{-20pt}
\end{figure}

\paraem{Finding 2: \zv helped retrieve 20\% and 35\% more accurate results than the baseline for drag and drop and custom query builder interfaces, respectively.} 
Comparing the participants' responses against the ground truth responses, 
we found that participants achieved the lowest accuracy with the baseline tool ($\mu=69.9\%$, $\sigma=13.3$). 
\techreport{Further, any tasks that required comparing multiple
pairs of visualizations or trends ended up having lower accuracy than those that looked
for a pattern, such as a peak or an increasing trend, in a single visualization.}
The low accuracy could be attributed to the fact that the participants selected suboptimal answers before browsing through the entire list of results for better answers. Conversely, \zv is able to
accept more fine-grained user input, and
rank output visualizations. With \zv,
participants were able to retrieve accurate answers
with less effort.
Between the two interfaces in \zv,
while the drag and drop interface had a faster task completion time,
it also had a lower accuracy ($\mu=85.3\%$, $\sigma=7.61$).
The custom query builder interface had the
highest accuracy ($\mu=96.3\%$, $\sigma=5.82$).
The accuracy was greater than the baseline by over 20\% (for drag and drop)
and over 35\% (for custom builder).
Finally, looking at both accuracy and task completion times together in Figure~\ref{fig:accuracyvstime}, we see that \zv \emph{helps in ``fast-forwarding to desired visualizations'' with high accuracy.}

\paraem{Finding 3: The drag and drop interface and the custom query builder complemented each other.
The drag and drop interface was intuitive, while the custom builder was useful for complex queries.} 
The majority of participants (8/12) 
mentioned that the drag and drop 
and custom query builder interfaces complemented each other \techreport{in making \zv powerful}
when searching for specific insights.\techreport{ 25 percent (3/12) of the participants (all with backgrounds in machine learning and data mining) 
preferred using only the custom query builder interface, while just one participant (with no prior experience with query languages or programming languages) found only the drag and drop interface as useful.}
Participants stated that they would use drag and drop for exploration and simple pattern search and custom query builder for specific / complex queries. One stated: \textit{``When I know what I am looking for is exact, I want to use the query table. The other one [drag and drop] is for exploration when you are not so sure.'' (P9).} 
Participants liked the simplicity and intuitiveness of the drag and drop interface. 
One stated: \textit{``[The drawing feature] is very easy and intuitive, I can draw... and automatically get the results.''} (P2).
On a five-point Likert (5 is strongly agree), participants
found the drawing feature easy to use ($\mu=4.45$, $\sigma=0.674$). 
Despite the positive feedback, participants did identify limitations,
specifically, that the functionality was restricted to identifying trends
similar to a single hand-drawn trend.

The majority of the participants (11/12) 
stated that the custom query builder 
allowed them to search for complex insights that might be 
difficult to answer via drag and drop. 
When asked if the custom query builder was difficult to use compared to
the drag-and-drop interface, participants tended to agree ($\mu=3.73$, $\sigma=0.866$ on a five-point Likert scale where five is strongly agree).
At the same time, participants acknowledged 
the flexibility and the efficacy supported by the custom query builder interface. 
One participant explicitly stated why they 
preferred the custom query builder over the drag and drop interface: 
\textit{``When the query is very complicated, [the drag and drop] has its limitations, 
so I think I will use the query table most of the times unless I want to get a brief impression.''} (P6).
Another participant stated that the custom query builder interface \textit{``has more functionality and it lets me express what I want to do clearly more than the sketch''} (P4). This was consistent with other responses. 
Participants with a background in machine learning, 
data mining, and statistics (3/12) preferred
using the custom query builder over the drag and drop interface.

\paraem{Finding 4: \zv was more effective (4.27 vs. 2.67 on a five-point Likert scale)
than the baseline tool at visual data exploration, and would be
a valuable addition to the participants' analytics workflow.} 
In response to the survey question \emph{``I found the tool to be effective in visualizing the data I want to see''}, the participants rated \zv (the drag and drop interface and the custom query builder interface) higher ($\mu=4.27$, $\sigma=0.452$) than the baseline tool ($\mu=2.67$, $\sigma=0.890$) on a five-point Likert scale.
A participant experienced in Tableau commented: \textit{``In Tableau, there is no pattern searching. If I 
see some pattern in Tableau, such as a decreasing pattern, and I want to see if any other variable is 
decreasing in that month, I have to go one by one to find this trend. But here I can find this through the query table.''} (P10).

\paraem{Finding 5: \zv complements existing database and data mining systems, and programming languages.}
When explicitly asking participants about comparing \zv with the
tools they use on a regular basis for data analysis, 
all participants acknowledged that \zv adds 
value in data exploration not encompassed by their tools. 
\zq augmented with inputs from the sketching canvas proved to be extremely effective.
For example P8 stated: \emph{``you can just [edit] and draw to find out similar patterns. You\textquotesingle ll 
need to do a lot more through Matlab to do the same thing.''}
Another experienced participant mentioned the benefits of not needing to know much programming to accomplish certain tasks: \emph{``The obvious good thing is that you can do complicated queries, and you don\textquotesingle t have to write SQL queries... I can imagine a non-cs student [doing] this.''} (P9).
When asked about the specific tools they would use to solve the user study tasks,
all participants reported a programming language like Matlab or Python. 
This is despite half of the participants reporting using a relational database regularly,
and a smaller number of participants (2) reporting using a data mining tool regularly.
Additionally, multiple participants even with extensive programming experience
reported that \zv would take less time and fewer lines of code for certain data exploration tasks.
(Indeed, we found that all participants were able to complete the user study tasks in under 2 minutes.)
In follow-up email interviews, we asked a few participants to respond with code from their
favorite data analytics tool for the user study tasks.
Two participants responded --- one with Matlab code, one with Python code.
Both these code snippets were much longer than \zq:
as a concrete example, the participant accomplished the same task with 
38 lines of Python code compared to 4 lines of \zq.
While comparing code may not be fair, the roughly order of magnitude
difference demonstrates the power of \zv over existing systems.

\paraem{Finding 6: \zv can be improved.} While the participants looked forward to using custom query builder in their own workflow, a few of them were interested in directly exposing the commonly-used trends/patterns such as outliers, through the drag and drop interface. Some were interested in knowing how they could integrate custom functional primitives (we could not cover it in the tutorial due to time constraints). In order to improve the user experience, participants suggested adding instructions and guidance for new users as part of the interface. Participants also commented on the unrefined look and feel of the tool, as well as the lack of a diverse set of usability related features, such as bookmarking and search history, that are offered in existing systems.
}

\vspace{-5pt}
\section{Related Work}
\label{sec:related}
\vspace{-2pt}

We now discuss related prior work in a number of areas.
We begin with analytics tools --- visualization tools,
statistical packages and programming libraries, and relational databases.
Then, we talk about other tools that overlap somewhat with  \zv.

\paragraph{Visual Analytics Tools.} 
Visualization tools, such as ShowMe, Spotfire, 
and Tableau~\cite{DBLP:journals/cacm/StolteTH08, DBLP:journals/tvcg/MackinlayHS07,Ahlberg:1996:SIE:245882.245893},
along with similar tools from the 
database community~\cite{DBLP:conf/sigmod/GonzalezHJLMSSG10,DBLP:conf/sigmod/KeyHPA12,DBLP:conf/sigmod/LivnyRBCDLMW97,DBLP:conf/avi/KandelPPHH12}
have recently gained in popularity, catering to 
data scientists who lack programming skills.
Using these tools, these scientists can 
select and view one visualization at a time.
However, these tools do not operate on 
collections of visualizations at a time---and thus
they are much less powerful and the 
optimization challenges are minimal.
\zv, on the other hand, supports queries over collections
of visualizations, returning results not much slower than the
time to execute a single query (See Section~\ref{sec:experiments}). 
Since these systems operate one visualization at a time, users are 
also not able to directly identify desired patterns or needs. 

\begin{table}[t]
\scriptsize
\begin{verbatim}
with ranking    as (
with distances as (
with distance_ product_year  as (
with aggregate_ product_year as (
   select   product, year, avg(profit) as avg_profit 
   from   table group by  product, year) )
   select  s. product as source, d. product as destination, s.year, 
           power(s.avg_profit - d.avg_profit,2) as distance_year 
   from   aggregate_ product_year s, aggregate_ product_year d 
   where  s. product!=d. product and s.year=d.year )
   select  source, destination, sum(distance_year) as distance 
   from   distance_ product_year   groupby  source, destination )
   select  source, destination, distance, 
           rank() over (partition by source order by distance asc) 
           rank from distances )
   select  source, destination, distance 
   from   ranking   where  rank < 10;
\end{verbatim}
\vspace{-15pt}
\caption{Verbose \sql query}\label{tab:verbose}
\vspace{-20pt}
\end{table}

\paragraph{Statistical Packages and Programming Libraries:}
Statistical tools (e.g., KNIME, RapidMiner, SAS, SPSS)
support the easy application of 
data mining and statistical primitives---including
prediction algorithms and statistical tests. 
While these tools support the selection of a prediction
algorithm (e.g., decision trees) 
to apply, and the appropriate parameters,
they offer no querying capabilities, 
and as a result do not need
extensive optimization.
As a result, these tools cannot support user needs like 
those describe in the examples in the introduction. 
Similarly, programming libraries such as 
Weka~\cite{holmes1994weka}
and Scikit-learn~\cite{pedregosa2011scikit} embed machine learning
 within programs.
However, manually translating the user desired patterns
into code that uses these libraries
will require substantial user effort and hand-optimization.
In addition, writing new code and hand-optimization will need
to be performed every time the exploration needs change.
Additionally, for both statistical tools and programming libraries, 
there is a need for programming ability and understanding
of machine learning and statistics to be useful---something 
we cannot expect all data scientists to possess.

\begin{figure}[!h]
\vspace{-10pt}
\begin{center}
\includegraphics[width=0.48\columnwidth]{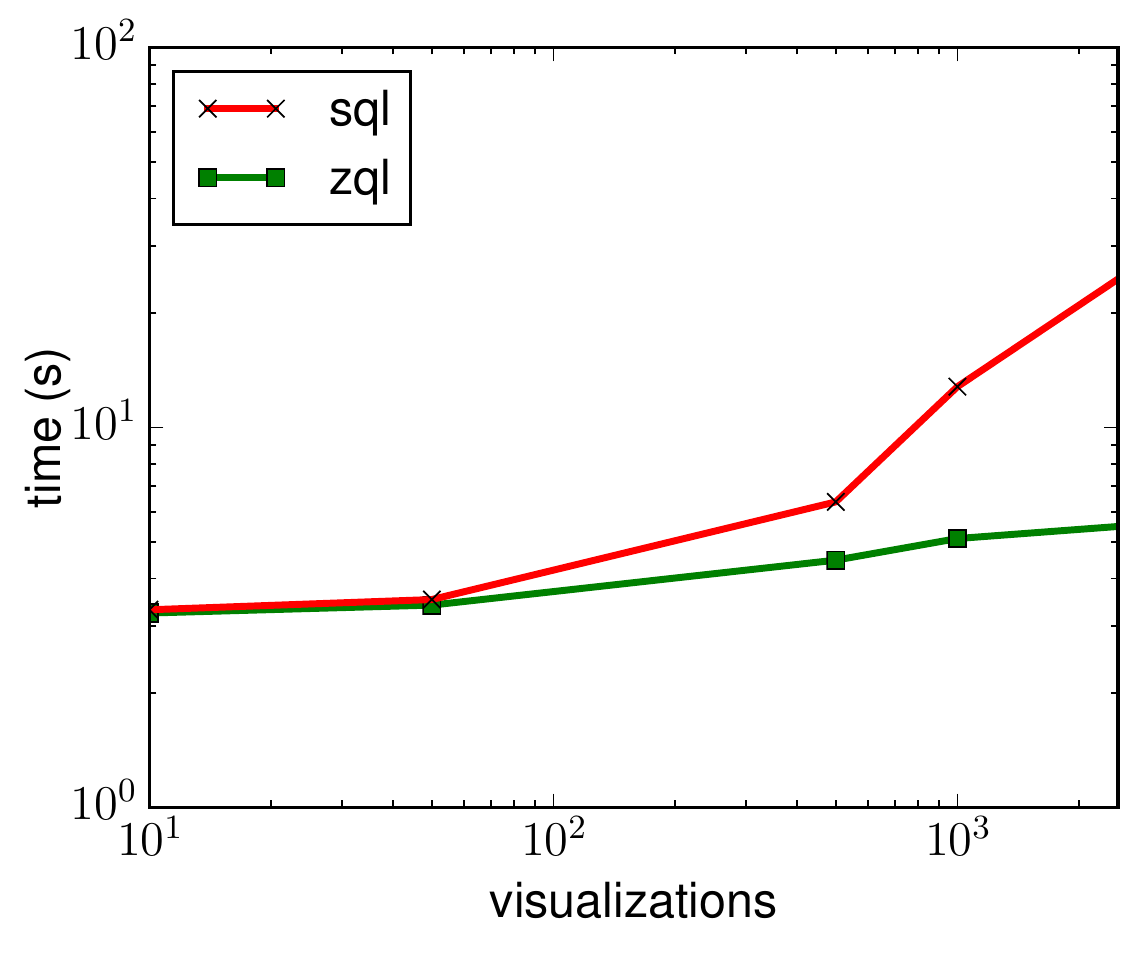}
\end{center}
\vspace{-15pt}
\caption{\zq vs \sql: we want to find top 10 similar products for every product
on varying the number of products from 10---5000.}
\vspace{-13pt}
\label{fig:opt_syn_zqlvssql}
\end{figure}

\paragraph{Relational Databases.}
Relational databases can certainly support interactive analytics
via \sql. \zv uses relational databases as a backend computational
component, augmented with an engine that uses \fuse to optimize
accesses to the database, along with efficient processing code.
Thus, one can certainly express some \zq queries 
by writing multiple \sql queries (via procedural \sql), 
using complex constructs only found in some databases, 
such as common table expressions (CTE) 
and window functions.
As we saw in Section~\ref{sec:userstudy}, these \sql queries are too cumbersome to write, and 
are not known to most users of databases---during our user study, we found that
all participants who had experience with \sql were not aware of these constructs; in fact,
they responded that they did not know of any way of issuing \zq queries in \sql, preferring
instead to express these needs in Python.
In Table~\ref{tab:verbose}, we list the verbose \sql query that
computes the following: for each product,
find 10 other products that have most similar profit over year trends. 
The equivalent 
\zq query takes two lines.
Further, we were able to write the \sql query only because the 
function $D$ is Euclidean distance: for other functions, we are unable
to come up with appropriate \sql rewritings.
On the other hand, for \zq, it is effortless to change the function by selecting it from a drop-down menu.
Beyond being cumbersome to write, the constructs required 
lead to severe performance penalties on most databases---for instance,
PostgreSQL materializes intermediate results when executing queries with CTEs.
To illustrate, we took the \sql query in Table~\ref{tab:verbose}, 
and compared its execution with the execution of the equivalent \zq. 
As depicted in Figure \ref{fig:opt_syn_zqlvssql}, 
the time taken by PostgreSQL increases sharply as 
the number of visualizations increases, taking up to 
10$\times$ more time as compared to \zq query executor.
This indicates that \zv is still important even for the restricted
cases where we are able to correctly write the queries in \sql.

\paragraph{OLAP Browsing.} There has been some work on 
interactive browsing of data cubes~\cite{DBLP:conf/vldb/Sarawagi99,DBLP:conf/vldb/SatheS01}.
The work focuses on suggestions for raw aggregates to examine
that are informative given past browsing, or those that show a generalization 
or explanation of a specific cell---an easier problem meriting simpler techniques---not
addressing the full exploration capabilities provided by \zq.

\paragraph{Data Mining Languages:} 
There has been some limited work in data mining query languages, all from the early 90s, 
on association rule mining (DMQL~\cite{han1996dmql}, MSQL~\cite{imielinski2000query}),
or on storing and retrieving models on data
(OLE DB~\cite{netz2001integrating}),
as opposed to a general-purpose visual data exploration language
aimed at identifying visual trends. 

\paragraph{Visualization Suggestion Tools:}
There has been some recent work on building systems that 
suggest visualizations.
Voyager~\cite{wongvoyager2015} recommends
visualizations based on aesthetic properties of the
visualizations, as opposed to queries.
SeeDB~\cite{Vartak:2015:SED:2831360.2831371} 
recommends visualizations that best display the 
difference between two sets of data. 
SeeDB and Voyager can be seen to be special cases of \zv.
The optimization techniques outlined are a substantial generalization
of the techniques described in SeeDB; 
while the techniques in SeeDB are special-cased to one setting 
(a simple comparison),
here, our goal is to support and optimize all \zq queries.

\techreport{
\paragraph{Multi-Query Optimization:} 
There has been a lot of work on Multi-Query Optimization (MQO), both classic, 
e.g.,~\cite{DBLP:journals/tods/Sellis88,shim1994improvements,roy2000efficient}, and
recent work, e.g.,~\cite{giannikis2013workload,psaroudakis2013sharing,DBLP:journals/pvldb/KementsietsidisNCV08,giannikis2014shared}.
Overall, the approach adopted is to batch queries, decompose into operators,
and build ``meta''-query plans that process multiple queries at once,
with sharing at the level of scans, or at the level of higher level operators
(either via simultaneous pipelining or a true global query plan~\cite{psaroudakis2013sharing}).
Unlike these techniques which require significant modifications
to the underlying database engine---indeed, some of these systems do not even provide
full cost-based optimization and only support hand-tuned plans~\cite{giannikis2013workload}, 
in this paper, we adopted two 
syntactic rewriting techniques that operate outside of any relational database as a backend
without 
requiring any modification, and can thus seamlessly leverage improvements
to the database. 
Our third optimization is tailored to the \zq setting and does not apply more broadly.
}


\techreport{
\paragraph{Anomaly Discovery:}
Anomaly detection is a well-studied topic~\cite{Chandola:2009:ADS:1541880.1541882,Agyemang:2006:CSN:1609942.1609946,Patcha:2007:OAD:1269989.1270131}. Our goal in that
\zv is expected to be interactive, especially on large datasets; most work in anomaly detection focuses on accuracy at the cost of latency and is typically a batch operation. In our case, since interactiveness is of 
the essence, and requests can come at any time, the emphasis is on scalable on-the-fly data processing aspects.
}

\techreport{
\paragraph{Time Series Similarity and Indexing:}
There has been some work on indexing of
of time series data, e.g., \cite{timeseries1,timeseries2,timeseries3,timeseries4,timeseries5,
timeseries6,timeseries7};
for the attributes that are queried frequently, 
we plan to reuse these techniques for similarity search.
For other attributes, indexing and maintaining all trends
is impossible, since the number of trends grows exponentially with the
number of indexed attributes.
}

\later{
\tar{Previous text}
Our work is related to prior work in a number of areas.
\paragraph{Visual Analytics Tools:} A number of 
interactive data analytics tools have been developed, such as ShowMe, Polaris, Spotfire, and
Tableau~\cite{DBLP:journals/cacm/StolteTH08, DBLP:journals/tvcg/MackinlayHS07,Ahlberg:1996:SIE:245882.245893}.
Here, the onus is on the user to specify the visualization to be generated,
and then repeat until the desired visualization is identified. 
Similar specification tools have also been introduced
by the database community\cite{DBLP:conf/sigmod/GonzalezHJLMSSG10,DBLP:conf/sigmod/KeyHPA12,DBLP:conf/sigmod/LivnyRBCDLMW97,DBLP:conf/avi/KandelPPHH12}.
\zv is related to work on browsing data cubes---finding explanations, getting suggestions for cubes to visit,
or identifying generalizations from a single cube, 
e.g., \cite{DBLP:conf/vldb/Sarawagi99, 
DBLP:conf/vldb/SatheS01}. 
While we may be able to reuse the metrics from that line of work,
the same techniques will not directly apply to the identification of
trends, patterns, and anomalies for interactive visualizations.  
Specifically, no OLAP techniques capture the interestingness of slices
of the data cube that can be visualized via two (or more) dimensional
visualizations: instead, the focus is more on 
raw aggregate data, on interestingness of cells of the cube, 
displayed within multidimensional tables.  
Thus, the emphases are different. 
Indeed, we believe that, given the popularity of tools like Tableau,
users prefer to browse and explore data visually rather than in tables.
There has been some work sketch-based interfaces for visualization~\cite{visual1, visual2,visual3, visual4}: however, none of this work addresses
the deep scalability issues at play if these techniques are to be applied on large
datasets.
There is limited work on visual query interfaces~\cite{gesturedb, dbtouch, qbe, dataplay}, but the goal there is to make it easier to pose SQL queries.

\paragraph{Time Series Similarity and Indexing:}
There has been some work on indexing of
of time series data, e.g., \cite{timeseries1,timeseries2,timeseries3,timeseries4,timeseries5,
timeseries6,timeseries7};
for the attributes that are queried frequently, 
we plan to reuse these techniques for similarity search.
For other attributes, indexing and maintaining all trends
is impossible, since the number of trends grows exponentially with the
number of indexed attributes.
\paragraph{Visualization Recommendation Tools:}
There has been some recent work on building systems that 
recommend visualizations
to analysts.
Voyager~\cite{wongvoyager2015} recommends
visualizations based on aesthetic properties of the
visualizations, as opposed to user queries.
SeeDB~\cite{tr} 
recommends visualizations that best display the 
difference between two sets of data. 
Both SeeDB and Voyager can be seen to be special cases of \zv,
since \zv can be viewed to be a user-driven visualization
recommendation engine as well.
The optimization techniques outlined are a generalization
of the techniques described in SeeDB; 
while the techniques in SeeDB are special-cased to the 
query considered (a simple comparative query),
here, our goal is to support and optimize all \zq queries.
\paragraph{Data Mining Query Languages:}
There has been some limited work in the space of data mining query languages, all from the early 90s, 
focusing
either on
association rule mining (languages such as DMQL~\cite{han1996dmql} and MSQL~\cite{imielinski2000query}),
or on storing and retrieving models on data
(languages such as OLE DB~\cite{netz2001integrating}),
as opposed to a visual exploration language like \zq.
\paragraph{Multi-Query Optimization:} 
Our proposed primitive query optimization techniques are currently 
tailored to the \zv setting, specifically trying to identify 
opportunities for batching of 
individual visualization generation \sql queries 
so as to ``combine work''
as much as possible, even before they are issued to the database. 
These optimization techniques are external to the database system,
are completely interpreted based on the rows of \zq, and do not 
require any modifications to the underlying database.
Indeed, there are very powerful multi-query optimization techniques
in related work, e.g.,~\cite{DBLP:journals/tods/Sellis88,shim1994improvements,dalvi2003pipelining,roy2000efficient,DBLP:journals/pvldb/KementsietsidisNCV08},
that we plan to leverage as future work to further improve the performance
of \zv, when we attempt to holistically
process \zq queries by making modifications to the underlying database
query processor, rather than simply operating as a ``wrapper'' on top
of a database.
}


\vspace{-2pt}
\vspace{-5pt}
\section{Conclusion}
\label{subsec:conclusion}
\vspace{-2pt}
We propose \zv, a visual analytics tool for effortlessly identifying desired
visual patterns from large datasets.
We described the formal syntax of the query language \zq, motivated by many
real-world use-cases, and demonstrated that \zq is \vzalg-complete\papertext{ (See~\cite{techreport})}.
\zv enables users
to effectively and accurately perform visual exploration tasks, 
as shown by our user study, and complements other tools.
In addition, we show that our
optimizations for \zq execution lead to considerable improvements over
leveraging the parallelism inherent in databases.
Our work is a promising first step towards substantially simplifying
and improving the process of interactive data exploration
for novice and expert analysts alike.


\paragraph{Acknowledgements.} We thank the anonymous reviewers for their valuable feedback. We acknowledge support from grant IIS-1513407 and IIS-1633755 awarded by the National Science Foundation, grant 1U54GM114838 awarded by NIGMS and 3U54EB020406-02S1 awarded by NIBIB through funds provided by the trans-NIH Big Data to Knowledge (BD2K) initiative (www.bd2k.nih.gov), and funds from Adobe, Google, and the Siebel Energy Institute. 
The content is solely the responsibility of the authors and does not necessarily represent the views of the funding organizations.

\vspace{-8pt}
{\scriptsize
\bibliographystyle{abbrv}
\bibliography{zenvisageql}
}

\iftechreportcode
\appendix
Here, we provide additional details on X, Y, Z, and Process columns (Appendix
\ref{querylang-appendix}),
four real-world complex examples (Appendix \ref{sec:examples}),
 and provide a participant's python code implementation (Appendix
 \ref{sec:userstudy-appendix}) for a \zq task to support the finding 2 in the
user-study (Section \ref{sec:userstudy}).    
 
\begin{figure*}
    \centering
    \begin{minipage}{0.40\textwidth}
	\vspace{-10pt}
	\fbox{
        \includegraphics[width=\textwidth]{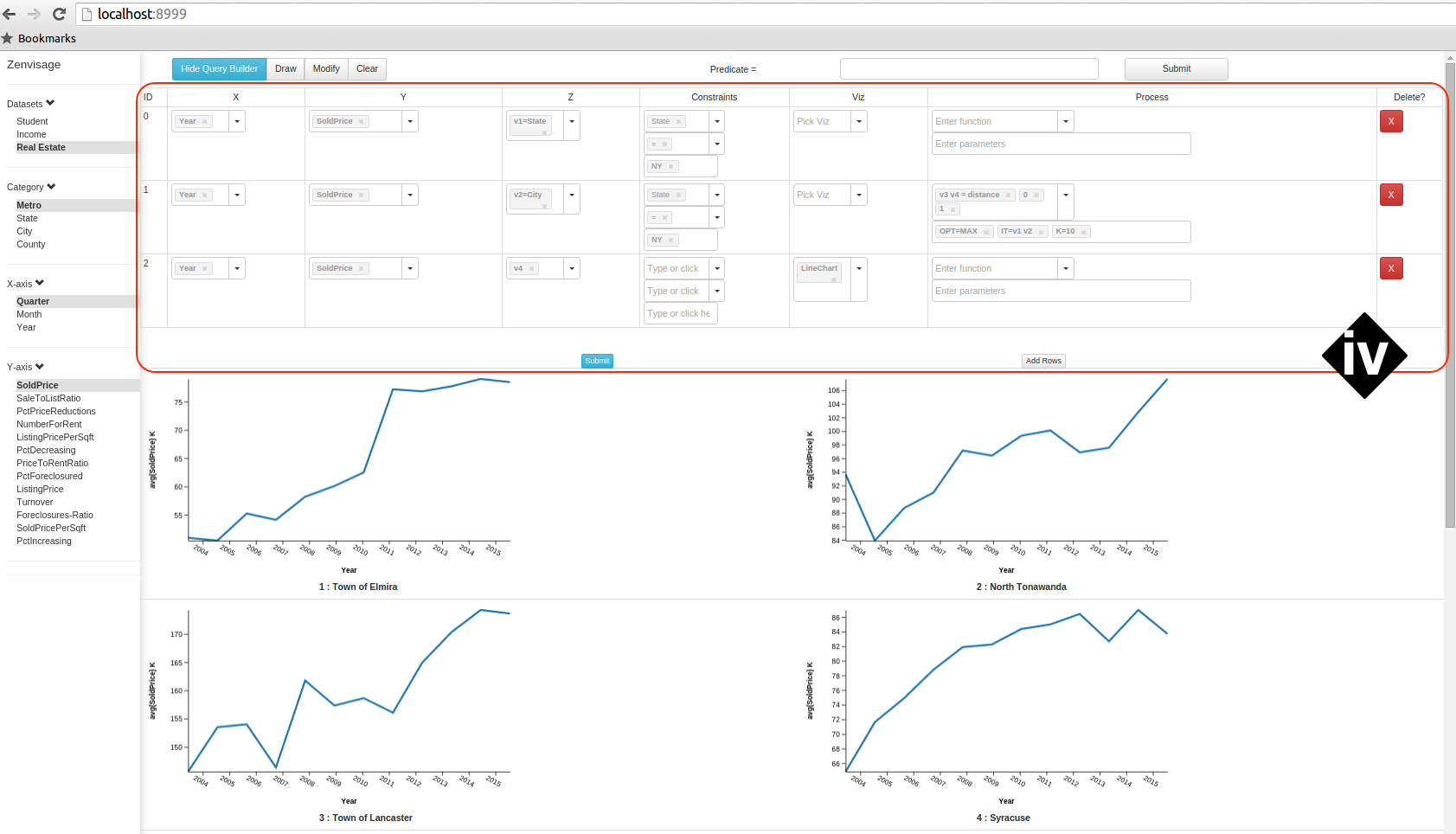}}
	\vspace{-15pt}
        \caption{Custom query builder with bar charts}
	\label{fig:querybuilder_linechart}
	\vspace{-10pt}
    \end{minipage}
    \qquad 
    \begin{minipage}{0.40\textwidth}
	\vspace{-10pt}
	\fbox{
        \includegraphics[width=\textwidth]{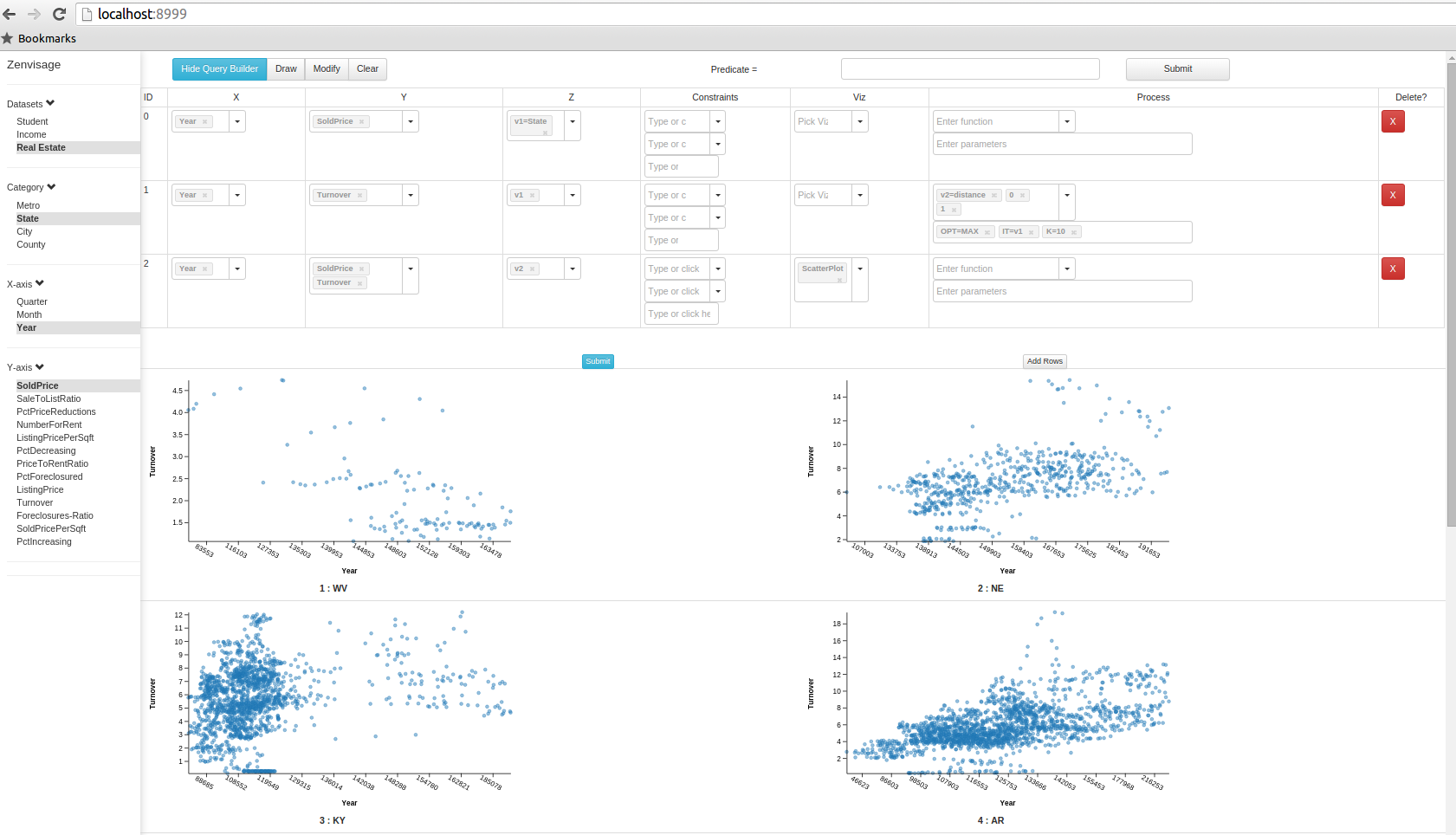}}
	\vspace{-15pt}
        \caption{Custom query builder with scatterplots}
        \label{fig:querybuilder_scatter}
	\vspace{-10pt}
    \end{minipage}
    \label{fig:scatterbar}
\end{figure*}
\techreport{

\section{Query Language Formalization: Additional Details}
\label{querylang-appendix}

In this section, we present some additional details on our formalization
that we did not cover in the main body of the paper.

\subsection{Additional Details on X and Y Columns}
In addition to using a single attribute for an X or Y column,
\zq also allows the use of the Polaris
table algebra~\cite{tableau} in the X and Y columns to to arbitrarily compose multiple
attributes into a single attribute;
all three operators are supported: $+$, $\times$,
$/$. Table~\ref{tab:polaris-algebra-plus} shows an example of using the $+$ operator to
visualize both profits and sales on a single y-axis. 
Note that this is different
from the example given in Table~\ref{tab:xy-coll}, which generates
two visualizations, as opposed to a single visualization.
An example using both table algebra and sets is given in
Table~\ref{tab:polaris-algebra-times}, which uses the $\times$ operator to return
the set of visualizations which measures the sales for the Cartesian product of `product'
and one of `county', `state', and `country'.

\begin{table}[H]
 \centering\scriptsize
  {\tt   
    \begin{tabular}{|r|l|l|l|} \hline
      \textbf{Name} & \textbf{X} & \textbf{Y} & \textbf{Z} \\ \hline
      *f1 & `product' & `profit' + `sales' & location.`US' \\ \hline
    \end{tabular}
  }
  \vspace{-10pt}\caption{A \zq query for a visualization which depicts both profits and sales
  on the y-axis for products in the US.}
  \label{tab:polaris-algebra-plus}
\vspace{-10pt}\end{table}

\begin{table}[H]
  \centering\scriptsize
  {\tt   
    \begin{tabular}{|r|l|l|} \hline
      \textbf{Name} & \textbf{X} & \textbf{Y} \\ \hline
      *f1 & `product' $\times$ \{`county', `state', `country'\}
      & `sales' \\ \hline
    \end{tabular}
  }
  \vspace{-10pt}\caption{A \zq query for the set of visualizations which measures the sales
    for one of (`product', `county'), (`product', `state'), and
  (`product', `country').}
  \label{tab:polaris-algebra-times}
\vspace{-10pt}\end{table}

\later{
\alkim{Mention 1-D visualizations}\agp{probably not critical}

\alkim{Simple functions which change only the value of an attribute are
  allowed (e.g., extracting year from date), but functions are only allowed to
be called constant atoms; this can extend to Z}
\agp{good to know but probably not critical. keep for TR.}
}

\subsection{Additional Details on the Z Column}
\zq also allows the iteration over attributes in the Z column as shown in
Table~\ref{tab:z-3}. The result of this query is the set of all sales over time
visualizations for every possible slice in every dimension except `time' and
`sales'. Since both attribute and attribute value can vary in this case, we
need separate variables for each component, and the full attribute name, value
pair (z1.v1) must be specified. Note that the resulting set of visualizations
comes from the Cartesian product of possible attribute and attribute value pairs.
The first * symbol refers to all possible attributes, while the second * symbol
refers to all possible attribute values given an
attribute. If the user wishes to specify specific subsets of attribute values
for attributes, she must name them individually.

\begin{table}[H]
  \centering\scriptsize
  {\tt  
    \begin{tabular}{|r|l|l|l|} \hline
      \textbf{Name} & \textbf{X} & \textbf{Y} & \textbf{Z} \\ \hline
      *f1 & `year' & `sales' & z1.v1 \IN (* $\setminus$ \{`year', `sales'\}).* \\ \hline
    \end{tabular}
  }
  \vspace{-10pt}\caption{A \zq query which returns the set of sales over year visualizations
  for each attribute that is not time or sales.}
  \label{tab:z-3}
\vspace{-10pt}\end{table}



\later{
\agp{Suggest commenting out:
Because attribute
values must be specified, data types other than string literals, such as
numbers, are also valid in this column\footnote{While \zq currently supports
  only string literals, numbers, and datetimes as primitive data types, nothing
  precludes us from including the other standard data types from \sql
  databases.}.
}
}

\later{\alkim{Ways to do arbitrary binning? (e.g., strings binned by the first letter
of alphabet)}}

\subsection{Additional Details on the Name Column}
\label{name-appex}


For deriving visualization collections based on other
collections in the Name column, in addition to + operation, \zq supports following operations.
\begin{inparaenum}[\itshape (i)\upshape]
\item
  {\tt f3 \IN f1-f2}: where {\tt f3} refers to the list of visualizations in {\tt f1}
  with the exception of the visualizations which appear in {\tt f2},
\item
  {\tt f2 \IN [f1[i]]}: where {\tt f2} refers to the {\tt i}th visualization in {\tt f1},
\item
  {\tt f2 \IN f1[i:j]}: where {\tt f2} refers to the list of visualizations starting from
  {\tt i}th visualization to the {\tt j}th visualization in {\tt f1},
\item
  {\tt f2 \IN f1.uniq}: where {\tt f2} refers to the set of visualizations derived
  from {\tt f1} by removing duplicate visualizations (only the first appearance
  of each visualization is kept), and
\item
  {\tt f3 IN f1\^{}f2}: where {\tt f3} refers to the list of visualizations in
  {\tt f1} which also appear in {\tt f2}; an ``intersection'' between lists of
  visualizations in some sense. \alkim{Should I mention here that {\tt f3} is in
  same ordering as {\tt f1} or is that already sort of implied by this.}
\end{inparaenum}
These operations are useful if the user wants to throw away some visualizations,
or create a new larger set of visualizations from smaller sets of visualizations.

After a visualization collection has been derived using the Name column, the user may
also define axis variables in the X, Y, and Z columns using the special {\tt
\_} symbol to bind to the derived collection.
For example in Table~\ref{tab:name-extra}, {\tt v2}
is defined to be the iterator which iterates over the set of product values
which appear in derived collection {\tt f3}; in this case, {\tt v2}
iterates over all possible products. {\tt y1} is defined to be the iterator
over all the values in the Y column of {\tt f3}. Although in the case of
Table~\ref{tab:name-extra}, the only value {\tt y1} takes on is `sales', 
{\tt y1} and {\tt v2} are considered to be declared together, so the iterations
for {\tt y1}, {\tt v2} will look like: [(`sales', `chair'), (`sales', `table'),
...].
Also in this case, the variable {\tt y1} is not used, however, there may be 
other cases where it may be useful to iterate over multiple axis variables. 
The defined axis variables can then be used to create other visualization
collections
or within the Process column as shown in the 4th row of Table~\ref{tab:name-extra}.


  Finally, visualization collections may also be ordered based on the values of axis
  variables: {\tt f2 \IN f1.order}. Here, {\tt f1} is ordered based on the axis
  variables which appear together with the \TO symbol. Table~\ref{tab:order-ex}
  shows an example of such an operator in use.

  \alkim{Should we mention what happens if there exists more or less than one
  match in {\tt f1} with the axis variables which appear with \TO?}

  \begin{table*}[!htbp]
    \centering\scriptsize
    {\tt   
      \begin{tabular}{|r|l|l|l|l|} \hline
        \textbf{Name} & \textbf{X} & \textbf{Y} & \textbf{Z} & \textbf{Process} \\ \hline
        f1 & `year' & `sales' & v1 \IN `product'.* & u1 \IN
        $argmin_{v1}[k=\infty] T(f1)$\\
        *f2=f1.order & & & u1 \TO & \\
        \hline
      \end{tabular}
    }
    \vspace{-10pt}
    \caption{A \zq query which reorders the set of sales over years
      visualizations for different products based on increasing overall trend.
    }
    \label{tab:order-ex}
    \vspace{-10pt}
  \end{table*}

\begin{table*}[!htbp]
  \centering\scriptsize

  {\tt
      \begin{tabular}{|r|l|l|l|l|} \hline
        \textbf{Name} & \textbf{X} & \textbf{Y} & \textbf{Z} & \textbf{Process} \\ \hline
        f1 & `year' & `sales' & v1 \IN `product'.(* - `stapler') & \\
        f2 & `year' & `sales' & `stapler' & \\ 
        f3=f1+f2 &  &  y1 \IN \_ & v2 \IN `product'.\_ & \\
        f4 & `year' & `profit' & v2 & v3 \IN $argmax_{v2}k=10] D(f3,f4)$ \\
        *f5 & `year' & `sales' & v3 & \\
        \hline
      \end{tabular}
  }
  \vspace{-10pt}\caption{A \zq query which returns the sales over years visualizations for the
    top 10 products which have the most different sales over years
    visualizations and profit over years visualizations.
  }
  \label{tab:name-extra}
\vspace{-10pt}\end{table*}


\subsection{Additional Details on the Process Column}

\begin{table*}[!htbp]
  \centering\scriptsize

  {\tt
      \begin{tabular}{|r|l|l|l|l|} \hline
        \textbf{Name} & \textbf{X} & \textbf{Y} & \textbf{Z} & \textbf{Process}
        \\ \hline
        f1 & - & - & - & \\
        f2 & `year' & `sales' & v1 \IN `product'.* & (v2 \IN $argmax_{v1}[k=1]
        D(f1,f2)$), (v3 \IN $argmin_{v1}[k=1] D(f1,f2)$) \\
        *f3 & `year' & `sales' & v2 & \\
        *f4 & `year' & `sales' & v3 & \\ \hline
      \end{tabular}
  }
  \vspace{-10pt}\caption{A \zq query which returns the sales over years visualizations for the
    product that looks most similar to the user-drawn input and most dissimilar
  to the user-drawn input.}
  \label{tab:process-3}
\vspace{-10pt}\end{table*}



Although visualization collections typically outnumber processes, there may
occur cases in which the user would like to specify multiple processes in one
line. To accomplish this, the user simply delimits each process with a comma
and surrounds each declaration of variables with parentheses.
Table~\ref{tab:process-3} gives an example of this.

\later{\alkim{Should we mention how we do double for loops over the same variable?}}

}
\section{Additional Complete Examples}
\label{sec:examples}
To demonstrate the full expressive power of \zq, we present four realistic,
complex example queries. We show that even with complicated scenarios, the user
is able to capture the insights she wants with a few meaningful lines of \zq.

\later{
of a fictional GlobalMart company as a running example. It consists of the
following schema: \emph{Sales}\textit{(product, location, quantity, profit,
revenue, sales)} \alkim{And different time columns}. The Vice President of
Sales Department wants to develop a successful marketing strategy. So, he comes
up a list of interesting queries that can help him understand the product
trends.  In this section, we will discuss about how these queries can be easily
specified using our query language.
}

\paragraph{Query 1.} The stapler has been one of the most profitable products
in the last years for GlobalMart. The Vice President is interested in
learning about other products which have had similar profit trends. She wishes
to see some representative sales over the years visualizations for these
products.


\begin{table*}
\centering\scriptsize
{\tt
  \begin{tabular}{|r|l|l|l|l|l|} \hline
    \textbf{Name} &  \textbf{X} &  \textbf{Y} &  \textbf{Z} &  {\tt Viz} & \textbf{Process} \\ \hline
    f1 & `year' & `profit' & `product'.`stapler' & bar.(y=agg(`sum')) & \\
    f2 & `year' & `profit' & v1 \IN `product'.(* $\setminus$ \{`stapler'\}) & bar.(y=agg(`sum')) &
    v2 \IN $argmin_{v1}[k=100] D(f1,f2)$ \\
    f3 & `year' & `sales' & v2 & bar.(y=agg(`sum')) &v3 \IN $R(10, v2, f3)$ \\
    *f4 & `year' & `sales' & v3 & bar.(y=agg(`sum')) & \\ \hline
  \end{tabular}
}
\vspace{-10pt}
\caption{The \zq query which returns 10 most representative sales over year visualizations for
  products which have similar profit over year visualizations to
that of the stapler's.}
\label{tab:ex-1}
\vspace{-10pt}
\end{table*}

Table~\ref{tab:ex-1} shows what the query that the Vice President would write
for this scenario. She first filters down to the top 100 products which have
the most similar to profit over year visualizations to that of the stapler's
using the $argmin$ in the second row. Then, from the resulting set of products,
{\tt v2}, she picks the 10 most representative set of sales over visualizations
using $R$, and displays those visualizations in the next line with {\tt f4}.
Although the Vice President does not specify the exact distance metric for $D$
or specify the exact algorithm for $R$, she knows \zv will select the most
reasonable default based on the data.

\paragraph{Query 2.} The Vice President, to her surprise, sees that there a few
products whose sales has gone up over the last year, yet their profit has
declined. She also notices some product's sales have gone down, yet their
profit has increased. To investigate, the Vice President would like to know
about the top 10 products who have the most discrepancy in their sales and
profit trends, and she would like to visualize those trends.

%
\begin{table*}
\centering\scriptsize
{\tt
  \begin{tabular}{|r|l|l|l|l|l|l|} \hline
    \textbf{Name} &  \textbf{X} &  \textbf{Y} &  \textbf{Z} &  {\tt
    Z2} & {\tt Viz} & \textbf{Process} \\ \hline
    f1 & `month' & `profit' & v1 \IN `product'.* & `year'.2015 & bar.(y=agg(`sum')) & \\
    f2 & `month' & `sales' & v1 & `year'.2015 & bar.(y=agg(`sum')) &
    v2 \IN $argmax_{v1}[k=10] D(f1,f2)$ \\
    *f3 & `month' & y1 \IN \{`sales', `profit'\} & v2 & `year'.2015 & bar.(y=agg(`sum')) & \\ \hline
  \end{tabular}
}
\vspace{-10pt}
\caption{
  The \zq query which returns the sales over month and profit over month
  visualizations for 2015 for the top 10 products which have the biggest
  discrepancies in their sales and profit trends.
}
\label{tab:ex-2}
\vspace{-10pt}
\end{table*}

This scenario can be addressed with the query in Table~\ref{tab:ex-2}. The Vice
President names the set of visualizations for profit over month {\tt f1} and
the sales over month visualizations {\tt f2}. She then compares the
visualizations in the two set using the $argmax$ and retrieves the top 10
products whose visualizations are the most different. For these visualizations,
she plots both the sales and profit over months; {\tt y1 \IN \{`sales',
`profit'\}} is a shortcut to avoid having to separates rows for sales and
profit. Note that the Vice President was careful to constrain \zq to only look
at the data from 2015.


\later{
\textbf{\emph{Query 3}}:  The Vice-President notices there are several top performing products in the current year which were not among the well performing products last year and vice versa. Thus, he is interested to see top 20 such products whose sales have changed drastically over the previous year.

\begin{table}
\centering
\resizebox{\linewidth}{!}{%
\begin{tabular}{|l|l|l|l|l|} \hline
 \textbf{viz} &  \textbf{x} &  \textbf{y} &  \textbf{z} &  \textbf{Opt} \\ \hline
& 2014 & sales & product.$v_1$ in chairs,... &  \\ 
$*$ & 2013  & sales & product.$v_1$ & $argmax_{v_1}D(f_1, f_2)$ \\ \hline
\end{tabular}%
}
\end{table}

The above query is similar to query 2 except that the two trends of a product differ in x axis  rather than y axis. \\

}

\later{
\textbf{\emph{Query 4}}: The Vice-President draws an increasing trend over month similar to the revenue trend for the last year. He wants to know the attributes among profit, sales and quantity sold that result in the most similar trend to the drawn trend.

\begin{table}
\centering
\resizebox{\linewidth}{!}{%
\begin{tabular}{|l|l|l|l|l|} \hline
 \textbf{viz} &  \textbf{x} &  \textbf{y} &  \textbf{z} &  \textbf{Opt} \\ \hline
& month & revenue & product.staplers &  \\ 
$*$ & month  & $y_1$ & product.staplers & $argmin_{y_1}D(f_1, f_2)$ \\ \hline
\end{tabular}%
}
\end{table}

Here, we find another Y axis attribute which is most correlated with the revenue trend. A similar example for the x axis is to find one of the previous years where the revenue trend was similar to the current year’s trend. \\
}

\later{
\textbf{\emph{Query 5}}: The Vice-President has been informed that US and European markets have been behaving differently over last several months. So, he is interested in knowing the products which have been doing well in USA but performing poorly in Europe.

\begin{table}
\centering
\resizebox{\linewidth}{!}{%
\begin{tabular}{|l|l|l|l|l|} \hline
 \textbf{viz} &  \textbf{x} &  \textbf{y} &  \textbf{z} & \textbf{Opt} \\ \hline
& month & profit & product.$v_1$  &  \\ 
& month & profit & product.$v_2$ and Location=US & $f_2=argmin_{v_2}D(f_1,f_2)$ \\ 
$*$ & month & profit & product.$v_3$ and Location=Europe & $argmax_{v_3} D(f_3, f_1)$ \\ \hline
\end{tabular}%
}
\end{table}

Line 1 represents an input trend which is representative of a well-performing revenue over months trend for any product. In Line 2, we find the products in USA which are most similar to the drawn trend while in line 3 we find the products in Europe which are most dissimilar to the input trend. Finally in line 4, we return the top results which are part of both line 3 and line 4. Another way of doing this could be to take the output of US into input of Europe.\\
}

\paragraph{Query 3.} The Vice President would like to know more about
the differences between a product whose sales numbers do not change over the
year and a
product that has the largest growth in the number of sales. To address this question, she writes
the query in Table~\ref{tab:ex-3}. The first $R$ function call returns the
one product whose sales over year visualization is most representative for all
products; in other words, {\tt v2} is set to the product that has the most
average number of sales. The task in the second row selects the product {\tt
v3} which has the greatest upward trending slope $T$ for sales. Finally, the
Vice President tries to finds the y-axes which distinguish the two products the
most, and visualizes them. Although we know {\tt v2} and {\tt v3} only contain
one value, they are still sets, so $argmax$ must iterate over them and output
corresponding values
{\tt v4} and {\tt v5}.

\begin{table*}
\centering\scriptsize
{\tt
  \begin{tabular}{|r|l|l|l|l|l|} \hline
    \textbf{Name} &  \textbf{X} &  \textbf{Y} &  \textbf{Z} & 
     {\tt Viz} & \textbf{Process} \\ \hline
    f1 & `year' & `sales' & v1 \IN `product'.* & bar.(y=agg(`sum')) & v2 \IN
    $R(1, v1, f1)$ \\
    f2 & `year' & y1 \IN M & v2 & bar.(y=agg(`sum')) & v3 \IN $argmax_{v1}[k=1]
    T(f1)$ \\
    f3 & `year' & y1 & v3 & bar.(y=agg(`sum')) & y2,v4,v5 \IN
    $argmax_{y1,v2,v3}[k=10] D(f2,f3)$ \\
    *f4 & `year' & y2 & v6 \IN (v4 | v5) & bar.(y=agg(`sum')) & \\ \hline
  \end{tabular}
}
\vspace{-10pt}
\caption{
  The \zq query which returns varying y-axes visualizations where the following two products differ the most: one whose
  sales numbers do not change over the year
  and another which has the largest growth in the number of sales.
}
\label{tab:ex-3}
\vspace{-10pt}
\end{table*}

\noindent \textbf{Query 4}:   
Finally, the Vice President wants to see a pair of dimensions 
whose correlation pattern (depicted as a scatterplot) 
is the most unusual, compared to correlation patterns of other pairs of attributes. 
To address this question, she writes the query in Table~\ref{tab:ex-4}. 
She keeps the {\tt Z} column empty as she does not want to slice the data. 
Both {\tt X} and {\tt Y} refer to a set {\tt M} consisting of all the attributes in the dataset
she wishes to explore. 
The task in the second row selects the {\tt X} and {\tt Y} attributes 
whose sum of distances from other visualizations (generated by considering all pairs of attributes)is the maximum.

\begin{table*}
\centering\scriptsize
{\tt
  \begin{tabular}{|r|l|l|l|l|l|} \hline
    \textbf{Name} &  \textbf{X} &  \textbf{Y} &  \textbf{Z} & 
     {\tt Viz} & \textbf{Process} \\ \hline
    f1 & x1 \IN M & y1 \IN M  &  &  &  \\

    f2 & x2 \IN M & y2 \IN M &  &  & x3,y3 \IN $argmax_{x1,y1}[k=1]sum_{x2,y2}D(f1,f2)$ \\

    *f3 & x3 & y3 &  & scatterplot &  \\ \hline
   
  \end{tabular}
}
\vspace{-10pt}
\caption{
  The \zq query which returns scatter plot visualization between a pair of attributes whose pattern is most unusual, i.e very different from the patterns made by any other pair of attributes in M. 
}
\label{tab:ex-4}
\vspace{-10pt}
\end{table*}

%
%
%
%

\later{
\textbf{\emph{Query 7}}:  Further, the Vice President wants to know about two dimensions where the trends of products in Europe and US behave most differently. In other, he wants to figure out two dimensions which can visually separate the two products the most.

\begin{table}[H]
\centering
\resizebox{\linewidth}{!}{%
\begin{tabular}{|l|l|l|l|l|} \hline
 \textbf{viz} &  \textbf{x} &  \textbf{y} &  \textbf{z} & \textbf{Opt} \\ \hline
& x1=... & y1=... & Loc.$v_1=US$  &  \\ 
& x1 & y1 &  Loc.$v_2=Eur$ &  \\ 
$*$ & x1 & y1 & Loc.${v_1,v2}$ & $argmax_{x_1,y_1} D(f_1, f_2)$ \\ \hline
\end{tabular}%
}
\end{table}

Here, we need to find both X and Y axes which can visually separate two subsets of data the most. Line 3 represents the optimisation constraint for finding such pairs of X and Y dimensions.
}

\balance

\section{User Study: Additional details on finding 2}
\label{sec:userstudy-appendix}

\iftechreportcode
Participant P6\textquotesingle s python code implementation for a task where we want to find a pair of X and Y axes where two states 'CA' and 'NY' differ the most:
\begin{verbatim}[language=python]
import pandas
import numpy as np
def generate_maps(date_list, d, Y, Z):
    d = d[d['State']==Z][np.append(date_list, Y)]
    maps = {}
    for id, item in d.iterrows():
        date = ""
        for k in date_list:
            date += str(item[k])
        if date not in maps:
            maps[date] = []
        maps[date].append(item[Y])
    maps = dict([(k, np.mean(v)) for k, v in maps.items()])
    return maps
def filter(d, X, Y, Z):
    '''
    X : Month, Year, Quater
    Y : SoldPrice, ListingPrice, Turnover_rate
    Z : State Name such as CA
    '''
    maps = {}
    if X =='Year':
        date_list = ['Year']
    elif X =='Quater':
        date_list = ['Year', "Quater"]
    elif X =='Month':
        date_list = ['Year', "Quater", "Month"]
    return generate_maps(date_list, d, Y, Z)
def mapping(map1, map2):
    ''' calculate distance'''
    t = 0.0
    for k, v in map1.items():
        t += (map2[k] - v) * (map2[k] -v)
    return t

if __name__=="__main__":
    import matplotlib.pyplot as plt
    import numpy.linalg as LA
    d = pandas.read_csv("./tarique_data")
    XSet = ["Year", "Quater", "Month"]
    YSet = ["SoldPrice", "ListingPrice", "Turnover_rate"]
    result = [(X, Y, mapping(filter(d, X, Y, 'CA'),
	 filter(d, X, Y, 'NY'))) for X in XSet for Y in YSet]
    best_x, best_y, difference = sorted(result, 
	cmp=lambda x, y: -cmp(x[2],y[2]))[0]
    CA, NY = filter(d, best_x, best_y, 'CA'),
	 filter(d, best_x, best_y, "NY")
    xset = CA.keys()
    xset.sort()
    y_CA, y_NY = [CA[x] for x in xset],
	 [NY[x] for x in xset]
    plt.plot(range(len(xset)), y_CA, label='CA')
    plt.plot(range(len(xset)), y_NY, label='NY')
    plt.legend()
    plt.show()
\end{verbatim}
\fi

\fi

\end{document}